\documentclass{aa}
\usepackage{graphicx}
\usepackage{tikz}
\usepackage{pgf}
\usepackage{pgfplots}
\usepgfplotslibrary{external}
\usetikzlibrary{shapes,arrows,shapes.multipart,positioning,fit,calc}
\usepackage{xcolor}
\usepackage{color}

\usepackage{txfonts}
\usepackage{textgreek}
\usepackage{wasysym}
\usepackage[
        colorlinks=true,
        linkcolor=black,
        citecolor=black,
        filecolor=black,
        urlcolor=black]{hyperref}

\usepackage[ugly]{units}
\usepackage[english=american]{csquotes}
\usepackage{longtable}
\usepackage{threeparttable}
\usepackage{booktabs}
\usepackage{subcaption}
\usepackage{blindtext}
\usepackage{natbib}
\bibpunct{(}{)}{;}{a}{}{,}

\begin{document} 

\title{The EXOTIME project: Signals in the $ O-C $ diagrams of the rapidly pulsating subdwarfs DW Lyn, V1636 Ori, QQ Vir, and V541 Hya\thanks{Based on observations obtained at the $ \unit[0.9]{m} $ SARA-KP telescope, which is operated by the Southeastern Association for Research in Astronomy (\url{saraobservatory.org}).}\fnmsep\thanks{Photometric data of Fig.~\ref{fig:lc}, results in Figs.~\ref{fig:dwlyn-oc}, \ref{fig:v1636ori-oc}, \ref{fig:qqvir-oc}, and \ref{fig:v541hya-oc}, and figures in the appendix are available in electronic form at the CDS via anonymous ftp to cdsarc.u-strasbg.fr (130.79.128.5) or via .}}
\titlerunning{EXOTIME: Signals in the $ O-C $-diagrams of rapid pulsating subdwarfs}
\author{F. Mackebrandt \inst{1,2}\fnmsep\thanks{\email{mackebrandt@mps.mpg.de}}
                \and S. Schuh \inst{1}
                \and R. Silvotti \inst{3}
                \and S.-L. Kim \inst{4}
                \and D. Kilkenny \inst{5}
                \and E. M. Green \inst{6}
                \and R. Lutz \inst{7}
                \and T. Nagel \inst{8}
                \and J. L. Provencal \inst{9,10}
                \and T. Otani \inst{11,12}
                \and T. D. Oswalt \inst{11,12}
                \and S. Benatti \inst{13}
                \and L. Lanteri \inst{3}
                \and A. Bonanno \inst{14}
                \and A. Frasca \inst{14}
                \and R. Janulis \inst{15}
                \and M. Papar\'{o} \inst{16}
                \and L. Moln\'ar \inst{16,17}
                \and R. Claudi \inst{13}
                \and R. H. \O stensen\inst{18}
                }
\institute{
        Max-Planck-Institut für Sonnensystemforschung, Justus-von-Liebig-Weg 3, 37077 G\"ottingen, Germany
        \and Institut f\"ur Astrophysik, Georg-August-Universit\"at G\"ottingen, Friedrich-Hund-Platz 1, 37077 G\"ottingen, Germany
        \and INAF -- Osservatorio Astrofisico di Torino, strada dell'Osservatorio 20, 10025 Pino Torinese, Italy
        \and Korea Astronomy and Space Science Institute, Daejeon 34055, South Korea
        \and Department of Physics and Astronomy, University of the Western Cape, Private Bag X17, Bellville 7535, South Africa
        \and Steward Observatory, University of Arizona, 933 N. Cherry Avenue, Tucson, AZ 85721, USA
        \and German Aerospace Center (DLR), Remote Sensing Technology Institute, Münchener Str. 20, 82234 Weßling, Germany
        \and Institute for Astronomy and Astrophysics, Kepler Center for Astro and Particle Physics, University of T\"ubingen, 72076 T\"ubingen, Germany
        \and University of Delaware, Department of Physics and Astronomy Newark, DE 19716, USA
        \and Delaware Asteroseismic Research Center, Mt. Cuba Observatory, Greenville, DE 19807, USA
        \and Embry-Riddle Aeronautical University, Department of Physical Science and SARA, Daytona Beach, FL 32114, USA
        \and Florida Institute of Technology, Department of Physics \& Space Sciences, Melbourne, FL 32901, USA
        \and INAF -- Osservatorio Astronomico di Padova, Vicolo dell’Osservatorio 5, 35122 Padova, Italy
        \and INAF -- Osservatorio Astrofisico di Catania, Via S. Sofia 78, 95123 Catania, Italy\newpage
        \and Institute of Theoretical Physics and Astronomy, Vilnius University, Gostauto 12, Vilnius 01108, Lithuania
        \and Konkoly Observatory, Research Centre for Astronomy and Earth Sciences, Konkoly-Thege M. \'ut 15-17, 1121 Budapest, Hungary
        \and MTA CSFK Lend\"ulet Near-Field Cosmology Research Group, Konkoly Observatory
        \and Department of Physics, Astronomy, and Materials Science, Missouri State University, Springfield, MO 65897, USA
        }
\date{Received ; accepted}
\abstract
{}
{We aim to investigate variations in the arrival time of coherent stellar pulsations due to the light-travel time effect to test for the presence of sub-stellar companions. Those companions are the key to one possible formation scenario of apparently single sub-dwarf B stars.}
{We made
use of an extensive set of ground-based observations of the four large amplitude p-mode pulsators DW Lyn, V1636 Ori, QQ Vir, and V541 Hya. Observations of the TESS space telescope are available on two of the targets. The timing method compares the phase of sinusoidal fits to the full multi-epoch light curves with phases from the fit of a number of subsets of the original time series.}
{Observations of the TESS mission do not sample the pulsations well enough to be useful due to the (currently) fixed two-minute cadence. From the ground-based observations, we infer evolutionary parameters from the arrival times. The residual signals show many statistically significant periodic signals, but no clear evidence for changes in arrival time induced by sub-stellar companions. The signals can be explained partly by mode beating effects. We derive upper limits on companion masses set by the observational campaign.}
{}

\definecolor{cb-red}{HTML}{E41A1C}
\definecolor{cb-blue}{HTML}{377EB8}
\definecolor{cb-green}{HTML}{4DAF4A}
\definecolor{cb-purple}{HTML}{984EA3}
\definecolor{cb-orange}{HTML}{FF7F00}

\maketitle
\section{Introduction}
Subdwarf B stars (sdBs) are sub-luminous stars with a mass of about $ \unit[0.5]{M_{\odot}} $ located at the blue end of the horizontal branch, which is the so-called extreme horizontal branch \citep[EHB,][]{heber_atmosphere_1986}. They maintain a helium burning core, but their thin hydrogen envelope ($ M_{\text{env}} < \unit[0.01]{M_{\odot}}$) cannot sustain hydrogen shell burning, which identifies sdBs with stripped cores of red giants \citep{heber_hot_2016}. Binary evolution with a common envelope (CE) is the favoured formation scenario for most sdBs. The sdB progenitor fills its Roche lobe near the tip of the RGB. A CE is formed when the mass transfer rate is sufficiently high and the companion star cannot accrete all the matter. 
For close binary systems with small initial mass ratios $ q < 1.2 \text{--} 1.5 $, two phases of mass transfer occur. The first Roche-lobe overflow is stable, whereas the second one is unstable, leading to the ejection of the CE. The resulting binary consists of an sdB star and a white dwarf in a short-period orbit. For initial mass ratios $ q > 1.2 \text{--} 1.5, $ the first mass-transfer phase is unstable and the CE is ejected, producing an sdB star with a non-degenerate (e.g. main sequence star) companion. A more detailed review of formation and evolution of compact binary systems can be found in \citet{podsiadlowski_evolution_2008} and \citet{postnov_evolution_2014}. These formation scenarios cannot explain the observed additional occurrence of apparently single sdBs \citep{maxted_binary_2001}.
Among the proposed formation scenarios is the proposal by \citet{webbink_double_1984} that they could be formed by a merger of two helium white dwarfs. But such mergers are problematic, as they are expected to retain very little  hydrogen \citep{han_origin_2002} and be left with higher rotation rates than what has been observed \citep{charpinet_rotation_2018}.
Moreover, the overall observed mass distribution of single sdB stars is not consistent with that expected from the proposed formation scenario. Sub-stellar companions could resolve this disagreement between theory and observations. Planetary-mass companions like the candidates V391 Peg\,b \citep{silvotti_giant_2007}, KIC 05807616\,b,c \citep{charpinet_compact_2011}, KIC 10001893\,b,c,d \citep{silvotti_kepler_2014}, or brown dwarf companions like V2008-1753\,B \citep{schaffenroth_eclipsing_2015} or CS 1246 \citep{barlow_fortnightly_2011} indicate the existence of a previously undiscovered population of companions to apparently single sdBs.

Due to the high surface gravity and effective temperature (leading to few, strongly broadened spectral lines in the optical) and the small radii of sdBs, the detection efficiency for companions via methods like radial velocity variations or transits is small. The timing of stellar pulsations offers a complementary detection method, sensitive to large orbital separations. 

A small fraction of sdB stars shows pulsational variations in the p- (pressure-) and g- (gravity-) mode regimes.
Rapid p-mode pulsators (sdBV\textsubscript{r}), discovered by \citet{kilkenny_new_1997}, show periods of the order of minutes and amplitudes of a few tens of $ \unit{mmag} $. Such pulsations were predicted by \citet[][et seq.]{charpinet_driving_1997} to be driven by the \textkappa-mechanism due to a $ Z $-opacity bump. For slow pulsators (sdBV\textsubscript{s}) the periods range from 30 to 80 $ \unit{min} $ with small amplitudes of a few $ \unit{mmag} $. This class was  discovered by \citet{green_discovery_2003} and the pulsations are explained by the \textkappa-mechanism as well \citep{fontaine_driving_2003}. Some sdB stars show both types of pulsation modes simultaneously (sdBV\textsubscript{rs}). These hybrid pulsators lie at the temperature boundary near $ \unit[28000]{K} $ between the two classes of pulsating stars, for example, the prototype for this class DW~Lyn \citep{schuh_hs_2006}, which is also addressed in this work, or Balloon~090100001 \citep{baran_multicolour_2005}.

Pulsations driven by the \textkappa-mechanism are coherent, which qualifies these objects for the timing method to search for sub-stellar companions. This method is based on the light-travel time effect, with the host star acting as a stable 'clock' Spatial movements of the star around the barycentre induced by a companion result in time delays of the stellar light measured by the observer. Examples of detections using this method are 'pulsar planets' \citep[e.g.][]{wolszczan_planetary_1992}, planets detected by transit timing variations \citep[e.g. Kepler 19\,c,][]{ballard_kepler-19_2011}, planets orbiting \textdelta Scuti stars \citep{murphy_planet_2016}, or eclipsing binaries \citep[e.g. V2051 Oph (AB)\,b,][]{qian_long-term_2015}. In particular, the detection of a late-type main sequence star companion to the sdB CS 1246 by \citet{barlow_fortnightly_2011}, subsequently confirmed with radial velocity data \citep{barlow_radial_2011}, or other studies like \citet{otani_orbital_2018}, demonstrate the viability of this method in sdB systems. On the other hand, the particular example of V391 Peg\,b is currently under discussion \citep{silvotti_sdb_2018} because of possible non-linear interactions between different pulsation modes that change arrival times
(see \citealt{zong_oscillation_2018} for a detailed study of amplitude/frequency variations related to non-linear effects). Stochastically driven pulsations, are suspected by \citet{reed_followup_2007,kilkenny_amplitude_2010} and their nature confirmed by \citet{ostensen_stochastic_2014}. 
Also, the candidate detections of KIC 05807616 and KIC 10001893 are uncertain, since other sdBs observed within the \textit{Kepler K2} mission exhibit $ g $-modes with long periods up to a few hours. They question the interpretation of the low-frequency variations for KIC 05807616 and KIC 10001893 \citep[e.g.][]{krzesinski_planetary_2015,blokesz_analysis_2019}.

In order to detect sub-stellar companions orbiting rapidly pulsating sdB stars, the \textit{EXOTIME} observational programme (EXOplanet search with the TIming MEthod) has been taking long-term data since 1999. \textit{EXOTIME} conducted a long-term monitoring programme of five rapidly pulsating sdB stars. V391 Peg has been discussed by \citet{silvotti_giant_2007,silvotti_sdb_2018}. In this paper, we present the observations of DW Lyn and V1636 Ori, previously discussed in \citet{lutz_light_2008,schuh_exotime_2010,lutz_search_2011,lutz_exotime_2011}, and re-evaluate their findings using an   extended set of observations. In addition, the observations of QQ Vir and V541 Hya are presented and analysed. In the beginning of the programme, the mode stability was tested for all targets over a timespan of months in order to ensure the pulsation modes were\ coherent.

For the DW Lyn observations, \citet{lutz_exotime_2011} found no significant signals in a periodogram of the $ O-C $ data of the two analysed pulsation frequencies, which would indicate sub-stellar companions. A tentative signal in the second frequency (in this work labelled $ f_2 $, as well), formally corresponding to an 80-day companion orbit, is concluded to arise from mode beating of an unresolved frequency doublet. The analysis of V1636 Ori revealed a signal at $ \unit[160]{d} $ in the periodogram of the main frequency $ O-C $ data \citep{lutz_exotime_2011}. Although this periodicity showed a significance of only $ \unit[1]{\sigma} $, \citet{lutz_exotime_2011} predicted an increase of significance with follow-up observations. We are using this extended data set in our work, now incorporating observations up to 2015.


This paper is organised as follows. Section~\ref{sec:observations} describes the observational aspects within the \textit{EXOTIME} programme and the data reduction, followed by a description of our analysis in Section~\ref{sec:analysis}. Our results are presented in Section~\ref{sec:results}, together with a discussion.


\section{Observations and data reduction} \label{sec:observations}
The observational data necessary for the analysis are comprised of many individual data sets gathered over the course of up to two decades. The detection method demands the observation of a target for a total time base at least as long as one orbit of a potential companion, which can span several years. This requires coordinated campaigns with observatories using \textasciitilde 1 to 4 $ \unit{m} $ telescopes.
In order to derive sufficient accuracy for the analysis, observations with at least three to four consecutive nights, each with a minimum of two to three hours per target are required. To resolve the short-period p-modes the cadence must be shorter than about $ \unit[30]{s} $ but still with a sufficient signal-to-noise ratio (S/N). All observations used the Johnson-Bessel \textit{B} band. The correct time stamps for each observation are of most importance for the timing analysis. Most observatories of this study already successfully contributed to the work of \citet[][Table 2]{silvotti_giant_2007,silvotti_subdwarf_2008}. The following list features some references where telescopes used for this study contributed successfully to other timing-relevant observations.
Konkoly RCC $ \unit[1.0]{m} $ Telescope: \citet{provencal_2006_2009,stello_multisite_2006};
Mt. Lemmon Optical Astronomy Observatory: \citet{bischoff-kim_gd358_2019,lee_pulsating_2014};
Serra la Nave $ \unit[0.9]{m} $: \citet{bonanno_pg_2003,bonanno_asteroseismology_2003};
SARA-KP $ \unit[0.9]{m} $ telescope: \citet{kilkenny_orbital_2014,baran_pulsations_2018}.
        
\begin{figure*}[tp]
        \centering
        \begin{subfigure}[b]{0.49\textwidth}
                \centering
\begingroup
  \makeatletter
  \providecommand\color[2][]{%
    \GenericError{(gnuplot) \space\space\space\@spaces}{%
      Package color not loaded in conjunction with
      terminal option `colourtext'%
    }{See the gnuplot documentation for explanation.%
    }{Either use 'blacktext' in gnuplot or load the package
      color.sty in LaTeX.}%
    \renewcommand\color[2][]{}%
  }%
  \providecommand\includegraphics[2][]{%
    \GenericError{(gnuplot) \space\space\space\@spaces}{%
      Package graphicx or graphics not loaded%
    }{See the gnuplot documentation for explanation.%
    }{The gnuplot epslatex terminal needs graphicx.sty or graphics.sty.}
  }%
  \providecommand\rotatebox[2]{#2}%
  \@ifundefined{ifGPcolor}{%
    \newif\ifGPcolor
    \GPcolortrue
  }{}%
  \@ifundefined{ifGPblacktext}{%
    \newif\ifGPblacktext
    \GPblacktexttrue
  }{}%
  \let\gplgaddtomacro\g@addto@macro
  \gdef\gplbacktext{}%
  \gdef\gplfronttext{}%
  \makeatother
  \ifGPblacktext
    \def\colorrgb#1{}%
    \def\colorgray#1{}%
  \else
    \ifGPcolor
      \def\colorrgb#1{\color[rgb]{#1}}%
      \def\colorgray#1{\color[gray]{#1}}%
      \expandafter\def\csname LTw\endcsname{\color{white}}%
      \expandafter\def\csname LTb\endcsname{\color{black}}%
      \expandafter\def\csname LTa\endcsname{\color{black}}%
      \expandafter\def\csname LT0\endcsname{\color[rgb]{1,0,0}}%
      \expandafter\def\csname LT1\endcsname{\color[rgb]{0,1,0}}%
      \expandafter\def\csname LT2\endcsname{\color[rgb]{0,0,1}}%
      \expandafter\def\csname LT3\endcsname{\color[rgb]{1,0,1}}%
      \expandafter\def\csname LT4\endcsname{\color[rgb]{0,1,1}}%
      \expandafter\def\csname LT5\endcsname{\color[rgb]{1,1,0}}%
      \expandafter\def\csname LT6\endcsname{\color[rgb]{0,0,0}}%
      \expandafter\def\csname LT7\endcsname{\color[rgb]{1,0.3,0}}%
      \expandafter\def\csname LT8\endcsname{\color[rgb]{0.5,0.5,0.5}}%
    \else
      \def\colorrgb#1{\color{black}}%
      \def\colorgray#1{\color[gray]{#1}}%
      \expandafter\def\csname LTw\endcsname{\color{white}}%
      \expandafter\def\csname LTb\endcsname{\color{black}}%
      \expandafter\def\csname LTa\endcsname{\color{black}}%
      \expandafter\def\csname LT0\endcsname{\color{black}}%
      \expandafter\def\csname LT1\endcsname{\color{black}}%
      \expandafter\def\csname LT2\endcsname{\color{black}}%
      \expandafter\def\csname LT3\endcsname{\color{black}}%
      \expandafter\def\csname LT4\endcsname{\color{black}}%
      \expandafter\def\csname LT5\endcsname{\color{black}}%
      \expandafter\def\csname LT6\endcsname{\color{black}}%
      \expandafter\def\csname LT7\endcsname{\color{black}}%
      \expandafter\def\csname LT8\endcsname{\color{black}}%
    \fi
  \fi
    \setlength{\unitlength}{0.0500bp}%
    \ifx\gptboxheight\undefined%
      \newlength{\gptboxheight}%
      \newlength{\gptboxwidth}%
      \newsavebox{\gptboxtext}%
    \fi%
    \setlength{\fboxrule}{0.5pt}%
    \setlength{\fboxsep}{1pt}%
\begin{picture}(5102.00,3152.00)%
    \gplgaddtomacro\gplbacktext{%
      \csname LTb\endcsname%
      \put(682,955){\makebox(0,0)[r]{\strut{}0.8}}%
      \put(682,1331){\makebox(0,0)[r]{\strut{}0.9}}%
      \put(682,1708){\makebox(0,0)[r]{\strut{}1.0}}%
      \put(682,2084){\makebox(0,0)[r]{\strut{}1.1}}%
      \put(682,2460){\makebox(0,0)[r]{\strut{}1.2}}%
      \put(1376,484){\makebox(0,0){\strut{}$54500$}}%
      \put(2208,484){\makebox(0,0){\strut{}$55000$}}%
      \put(3041,484){\makebox(0,0){\strut{}$55500$}}%
      \put(3873,484){\makebox(0,0){\strut{}$56000$}}%
      \put(4705,484){\makebox(0,0){\strut{}$56500$}}%
      \put(1321,2931){\makebox(0,0){\strut{}2008}}%
      \put(2538,2931){\makebox(0,0){\strut{}2010}}%
      \put(3753,2931){\makebox(0,0){\strut{}2012}}%
    }%
    \gplgaddtomacro\gplfronttext{%
      \csname LTb\endcsname%
      \put(176,1707){\rotatebox{-270}{\makebox(0,0){\strut{}relative flux $F$}}}%
      \put(2791,154){\makebox(0,0){\strut{}$\unit[t/]{MBJD_{TDB}}$}}%
    }%
    \gplbacktext
    \put(0,0){\includegraphics[scale=0.5]{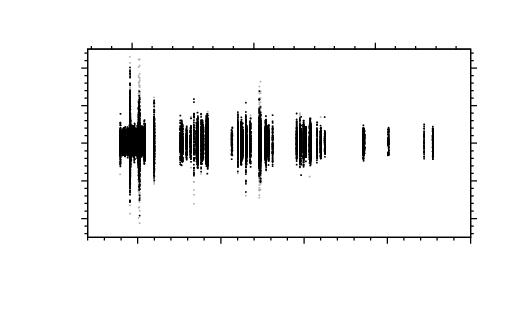}}%
    \gplfronttext
  \end{picture}%
\endgroup


                \caption{DW Lyn.}
                \label{fig:dwlyn-lc}
        \end{subfigure}
        \hfil
        \begin{subfigure}[b]{0.49\textwidth}
                \centering
\begingroup
  \makeatletter
  \providecommand\color[2][]{%
    \GenericError{(gnuplot) \space\space\space\@spaces}{%
      Package color not loaded in conjunction with
      terminal option `colourtext'%
    }{See the gnuplot documentation for explanation.%
    }{Either use 'blacktext' in gnuplot or load the package
      color.sty in LaTeX.}%
    \renewcommand\color[2][]{}%
  }%
  \providecommand\includegraphics[2][]{%
    \GenericError{(gnuplot) \space\space\space\@spaces}{%
      Package graphicx or graphics not loaded%
    }{See the gnuplot documentation for explanation.%
    }{The gnuplot epslatex terminal needs graphicx.sty or graphics.sty.}
  }%
  \providecommand\rotatebox[2]{#2}%
  \@ifundefined{ifGPcolor}{%
    \newif\ifGPcolor
    \GPcolortrue
  }{}%
  \@ifundefined{ifGPblacktext}{%
    \newif\ifGPblacktext
    \GPblacktexttrue
  }{}%
  \let\gplgaddtomacro\g@addto@macro
  \gdef\gplbacktext{}%
  \gdef\gplfronttext{}%
  \makeatother
  \ifGPblacktext
    \def\colorrgb#1{}%
    \def\colorgray#1{}%
  \else
    \ifGPcolor
      \def\colorrgb#1{\color[rgb]{#1}}%
      \def\colorgray#1{\color[gray]{#1}}%
      \expandafter\def\csname LTw\endcsname{\color{white}}%
      \expandafter\def\csname LTb\endcsname{\color{black}}%
      \expandafter\def\csname LTa\endcsname{\color{black}}%
      \expandafter\def\csname LT0\endcsname{\color[rgb]{1,0,0}}%
      \expandafter\def\csname LT1\endcsname{\color[rgb]{0,1,0}}%
      \expandafter\def\csname LT2\endcsname{\color[rgb]{0,0,1}}%
      \expandafter\def\csname LT3\endcsname{\color[rgb]{1,0,1}}%
      \expandafter\def\csname LT4\endcsname{\color[rgb]{0,1,1}}%
      \expandafter\def\csname LT5\endcsname{\color[rgb]{1,1,0}}%
      \expandafter\def\csname LT6\endcsname{\color[rgb]{0,0,0}}%
      \expandafter\def\csname LT7\endcsname{\color[rgb]{1,0.3,0}}%
      \expandafter\def\csname LT8\endcsname{\color[rgb]{0.5,0.5,0.5}}%
    \else
      \def\colorrgb#1{\color{black}}%
      \def\colorgray#1{\color[gray]{#1}}%
      \expandafter\def\csname LTw\endcsname{\color{white}}%
      \expandafter\def\csname LTb\endcsname{\color{black}}%
      \expandafter\def\csname LTa\endcsname{\color{black}}%
      \expandafter\def\csname LT0\endcsname{\color{black}}%
      \expandafter\def\csname LT1\endcsname{\color{black}}%
      \expandafter\def\csname LT2\endcsname{\color{black}}%
      \expandafter\def\csname LT3\endcsname{\color{black}}%
      \expandafter\def\csname LT4\endcsname{\color{black}}%
      \expandafter\def\csname LT5\endcsname{\color{black}}%
      \expandafter\def\csname LT6\endcsname{\color{black}}%
      \expandafter\def\csname LT7\endcsname{\color{black}}%
      \expandafter\def\csname LT8\endcsname{\color{black}}%
    \fi
  \fi
    \setlength{\unitlength}{0.0500bp}%
    \ifx\gptboxheight\undefined%
      \newlength{\gptboxheight}%
      \newlength{\gptboxwidth}%
      \newsavebox{\gptboxtext}%
    \fi%
    \setlength{\fboxrule}{0.5pt}%
    \setlength{\fboxsep}{1pt}%
\begin{picture}(5102.00,3152.00)%
    \gplgaddtomacro\gplbacktext{%
      \csname LTb\endcsname%
      \put(682,1081){\makebox(0,0)[r]{\strut{}0.9}}%
      \put(682,1708){\makebox(0,0)[r]{\strut{}1.0}}%
      \put(682,2335){\makebox(0,0)[r]{\strut{}1.1}}%
      \put(877,484){\makebox(0,0){\strut{}$54500$}}%
      \put(1586,484){\makebox(0,0){\strut{}$55000$}}%
      \put(2295,484){\makebox(0,0){\strut{}$55500$}}%
      \put(3004,484){\makebox(0,0){\strut{}$56000$}}%
      \put(3713,484){\makebox(0,0){\strut{}$56500$}}%
      \put(4421,484){\makebox(0,0){\strut{}$57000$}}%
      \put(1866,2931){\makebox(0,0){\strut{}2010}}%
      \put(2901,2931){\makebox(0,0){\strut{}2012}}%
      \put(3938,2931){\makebox(0,0){\strut{}2014}}%
    }%
    \gplgaddtomacro\gplfronttext{%
      \csname LTb\endcsname%
      \put(176,1707){\rotatebox{-270}{\makebox(0,0){\strut{}relative flux $F$}}}%
      \put(2791,154){\makebox(0,0){\strut{}$\unit[t/]{MBJD_{TDB}}$}}%
    }%
    \gplbacktext
    \put(0,0){\includegraphics[scale=0.5]{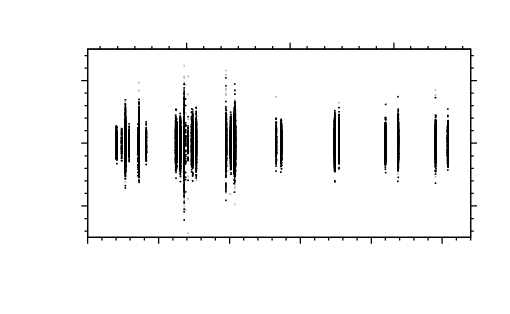}}%
    \gplfronttext
  \end{picture}%
\endgroup


                \caption{V1636 Ori.}
                \label{fig:v1636ori-lc}
        \end{subfigure}
        \hfil
        \begin{subfigure}[b]{0.49\textwidth}
                \centering
\begingroup
  \makeatletter
  \providecommand\color[2][]{%
    \GenericError{(gnuplot) \space\space\space\@spaces}{%
      Package color not loaded in conjunction with
      terminal option `colourtext'%
    }{See the gnuplot documentation for explanation.%
    }{Either use 'blacktext' in gnuplot or load the package
      color.sty in LaTeX.}%
    \renewcommand\color[2][]{}%
  }%
  \providecommand\includegraphics[2][]{%
    \GenericError{(gnuplot) \space\space\space\@spaces}{%
      Package graphicx or graphics not loaded%
    }{See the gnuplot documentation for explanation.%
    }{The gnuplot epslatex terminal needs graphicx.sty or graphics.sty.}
  }%
  \providecommand\rotatebox[2]{#2}%
  \@ifundefined{ifGPcolor}{%
    \newif\ifGPcolor
    \GPcolortrue
  }{}%
  \@ifundefined{ifGPblacktext}{%
    \newif\ifGPblacktext
    \GPblacktexttrue
  }{}%
  \let\gplgaddtomacro\g@addto@macro
  \gdef\gplbacktext{}%
  \gdef\gplfronttext{}%
  \makeatother
  \ifGPblacktext
    \def\colorrgb#1{}%
    \def\colorgray#1{}%
  \else
    \ifGPcolor
      \def\colorrgb#1{\color[rgb]{#1}}%
      \def\colorgray#1{\color[gray]{#1}}%
      \expandafter\def\csname LTw\endcsname{\color{white}}%
      \expandafter\def\csname LTb\endcsname{\color{black}}%
      \expandafter\def\csname LTa\endcsname{\color{black}}%
      \expandafter\def\csname LT0\endcsname{\color[rgb]{1,0,0}}%
      \expandafter\def\csname LT1\endcsname{\color[rgb]{0,1,0}}%
      \expandafter\def\csname LT2\endcsname{\color[rgb]{0,0,1}}%
      \expandafter\def\csname LT3\endcsname{\color[rgb]{1,0,1}}%
      \expandafter\def\csname LT4\endcsname{\color[rgb]{0,1,1}}%
      \expandafter\def\csname LT5\endcsname{\color[rgb]{1,1,0}}%
      \expandafter\def\csname LT6\endcsname{\color[rgb]{0,0,0}}%
      \expandafter\def\csname LT7\endcsname{\color[rgb]{1,0.3,0}}%
      \expandafter\def\csname LT8\endcsname{\color[rgb]{0.5,0.5,0.5}}%
    \else
      \def\colorrgb#1{\color{black}}%
      \def\colorgray#1{\color[gray]{#1}}%
      \expandafter\def\csname LTw\endcsname{\color{white}}%
      \expandafter\def\csname LTb\endcsname{\color{black}}%
      \expandafter\def\csname LTa\endcsname{\color{black}}%
      \expandafter\def\csname LT0\endcsname{\color{black}}%
      \expandafter\def\csname LT1\endcsname{\color{black}}%
      \expandafter\def\csname LT2\endcsname{\color{black}}%
      \expandafter\def\csname LT3\endcsname{\color{black}}%
      \expandafter\def\csname LT4\endcsname{\color{black}}%
      \expandafter\def\csname LT5\endcsname{\color{black}}%
      \expandafter\def\csname LT6\endcsname{\color{black}}%
      \expandafter\def\csname LT7\endcsname{\color{black}}%
      \expandafter\def\csname LT8\endcsname{\color{black}}%
    \fi
  \fi
    \setlength{\unitlength}{0.0500bp}%
    \ifx\gptboxheight\undefined%
      \newlength{\gptboxheight}%
      \newlength{\gptboxwidth}%
      \newsavebox{\gptboxtext}%
    \fi%
    \setlength{\fboxrule}{0.5pt}%
    \setlength{\fboxsep}{1pt}%
\begin{picture}(5102.00,3152.00)%
    \gplgaddtomacro\gplbacktext{%
      \csname LTb\endcsname%
      \put(814,955){\makebox(0,0)[r]{\strut{}0.80}}%
      \put(814,1331){\makebox(0,0)[r]{\strut{}0.90}}%
      \put(814,1708){\makebox(0,0)[r]{\strut{}1.00}}%
      \put(814,2084){\makebox(0,0)[r]{\strut{}1.10}}%
      \put(814,2460){\makebox(0,0)[r]{\strut{}1.20}}%
      \put(1009,484){\makebox(0,0){\strut{}$52000$}}%
      \put(1779,484){\makebox(0,0){\strut{}$53000$}}%
      \put(2549,484){\makebox(0,0){\strut{}$54000$}}%
      \put(3319,484){\makebox(0,0){\strut{}$55000$}}%
      \put(4089,484){\makebox(0,0){\strut{}$56000$}}%
      \put(1221,2931){\makebox(0,0){\strut{}2002}}%
      \put(1783,2931){\makebox(0,0){\strut{}2004}}%
      \put(2346,2931){\makebox(0,0){\strut{}2006}}%
      \put(2908,2931){\makebox(0,0){\strut{}2008}}%
      \put(3471,2931){\makebox(0,0){\strut{}2010}}%
      \put(4033,2931){\makebox(0,0){\strut{}2012}}%
      \put(4596,2931){\makebox(0,0){\strut{}2014}}%
    }%
    \gplgaddtomacro\gplfronttext{%
      \csname LTb\endcsname%
      \put(176,1707){\rotatebox{-270}{\makebox(0,0){\strut{}relative flux $F$}}}%
      \put(2857,154){\makebox(0,0){\strut{}$\unit[t/]{MBJD_{TDB}}$}}%
    }%
    \gplbacktext
    \put(0,0){\includegraphics[scale=0.5]{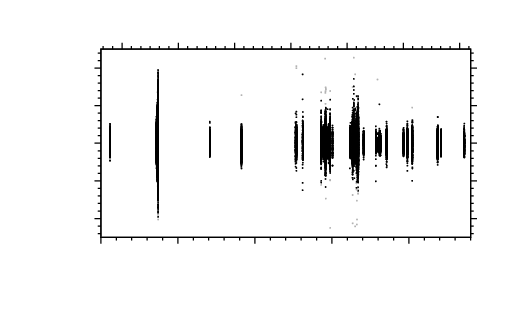}}%
    \gplfronttext
  \end{picture}%
\endgroup


                \caption{QQ Vir.}
                \label{fig:qqvir-lc}
        \end{subfigure}
        \hfil
        \begin{subfigure}[b]{0.49\textwidth}
                \centering
\begingroup
  \makeatletter
  \providecommand\color[2][]{%
    \GenericError{(gnuplot) \space\space\space\@spaces}{%
      Package color not loaded in conjunction with
      terminal option `colourtext'%
    }{See the gnuplot documentation for explanation.%
    }{Either use 'blacktext' in gnuplot or load the package
      color.sty in LaTeX.}%
    \renewcommand\color[2][]{}%
  }%
  \providecommand\includegraphics[2][]{%
    \GenericError{(gnuplot) \space\space\space\@spaces}{%
      Package graphicx or graphics not loaded%
    }{See the gnuplot documentation for explanation.%
    }{The gnuplot epslatex terminal needs graphicx.sty or graphics.sty.}
  }%
  \providecommand\rotatebox[2]{#2}%
  \@ifundefined{ifGPcolor}{%
    \newif\ifGPcolor
    \GPcolortrue
  }{}%
  \@ifundefined{ifGPblacktext}{%
    \newif\ifGPblacktext
    \GPblacktexttrue
  }{}%
  \let\gplgaddtomacro\g@addto@macro
  \gdef\gplbacktext{}%
  \gdef\gplfronttext{}%
  \makeatother
  \ifGPblacktext
    \def\colorrgb#1{}%
    \def\colorgray#1{}%
  \else
    \ifGPcolor
      \def\colorrgb#1{\color[rgb]{#1}}%
      \def\colorgray#1{\color[gray]{#1}}%
      \expandafter\def\csname LTw\endcsname{\color{white}}%
      \expandafter\def\csname LTb\endcsname{\color{black}}%
      \expandafter\def\csname LTa\endcsname{\color{black}}%
      \expandafter\def\csname LT0\endcsname{\color[rgb]{1,0,0}}%
      \expandafter\def\csname LT1\endcsname{\color[rgb]{0,1,0}}%
      \expandafter\def\csname LT2\endcsname{\color[rgb]{0,0,1}}%
      \expandafter\def\csname LT3\endcsname{\color[rgb]{1,0,1}}%
      \expandafter\def\csname LT4\endcsname{\color[rgb]{0,1,1}}%
      \expandafter\def\csname LT5\endcsname{\color[rgb]{1,1,0}}%
      \expandafter\def\csname LT6\endcsname{\color[rgb]{0,0,0}}%
      \expandafter\def\csname LT7\endcsname{\color[rgb]{1,0.3,0}}%
      \expandafter\def\csname LT8\endcsname{\color[rgb]{0.5,0.5,0.5}}%
    \else
      \def\colorrgb#1{\color{black}}%
      \def\colorgray#1{\color[gray]{#1}}%
      \expandafter\def\csname LTw\endcsname{\color{white}}%
      \expandafter\def\csname LTb\endcsname{\color{black}}%
      \expandafter\def\csname LTa\endcsname{\color{black}}%
      \expandafter\def\csname LT0\endcsname{\color{black}}%
      \expandafter\def\csname LT1\endcsname{\color{black}}%
      \expandafter\def\csname LT2\endcsname{\color{black}}%
      \expandafter\def\csname LT3\endcsname{\color{black}}%
      \expandafter\def\csname LT4\endcsname{\color{black}}%
      \expandafter\def\csname LT5\endcsname{\color{black}}%
      \expandafter\def\csname LT6\endcsname{\color{black}}%
      \expandafter\def\csname LT7\endcsname{\color{black}}%
      \expandafter\def\csname LT8\endcsname{\color{black}}%
    \fi
  \fi
    \setlength{\unitlength}{0.0500bp}%
    \ifx\gptboxheight\undefined%
      \newlength{\gptboxheight}%
      \newlength{\gptboxwidth}%
      \newsavebox{\gptboxtext}%
    \fi%
    \setlength{\fboxrule}{0.5pt}%
    \setlength{\fboxsep}{1pt}%
\begin{picture}(5102.00,3152.00)%
    \gplgaddtomacro\gplbacktext{%
      \csname LTb\endcsname%
      \put(682,767){\makebox(0,0)[r]{\strut{}0.9}}%
      \put(682,1707){\makebox(0,0)[r]{\strut{}1.0}}%
      \put(682,2648){\makebox(0,0)[r]{\strut{}1.1}}%
      \put(877,484){\makebox(0,0){\strut{}$53000$}}%
      \put(1728,484){\makebox(0,0){\strut{}$54000$}}%
      \put(2578,484){\makebox(0,0){\strut{}$55000$}}%
      \put(3429,484){\makebox(0,0){\strut{}$56000$}}%
      \put(4280,484){\makebox(0,0){\strut{}$57000$}}%
      \put(882,2931){\makebox(0,0){\strut{}2004}}%
      \put(1504,2931){\makebox(0,0){\strut{}2006}}%
      \put(2125,2931){\makebox(0,0){\strut{}2008}}%
      \put(2747,2931){\makebox(0,0){\strut{}2010}}%
      \put(3368,2931){\makebox(0,0){\strut{}2012}}%
      \put(3989,2931){\makebox(0,0){\strut{}2014}}%
      \put(4610,2931){\makebox(0,0){\strut{}2016}}%
    }%
    \gplgaddtomacro\gplfronttext{%
      \csname LTb\endcsname%
      \put(176,1707){\rotatebox{-270}{\makebox(0,0){\strut{}relative flux $F$}}}%
      \put(2791,154){\makebox(0,0){\strut{}$\unit[t/]{MBJD_{TDB}}$}}%
    }%
    \gplbacktext
    \put(0,0){\includegraphics[scale=0.5]{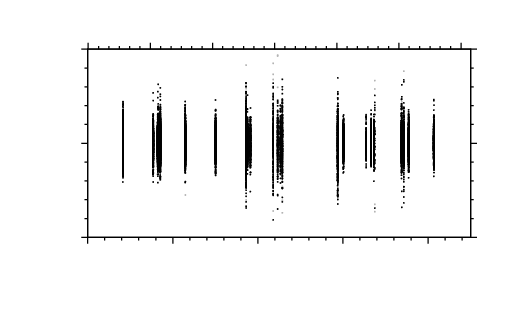}}%
    \gplfronttext
  \end{picture}%
\endgroup


                \caption{V541 Hya.}
                \label{fig:v541hya-lc}
        \end{subfigure}
        \caption{Light curves. Grey points are considered outliers and partially exceed the plotting range.}
        \label{fig:lc}
\end{figure*}

Table~\ref{tab:stellar} lists the atmospheric parameters of the stars, and Table~\ref{tab:photometric} summarises the photometric observations obtained at multiple medium-class telescopes. Fig.~\ref{fig:lc} summarises the observational coverage.
\begin{table}[tp]
        \centering
        \caption{Atmospheric parameters of the targets.}
        \label{tab:stellar}
        \begin{tabular}{lr@{$ \pm $}lr@{$ \pm $}lr@{$ \pm $}lc}
                \toprule
                Target & \multicolumn{2}{c}{$ \unit[T_{\text{eff}}/]{K} $} & \multicolumn{2}{c}{$ \log \left( g/\unitfrac{cm}{s} \right) $} & \multicolumn{2}{c}{$ \log \left( \frac{N \left(\element{He}\right)}{N \left(\element{H}\right)} \right) $} & Ref. \\
                \midrule
                DW Lyn & 28\,400 & 600 & 5.35 & 0.1 & -2.7 & 0.1 & 1 \\
                V1636 Ori & 33\,800 & 1\,000 & 5.60 & 0.15 & -1.85 & 0.20 & 2 \\
                QQ Vir & 34\,800 & 610 & 5.81 & 0.05 & -1.65 & 0.05 & 3 \\
                V541 Hya & 34\,806 & 230 & 5.794 & 0.044 & -1.680 & 0.056 & 4 \\
                \bottomrule
        \end{tabular}
        \tablebib{
                (1)~\citet{dreizler_hs0702+6043_2002}; (2)~\citet{ostensen_four_2001} ; (3) \citet{telting_radial-velocity_2004}; (4)~\citet{randall_observations_2009}
        }
\end{table}

\begin{table*}[tp!]
        \centering
        \caption{Summary of the observing time per target, per site in hours. Detailed tables, including observing dates and times per observatory are available online at the CDS, as are tables listing the allocation into the epochs.}
        \label{tab:photometric}
        \begin{tabular}{lrrrr}
                \toprule
                Site & DW Lyn & V1636 Ori & QQ Vir & V541 Hya \\
                \midrule
                Asiago $ \unit[1.8]{m} $ Copernico Telescope (Asi) & 20.25 &  &  &  \\
                Calar Alto Observatory $ \unit[2.2]{m} $ (CAHA) & 52.38 & 32.49 & 48.73 & 10.19 \\
                Baker $ \unit[0.4]{m} $ &  &  & 41.10 &  \\
                BAO $ \unit[0.85]{m} $ &  &  & 47.70 &  \\
                BOAO $ \unit[0.85]{m} $ &  &  & 25.60 &  \\
                G\"ottingen IAG $ \unit[0.5]{m} $ Telescope (Goe) & 52.53 &  &  &  \\
                Konkoly RCC $ \unit[1.0]{m} $ Telescope (Kon) & 14.27 &  & 4.76 &  \\
                La Palma $ \unit[0.6]{m} $ &  &  & 37.30 &  \\
                Mt. Lemmon Optical Astronomy Observatory $ \unit[1.0]{m} $ (LOAO) & 167.76 & 40.12 & 126.11 & 24.15 \\
                Loiano $ \unit[1.5]{m} $ Telescope (Loi) &  & 2.66 & 78.40 &  \\
                Lulin Observatory $ \unit[1]{m} $ Telescope (Lul) & 9.30 &  &  &  \\
                Moletai $ \unit[1.6]{m} $ Telescope (Mol) & 13.47 &  & 2.95 &  \\
                MONET/North Telescope $ \unit[1.2]{m} $ (M/N) & 138.90 & 41.29 & 34.24 &  \\
                Mt. Bigelow Kuiper Telescope $ \unit[1.5]{m} $ (MtB) & 440.85 &  &  &  \\
                Nordic Optical Telescope $ \unit[2.5]{m} $ (NOT) & &  & 3.86 &  \\
                SARA-KP $ \unit[0.9]{m} $ telescope  & 66.26 &  & 1.80 &  \\
                Serra la Nave $ \unit[0.9]{m} $ &  &  & 26.40 &  \\
                South African Astronomical Observatory $ \unit[1]{m} $ (SAAO) &  & 64.04 & 36.93 & 166.31 \\
                Steward Observatory Bok Telescope $ \unit[2.2]{m} $ (StB) & 12.00 &  &  &  \\
                Telescopio Nazionale Galileo $ \unit[3.6]{m} $ (TNG) &  & 7.00 & 3.37 &  \\
                T\"ubingen $ \unit[0.8]{m} $ Telescope (Tue) & 23.05 &  &  &  \\
                Whole Earth Telescope (WET) & & & 40.00 & \\
                Wise $ \unit[1]{m} $ &  &  & 9.00 &  \\
                \midrule
                $ \Sigma $ & 998.21 & 187.80 & 568.25 & 200.67 \\
                \bottomrule
        \end{tabular}
        \tablefoot{Observations at Baker Observatory, Mt. Bigelow Kuiper Telescope, Nordic Optical Telescope, Steward Observatory Bok Telescope were initially collected for other project(s) but also used for this work.}
\end{table*}

\subsection{DW Lyn}
\citet{dreizler_hs0702+6043_2002} identified DW Lyn (\object{HS 0702+6043}) as a p-mode pulsator. \citet{schuh_hs_2006} discovered additional g-mode pulsations making this star the prototype of hybrid sdB pulsators.

There are photometric data available from 1999. Large gaps make a consistent $ O-C $ analysis difficult. Regular monitoring within the \textit{EXOTIME} programme ran from 2007 until the beginning of 2010. Further observations cover a period up to the end of 2010. These multi-site observations are described in \citet{lutz_light_2008,lutz_long-term_2008,lutz_exotime_2011}. Here, we add observations made with the SARA-KP $ \unit[0.9]{m} $ telescope at Kitt Peak National Observatory in Arizona
, that used exposure times of $ \unit[30]{s} $.

\subsection{V1636 Ori}\label{sec:obs-v1636ori}
\citet{ostensen_four_2001} discovered V1636 Ori (\object{HS 0444+0458}) as a pulsating sdB star. \citet{reed_resolving_2007} conducted a frequency analysis, reporting one small and two large amplitude p-modes.

V1636 Ori was observed between August 2008 and January 2015 for the EXOTIME project. About a third of the data was obtained using the $ \unit[1]{m} $ South African Astronomical Observatory (SAAO) with the UCT and STE3 CCD instruments.
Observations at the $ \unit[1.2]{m} $ MONET/North telescope, equipped with an Apogee 1k$ \times $1k E2V CCD camera, were taken in 2x2 binnings, using $ \unit[20]{s} $ exposure times.
Observations at the $ \unit[2.2]{m} $ Calar Alto Observatory (CAHA) used the CAFOS instrument with $ \unit[10]{s} $ exposure time.
Two nights were obtained at the $ \unit[1.5]{m} $ telescope at Loiano observatory, using the BFOSC (Bologna Faint Object Spectrograph \& Camera) instrument and $ \unit[15]{s} $ exposure times.
Between October 2008 and December 2009, observations at the $ \unit[1]{m} $ Mt. Lemmon Optical Astronomy Observatory (LOAO) were conducted with a 2k$ \times $2k CCD camera with exposure times of $ \unit[12]{s} $ and $ \unit[20]{s} $.
The observations at the $ \unit[3.6]{m} $ Telescopio Nazionale Galileo (TNG) in August 2008 and 2010 were performed with the DOLORES instrument and $ \unit[5]{s} $ exposure times.

\subsection{QQ Vir}
The discovery of QQ Vir (\object{PG 1325+101}) as a multi-period pulsator was reported in \citet{silvotti_pg_2002}, followed by a frequency analysis and asteroseismological modelling by \citet{silvotti_rapidly_2006} and \citet{charpinet_rapidly_2006}, respectively.

Observations of QQ Vir in 2001 and 2003 are described in \citet{silvotti_pg_2002} and \citet{silvotti_rapidly_2006}, respectively. Between March 2008 and April 2010, the object was observed as part of the EXOTIME project \citep{benatti_exotime_2010}.
Additionally, one observation run in February 2005 was performed at the $ \unit[1.5]{m} $ telescope at Loiano observatory, using the BFOSC instrument.
Most of the observations were obtained in 2009, 2010, 2011, and 2012 at the LOAO, using an exposure time of $ \unit[10]{s} $.
The CAHA and MONET/North observations were conducted with $ \unit[10]{s} $ and $ \unit[20]{s} $ exposure times, respectively.
The Loiano observatory performed additional observations in 2009, 2010 and 2011 with the $ \unit[1.5]{m} $ telescope, using BFOSC and an exposure times of $ \unit[12]{s} $, $ \unit[15]{s,} $ and $ \unit[20]{s} $.
Observations at the Mol\.{e}tai Astronomical Observatory (Mol) in 2008 were performed using the $ \unit[1.6]{m} $ telescope and an Apogee 1k$ \times $1k E2V CCD camera using $ \unit[17.5]{s} $ of exposure time.
Observations at the SAAO used the same instrumental setup as described in Section~\ref{sec:obs-v1636ori}.
The TNG observed in 2010 and 2011.
A DARC-WET campaign on QQ Vir was performed in May 2010.


\subsection{V541 Hya}
V541 Hya (\object{EC 09582-1137}) was discovered by \citet{kilkenny_three_2006}. \citet{randall_observations_2009} conducted an asteroseismological analysis of this target.

Between 2005 and 2015, a large number of observations were obtained at the SAAO, using the same instrumentation noted in Section~\ref{sec:obs-v1636ori} and exposure times of $ \unit[10]{s} $. 
The LOAO conducted observations in 2009, 2012, and 2013 with exposure times of $ \unit[20]{s} $.
During March 2009 and February and March 2010, V541 Hya was observed at the CAHA, using an exposure time of $ \unit[10]{s} $. 


\subsection{TESS observations}
The primary goal of the NASA Transiting Exoplanet Survey Satellite (TESS) space telescope is to detect exoplanets transiting bright nearby stars \citep{ricker_transiting_2015}. However, the extensive time series photometry is valuable for asteroseimology and the TESS Asteroseismic Science Consortium (TASC) coordinates short cadence observations of pulsating evolved stars. TESS observed V1636 Ori and V541 Hya with a cadence of $ \unit[120]{s} $ between November 15, 2018 and December 11, 2018, and February 2, 2019 and February 27, 2019, respectively. We used the light curves provided by the MAST archive\footnote{\url{https://archive.stsci.edu/}} that had common instrumental trends removed by the Pre-Search Data Conditioning Pipeline \citep[PDC, ][]{stumpe_kepler_2012}. Light curves and amplitude spectra are presented in Fig.~\ref{fig:lc-tess} and \ref{fig:tess-power}. The two-minute ('short') cadence undersamples the p-modes at about $ \unit[140]{s} $. In combination with the large photometric scatter, the amplitude spectra show no evidence of the p-modes. Thus, we did not make use of the TESS observations in our study.

\subsection{Data reduction}
For the EXOTIME observations, the data reduction was carried out using the IDL software TRIPP \citep[Time Resolved Imaging Photometry Package, see][]{schuh_ccd_2000}. TRIPP performs bias-, dark-, flat-field corrections, and differential aperture photometry to calculate the relative flux of a target with respect to one or more comparison stars and extinction corrections (second order polynomial in time). In the presence of sub-stellar companions, we might expect variations in the arrival times of stellar pulsations on the order of seconds to tens of seconds. The corresponding uncertainties are expected to be about one second. These uncertainties rise from observational constraints, such as smearing and sampling effects due to the integration time. The accuracy of individual time stamps is better than $ \pm \unit[0.5]{s} $.
All time stamps were converted from GJD(UTC) to BJD(TDB), according to \citet{eastman_achieving_2010}, with an accuracy well below the expected observational uncertainty.

Typical S/N for our ground-based observations range from 60 for large amplitude pulsations, to 3 for the smallest pulsation amplitudes we investigate in this work. The amplitude spectra in Section~\ref{sec:results} also show pulsations with smaller S/N, but these are not suitable for timing analysis because the uncertainties are too large (see Table~\ref{tab:add-freq}).

\section{Analysis} \label{sec:analysis}

In order to detect variations in the arrival time of stellar pulsations, we developed a pipeline to process the reduced data. A schematic flowchart of our pipeline is presented in Fig.~\ref{fig:flowchart}. The input consists of the light curve (time series) and the dates of the observational epochs. In a light curve, typically spanning several years at a very low duty cycle of 0.2 to 1.7 per cent, an epoch consists of a few roughly consecutive nights of observation.



\begin{figure*}[t]
        \centering
        \caption{Flow chart representing time of arrival analysis. Light curve (LC), and start and end time of each observational epoch are provided as input. Each frequency is analysed, leading to an intermediate $ O-C $-diagram, and subtracted from the LC by itself before the sum of all sinusoidal functions is fitted simultaneously to the LC, resulting in the final $ O-C $-diagram.} 
        \label{fig:flowchart}
\end{figure*}

\paragraph{Outlier removal:}In case no uncertainty in the flux measurement $ F $ is provided, the root-mean-square of each observation is used as an approximate photometric error for the later analysis. We have used a running median filter to exclude $ \unit[5]{\sigma} $ flux-outliers. The length of the window size depends on the cadence of the observations. We constrained it to be not longer than half of the period of the main frequency.
The analysis is performed for each frequency individually before all frequencies were analysed simultaneously.

\paragraph{Full data fit:}For the individual fitting of pulsations, we first determined the frequency of the main signal. For this, we used the \verb|astropy| package to calculate the Lomb-Scargle periodogram \citep{astropy_collaboration_astropy_2013,astropy_collaboration_astropy_2018}. From this periodogram, we selected the frequency with the largest amplitude to continue. In the next step, we performed the fit of a sinusoidal function to the light curve, using
\begin{align}
        F(t) = A \sin \left( f t + \phi \right) + o
\end{align}
with amplitude $ A $, frequency $ f $, phase $ \phi $ and offset $ o $. The minimisation problem is solved using the \textit{scipy} implementation of the Trust Region Reflective algorithm \citep{jones_scipy_2011}. The selected frequency from the amplitude spectrum serves as initial value, the full width at half maximum of the corresponding peak in the periodogram is used as a boundary. The amplitude-guess is taken from the amplitude spectrum. In case of highly varying amplitudes, the initial value can be set manually. The fitting routine returns the parameters and their variance. 

\paragraph{Epoch fit:}Frequency and offset are all kept fixed for the following analysis of the individual observational epochs. The starting value of the current phase fit is determined by the average of the previous $ j $ phase values (or the global fit value from above in case there are no $ j $ previous values yet) in order to keep the fitting process stable and avoid 'phase-jumps'. For our target sample, a value of $ j = 3 $ has proven to be reasonable, except when observational gaps span over several years. 

The uncertainty in the phase measurement scales inverse with the length of the epochs. Thus, this length is chosen in a way to minimize the uncertainties of the fit but at the same time keep the epochs as short as possible to maximize the temporal resolution of the final $ O-C $ diagram. Often, the observations themselves constrain the length of the epochs (e.g. three consecutive nights of observations and a gap of several weeks before the next block of observations). If possible, we aimed for an epoch length such that the timing uncertainties are of the order of one second. The phase information of the global and the epoch fit result in a intermediate $ O-C $ diagram. 

As a last step in the single-frequency analysis, the fitted model is subtracted from the light curve. We noticed significant amplitude variations for some of our targets. Thus, we subtracted the model using the amplitude of the individual epochs. This pre-whitening procedure is repeated for every relevant pulsation in the data.

\paragraph{Multi frequency fit:}Close frequencies are likely to introduce artificial trends in the arrival times in such a step-by-step analysis. Thus, the sum of all sinusoidal functions,
\begin{align}
        F(t) = A_n \sin \left( f_n t + \phi_n \right) + o,
\end{align}
is fitted to the non-whitened light curve, where $ n $ is the number of investigated frequencies. The previously retrieved values for amplitude, frequency, phase and offset are used as initial values. 
We used the phase information $ \phi_n $ of the light curve as reference phase, namely \textit{calculated} phase $ C $ in the final $ O-C $ diagram.
Similar to the single-frequency analysis, the observational epochs are fitted individually using the sum of sinusoidal functions to yield the \textit{observed} phase information $ O $.

\begin{figure*}[tp]
        \centering
        \begin{subfigure}[b]{0.49\textwidth}
                \centering
\begingroup
  \makeatletter
  \providecommand\color[2][]{%
    \GenericError{(gnuplot) \space\space\space\@spaces}{%
      Package color not loaded in conjunction with
      terminal option `colourtext'%
    }{See the gnuplot documentation for explanation.%
    }{Either use 'blacktext' in gnuplot or load the package
      color.sty in LaTeX.}%
    \renewcommand\color[2][]{}%
  }%
  \providecommand\includegraphics[2][]{%
    \GenericError{(gnuplot) \space\space\space\@spaces}{%
      Package graphicx or graphics not loaded%
    }{See the gnuplot documentation for explanation.%
    }{The gnuplot epslatex terminal needs graphicx.sty or graphics.sty.}
  }%
  \providecommand\rotatebox[2]{#2}%
  \@ifundefined{ifGPcolor}{%
    \newif\ifGPcolor
    \GPcolortrue
  }{}%
  \@ifundefined{ifGPblacktext}{%
    \newif\ifGPblacktext
    \GPblacktexttrue
  }{}%
  \let\gplgaddtomacro\g@addto@macro
  \gdef\gplbacktext{}%
  \gdef\gplfronttext{}%
  \makeatother
  \ifGPblacktext
    \def\colorrgb#1{}%
    \def\colorgray#1{}%
  \else
    \ifGPcolor
      \def\colorrgb#1{\color[rgb]{#1}}%
      \def\colorgray#1{\color[gray]{#1}}%
      \expandafter\def\csname LTw\endcsname{\color{white}}%
      \expandafter\def\csname LTb\endcsname{\color{black}}%
      \expandafter\def\csname LTa\endcsname{\color{black}}%
      \expandafter\def\csname LT0\endcsname{\color[rgb]{1,0,0}}%
      \expandafter\def\csname LT1\endcsname{\color[rgb]{0,1,0}}%
      \expandafter\def\csname LT2\endcsname{\color[rgb]{0,0,1}}%
      \expandafter\def\csname LT3\endcsname{\color[rgb]{1,0,1}}%
      \expandafter\def\csname LT4\endcsname{\color[rgb]{0,1,1}}%
      \expandafter\def\csname LT5\endcsname{\color[rgb]{1,1,0}}%
      \expandafter\def\csname LT6\endcsname{\color[rgb]{0,0,0}}%
      \expandafter\def\csname LT7\endcsname{\color[rgb]{1,0.3,0}}%
      \expandafter\def\csname LT8\endcsname{\color[rgb]{0.5,0.5,0.5}}%
    \else
      \def\colorrgb#1{\color{black}}%
      \def\colorgray#1{\color[gray]{#1}}%
      \expandafter\def\csname LTw\endcsname{\color{white}}%
      \expandafter\def\csname LTb\endcsname{\color{black}}%
      \expandafter\def\csname LTa\endcsname{\color{black}}%
      \expandafter\def\csname LT0\endcsname{\color{black}}%
      \expandafter\def\csname LT1\endcsname{\color{black}}%
      \expandafter\def\csname LT2\endcsname{\color{black}}%
      \expandafter\def\csname LT3\endcsname{\color{black}}%
      \expandafter\def\csname LT4\endcsname{\color{black}}%
      \expandafter\def\csname LT5\endcsname{\color{black}}%
      \expandafter\def\csname LT6\endcsname{\color{black}}%
      \expandafter\def\csname LT7\endcsname{\color{black}}%
      \expandafter\def\csname LT8\endcsname{\color{black}}%
    \fi
  \fi
    \setlength{\unitlength}{0.0500bp}%
    \ifx\gptboxheight\undefined%
      \newlength{\gptboxheight}%
      \newlength{\gptboxwidth}%
      \newsavebox{\gptboxtext}%
    \fi%
    \setlength{\fboxrule}{0.5pt}%
    \setlength{\fboxsep}{1pt}%
\begin{picture}(5100.00,5100.00)%
    \gplgaddtomacro\gplbacktext{%
      \csname LTb\endcsname%
      \put(742,3998){\makebox(0,0)[r]{\strut{}$0.96$}}%
      \csname LTb\endcsname%
      \put(742,4201){\makebox(0,0)[r]{\strut{}$0.98$}}%
      \csname LTb\endcsname%
      \put(742,4405){\makebox(0,0)[r]{\strut{}$1$}}%
      \csname LTb\endcsname%
      \put(742,4608){\makebox(0,0)[r]{\strut{}$1.02$}}%
      \csname LTb\endcsname%
      \put(742,4811){\makebox(0,0)[r]{\strut{}$1.04$}}%
      \csname LTb\endcsname%
      \put(918,3636){\makebox(0,0){\strut{}4.70}}%
      \csname LTb\endcsname%
      \put(1349,3636){\makebox(0,0){\strut{}4.72}}%
      \csname LTb\endcsname%
      \put(1779,3636){\makebox(0,0){\strut{}4.74}}%
      \csname LTb\endcsname%
      \put(2210,3636){\makebox(0,0){\strut{}4.76}}%
      \csname LTb\endcsname%
      \put(2640,3636){\makebox(0,0){\strut{}4.78}}%
      \csname LTb\endcsname%
      \put(3071,3636){\makebox(0,0){\strut{}4.80}}%
      \csname LTb\endcsname%
      \put(3501,3636){\makebox(0,0){\strut{}4.82}}%
      \csname LTb\endcsname%
      \put(3932,3636){\makebox(0,0){\strut{}4.84}}%
      \csname LTb\endcsname%
      \put(4362,3636){\makebox(0,0){\strut{}4.86}}%
      \csname LTb\endcsname%
      \put(4793,3636){\makebox(0,0){\strut{}4.88}}%
    }%
    \gplgaddtomacro\gplfronttext{%
      \csname LTb\endcsname%
      \put(139,4404){\rotatebox{-270}{\makebox(0,0){\strut{}$F$}}}%
    }%
    \gplgaddtomacro\gplbacktext{%
      \csname LTb\endcsname%
      \put(742,2422){\makebox(0,0)[r]{\strut{}$0.96$}}%
      \csname LTb\endcsname%
      \put(742,2625){\makebox(0,0)[r]{\strut{}$0.98$}}%
      \csname LTb\endcsname%
      \put(742,2829){\makebox(0,0)[r]{\strut{}$1$}}%
      \csname LTb\endcsname%
      \put(742,3032){\makebox(0,0)[r]{\strut{}$1.02$}}%
      \csname LTb\endcsname%
      \put(742,3235){\makebox(0,0)[r]{\strut{}$1.04$}}%
      \csname LTb\endcsname%
      \put(918,2060){\makebox(0,0){\strut{}5.72}}%
      \csname LTb\endcsname%
      \put(1349,2060){\makebox(0,0){\strut{}5.74}}%
      \csname LTb\endcsname%
      \put(1779,2060){\makebox(0,0){\strut{}5.76}}%
      \csname LTb\endcsname%
      \put(2210,2060){\makebox(0,0){\strut{}5.78}}%
      \csname LTb\endcsname%
      \put(2640,2060){\makebox(0,0){\strut{}5.80}}%
      \csname LTb\endcsname%
      \put(3071,2060){\makebox(0,0){\strut{}5.82}}%
      \csname LTb\endcsname%
      \put(3501,2060){\makebox(0,0){\strut{}5.84}}%
      \csname LTb\endcsname%
      \put(3932,2060){\makebox(0,0){\strut{}5.86}}%
      \csname LTb\endcsname%
      \put(4362,2060){\makebox(0,0){\strut{}5.88}}%
      \csname LTb\endcsname%
      \put(4793,2060){\makebox(0,0){\strut{}5.90}}%
    }%
    \gplgaddtomacro\gplfronttext{%
      \csname LTb\endcsname%
      \put(139,2828){\rotatebox{-270}{\makebox(0,0){\strut{}$F$}}}%
    }%
    \gplgaddtomacro\gplbacktext{%
      \csname LTb\endcsname%
      \put(742,846){\makebox(0,0)[r]{\strut{}$0.96$}}%
      \csname LTb\endcsname%
      \put(742,1049){\makebox(0,0)[r]{\strut{}$0.98$}}%
      \csname LTb\endcsname%
      \put(742,1253){\makebox(0,0)[r]{\strut{}$1$}}%
      \csname LTb\endcsname%
      \put(742,1456){\makebox(0,0)[r]{\strut{}$1.02$}}%
      \csname LTb\endcsname%
      \put(742,1659){\makebox(0,0)[r]{\strut{}$1.04$}}%
      \csname LTb\endcsname%
      \put(918,484){\makebox(0,0){\strut{}6.72}}%
      \csname LTb\endcsname%
      \put(1349,484){\makebox(0,0){\strut{}6.74}}%
      \csname LTb\endcsname%
      \put(1779,484){\makebox(0,0){\strut{}6.76}}%
      \csname LTb\endcsname%
      \put(2210,484){\makebox(0,0){\strut{}6.78}}%
      \csname LTb\endcsname%
      \put(2640,484){\makebox(0,0){\strut{}6.80}}%
      \csname LTb\endcsname%
      \put(3071,484){\makebox(0,0){\strut{}6.82}}%
      \csname LTb\endcsname%
      \put(3501,484){\makebox(0,0){\strut{}6.84}}%
      \csname LTb\endcsname%
      \put(3932,484){\makebox(0,0){\strut{}6.86}}%
      \csname LTb\endcsname%
      \put(4362,484){\makebox(0,0){\strut{}6.88}}%
      \csname LTb\endcsname%
      \put(4793,484){\makebox(0,0){\strut{}6.90}}%
    }%
    \gplgaddtomacro\gplfronttext{%
      \csname LTb\endcsname%
      \put(139,1252){\rotatebox{-270}{\makebox(0,0){\strut{}$F$}}}%
      \csname LTb\endcsname%
      \put(2855,205){\makebox(0,0){\strut{}$\unit[t /]{d}$ since 2007-10-17}}%
    }%
    \gplbacktext
    \put(0,0){\includegraphics[scale=0.5]{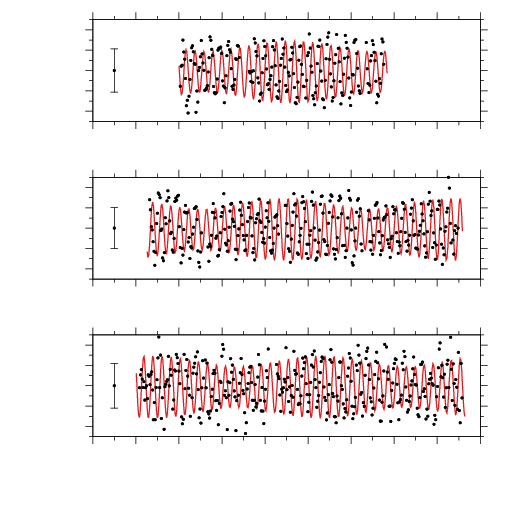}}%
    \gplfronttext
  \end{picture}%
\endgroup


        \end{subfigure}
        \begin{subfigure}[b]{0.49\textwidth}
                \centering
\begingroup
  \makeatletter
  \providecommand\color[2][]{%
    \GenericError{(gnuplot) \space\space\space\@spaces}{%
      Package color not loaded in conjunction with
      terminal option `colourtext'%
    }{See the gnuplot documentation for explanation.%
    }{Either use 'blacktext' in gnuplot or load the package
      color.sty in LaTeX.}%
    \renewcommand\color[2][]{}%
  }%
  \providecommand\includegraphics[2][]{%
    \GenericError{(gnuplot) \space\space\space\@spaces}{%
      Package graphicx or graphics not loaded%
    }{See the gnuplot documentation for explanation.%
    }{The gnuplot epslatex terminal needs graphicx.sty or graphics.sty.}
  }%
  \providecommand\rotatebox[2]{#2}%
  \@ifundefined{ifGPcolor}{%
    \newif\ifGPcolor
    \GPcolortrue
  }{}%
  \@ifundefined{ifGPblacktext}{%
    \newif\ifGPblacktext
    \GPblacktexttrue
  }{}%
  \let\gplgaddtomacro\g@addto@macro
  \gdef\gplbacktext{}%
  \gdef\gplfronttext{}%
  \makeatother
  \ifGPblacktext
    \def\colorrgb#1{}%
    \def\colorgray#1{}%
  \else
    \ifGPcolor
      \def\colorrgb#1{\color[rgb]{#1}}%
      \def\colorgray#1{\color[gray]{#1}}%
      \expandafter\def\csname LTw\endcsname{\color{white}}%
      \expandafter\def\csname LTb\endcsname{\color{black}}%
      \expandafter\def\csname LTa\endcsname{\color{black}}%
      \expandafter\def\csname LT0\endcsname{\color[rgb]{1,0,0}}%
      \expandafter\def\csname LT1\endcsname{\color[rgb]{0,1,0}}%
      \expandafter\def\csname LT2\endcsname{\color[rgb]{0,0,1}}%
      \expandafter\def\csname LT3\endcsname{\color[rgb]{1,0,1}}%
      \expandafter\def\csname LT4\endcsname{\color[rgb]{0,1,1}}%
      \expandafter\def\csname LT5\endcsname{\color[rgb]{1,1,0}}%
      \expandafter\def\csname LT6\endcsname{\color[rgb]{0,0,0}}%
      \expandafter\def\csname LT7\endcsname{\color[rgb]{1,0.3,0}}%
      \expandafter\def\csname LT8\endcsname{\color[rgb]{0.5,0.5,0.5}}%
    \else
      \def\colorrgb#1{\color{black}}%
      \def\colorgray#1{\color[gray]{#1}}%
      \expandafter\def\csname LTw\endcsname{\color{white}}%
      \expandafter\def\csname LTb\endcsname{\color{black}}%
      \expandafter\def\csname LTa\endcsname{\color{black}}%
      \expandafter\def\csname LT0\endcsname{\color{black}}%
      \expandafter\def\csname LT1\endcsname{\color{black}}%
      \expandafter\def\csname LT2\endcsname{\color{black}}%
      \expandafter\def\csname LT3\endcsname{\color{black}}%
      \expandafter\def\csname LT4\endcsname{\color{black}}%
      \expandafter\def\csname LT5\endcsname{\color{black}}%
      \expandafter\def\csname LT6\endcsname{\color{black}}%
      \expandafter\def\csname LT7\endcsname{\color{black}}%
      \expandafter\def\csname LT8\endcsname{\color{black}}%
    \fi
  \fi
    \setlength{\unitlength}{0.0500bp}%
    \ifx\gptboxheight\undefined%
      \newlength{\gptboxheight}%
      \newlength{\gptboxwidth}%
      \newsavebox{\gptboxtext}%
    \fi%
    \setlength{\fboxrule}{0.5pt}%
    \setlength{\fboxsep}{1pt}%
\begin{picture}(5100.00,5100.00)%
    \gplgaddtomacro\gplbacktext{%
      \csname LTb\endcsname%
      \put(640,3214){\makebox(0,0)[r]{\strut{}$0$}}%
      \csname LTb\endcsname%
      \put(640,3479){\makebox(0,0)[r]{\strut{}$0.5$}}%
      \csname LTb\endcsname%
      \put(640,3745){\makebox(0,0)[r]{\strut{}$1$}}%
      \csname LTb\endcsname%
      \put(640,4010){\makebox(0,0)[r]{\strut{}$1.5$}}%
      \csname LTb\endcsname%
      \put(640,4276){\makebox(0,0)[r]{\strut{}$2$}}%
      \csname LTb\endcsname%
      \put(640,4541){\makebox(0,0)[r]{\strut{}$2.5$}}%
      \csname LTb\endcsname%
      \put(1305,4801){\makebox(0,0){\strut{}$2500$}}%
      \csname LTb\endcsname%
      \put(2010,4801){\makebox(0,0){\strut{}$2600$}}%
      \csname LTb\endcsname%
      \put(2715,4801){\makebox(0,0){\strut{}$2700$}}%
      \csname LTb\endcsname%
      \put(3420,4801){\makebox(0,0){\strut{}$2800$}}%
      \csname LTb\endcsname%
      \put(4125,4801){\makebox(0,0){\strut{}$2900$}}%
      \csname LTb\endcsname%
      \put(4830,4801){\makebox(0,0){\strut{}$3000$}}%
      \csname LTb\endcsname%
      \put(1428,4276){\makebox(0,0)[l]{\strut{}full}}%
    }%
    \gplgaddtomacro\gplfronttext{%
      \csname LTb\endcsname%
      \put(165,3877){\rotatebox{-270}{\makebox(0,0){\strut{}$\unit[A_{full}/]{\%}$}}}%
      \csname LTb\endcsname%
      \put(2855,5079){\makebox(0,0){\strut{}$\unit[f /]{\mu Hz}$}}%
    }%
    \gplgaddtomacro\gplbacktext{%
      \csname LTb\endcsname%
      \put(640,1886){\makebox(0,0)[r]{\strut{}$0$}}%
      \csname LTb\endcsname%
      \put(640,2123){\makebox(0,0)[r]{\strut{}$0.5$}}%
      \csname LTb\endcsname%
      \put(640,2360){\makebox(0,0)[r]{\strut{}$1$}}%
      \csname LTb\endcsname%
      \put(640,2597){\makebox(0,0)[r]{\strut{}$1.5$}}%
      \csname LTb\endcsname%
      \put(640,2834){\makebox(0,0)[r]{\strut{}$2$}}%
      \csname LTb\endcsname%
      \put(640,3071){\makebox(0,0)[r]{\strut{}$2.5$}}%
      \csname LTb\endcsname%
      \put(1428,2948){\makebox(0,0)[l]{\strut{}epoch}}%
    }%
    \gplgaddtomacro\gplfronttext{%
      \csname LTb\endcsname%
      \put(165,2549){\rotatebox{-270}{\makebox(0,0){\strut{}$\unit[A_{epoch}/]{\%}$}}}%
    }%
    \gplgaddtomacro\gplbacktext{%
      \csname LTb\endcsname%
      \put(640,1885){\makebox(0,0)[r]{\strut{}}}%
      \csname LTb\endcsname%
      \put(640,558){\makebox(0,0)[r]{\strut{}$0$}}%
      \csname LTb\endcsname%
      \put(640,823){\makebox(0,0)[r]{\strut{}$0.2$}}%
      \csname LTb\endcsname%
      \put(640,1089){\makebox(0,0)[r]{\strut{}$0.4$}}%
      \csname LTb\endcsname%
      \put(640,1354){\makebox(0,0)[r]{\strut{}$0.6$}}%
      \csname LTb\endcsname%
      \put(640,1620){\makebox(0,0)[r]{\strut{}$0.8$}}%
      \csname LTb\endcsname%
      \put(816,298){\makebox(0,0){\strut{}$210$}}%
      \csname LTb\endcsname%
      \put(1632,298){\makebox(0,0){\strut{}$220$}}%
      \csname LTb\endcsname%
      \put(2448,298){\makebox(0,0){\strut{}$230$}}%
      \csname LTb\endcsname%
      \put(3263,298){\makebox(0,0){\strut{}$240$}}%
      \csname LTb\endcsname%
      \put(4079,298){\makebox(0,0){\strut{}$250$}}%
      \csname LTb\endcsname%
      \put(4895,298){\makebox(0,0){\strut{}$260$}}%
      \csname LTb\endcsname%
      \put(1428,1620){\makebox(0,0)[l]{\strut{}w}}%
    }%
    \gplgaddtomacro\gplfronttext{%
      \csname LTb\endcsname%
      \put(165,1221){\rotatebox{-270}{\makebox(0,0){\strut{}$A_{\text{window}}$}}}%
      \csname LTb\endcsname%
      \put(2855,19){\makebox(0,0){\strut{}$\unit[f /]{d^{-1}}$}}%
    }%
    \gplbacktext
    \put(0,0){\includegraphics[scale=0.5]{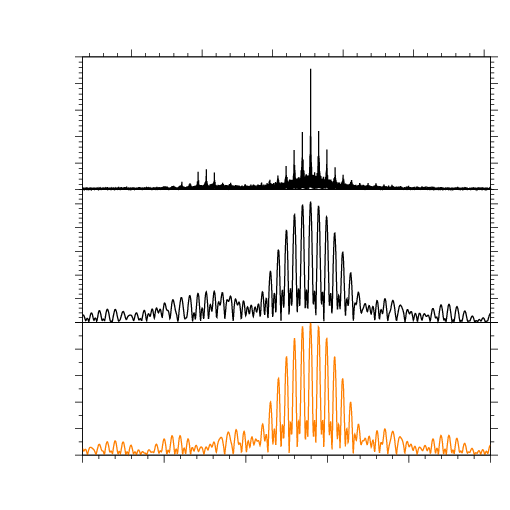}}%
    \gplfronttext
  \end{picture}%
\endgroup


        \end{subfigure}
        \caption{Example observations of DW~Lyn from October 21, 22, and 23, 2010 at the Lulin observatory (left, from top to bottom), combined used as one $ O-C $ measurement. The error bar on the left of each plot represents the photometric al uncertainty. The red line shows the simultaneous two frequency fits to the epoch data. The amplitude spectra on the right hand side show the spectrum of the full data set (top), the spectrum of this epoch (middle) and the respective window function computed at $ f_1 $.}
        \label{fig:dwlyn-lc-crop}
\end{figure*}

\begin{figure*}[tp]
        \centering
        \begin{subfigure}[b]{0.49\textwidth}
                \centering
\begingroup
  \makeatletter
  \providecommand\color[2][]{%
    \GenericError{(gnuplot) \space\space\space\@spaces}{%
      Package color not loaded in conjunction with
      terminal option `colourtext'%
    }{See the gnuplot documentation for explanation.%
    }{Either use 'blacktext' in gnuplot or load the package
      color.sty in LaTeX.}%
    \renewcommand\color[2][]{}%
  }%
  \providecommand\includegraphics[2][]{%
    \GenericError{(gnuplot) \space\space\space\@spaces}{%
      Package graphicx or graphics not loaded%
    }{See the gnuplot documentation for explanation.%
    }{The gnuplot epslatex terminal needs graphicx.sty or graphics.sty.}
  }%
  \providecommand\rotatebox[2]{#2}%
  \@ifundefined{ifGPcolor}{%
    \newif\ifGPcolor
    \GPcolortrue
  }{}%
  \@ifundefined{ifGPblacktext}{%
    \newif\ifGPblacktext
    \GPblacktexttrue
  }{}%
  \let\gplgaddtomacro\g@addto@macro
  \gdef\gplbacktext{}%
  \gdef\gplfronttext{}%
  \makeatother
  \ifGPblacktext
    \def\colorrgb#1{}%
    \def\colorgray#1{}%
  \else
    \ifGPcolor
      \def\colorrgb#1{\color[rgb]{#1}}%
      \def\colorgray#1{\color[gray]{#1}}%
      \expandafter\def\csname LTw\endcsname{\color{white}}%
      \expandafter\def\csname LTb\endcsname{\color{black}}%
      \expandafter\def\csname LTa\endcsname{\color{black}}%
      \expandafter\def\csname LT0\endcsname{\color[rgb]{1,0,0}}%
      \expandafter\def\csname LT1\endcsname{\color[rgb]{0,1,0}}%
      \expandafter\def\csname LT2\endcsname{\color[rgb]{0,0,1}}%
      \expandafter\def\csname LT3\endcsname{\color[rgb]{1,0,1}}%
      \expandafter\def\csname LT4\endcsname{\color[rgb]{0,1,1}}%
      \expandafter\def\csname LT5\endcsname{\color[rgb]{1,1,0}}%
      \expandafter\def\csname LT6\endcsname{\color[rgb]{0,0,0}}%
      \expandafter\def\csname LT7\endcsname{\color[rgb]{1,0.3,0}}%
      \expandafter\def\csname LT8\endcsname{\color[rgb]{0.5,0.5,0.5}}%
    \else
      \def\colorrgb#1{\color{black}}%
      \def\colorgray#1{\color[gray]{#1}}%
      \expandafter\def\csname LTw\endcsname{\color{white}}%
      \expandafter\def\csname LTb\endcsname{\color{black}}%
      \expandafter\def\csname LTa\endcsname{\color{black}}%
      \expandafter\def\csname LT0\endcsname{\color{black}}%
      \expandafter\def\csname LT1\endcsname{\color{black}}%
      \expandafter\def\csname LT2\endcsname{\color{black}}%
      \expandafter\def\csname LT3\endcsname{\color{black}}%
      \expandafter\def\csname LT4\endcsname{\color{black}}%
      \expandafter\def\csname LT5\endcsname{\color{black}}%
      \expandafter\def\csname LT6\endcsname{\color{black}}%
      \expandafter\def\csname LT7\endcsname{\color{black}}%
      \expandafter\def\csname LT8\endcsname{\color{black}}%
    \fi
  \fi
    \setlength{\unitlength}{0.0500bp}%
    \ifx\gptboxheight\undefined%
      \newlength{\gptboxheight}%
      \newlength{\gptboxwidth}%
      \newsavebox{\gptboxtext}%
    \fi%
    \setlength{\fboxrule}{0.5pt}%
    \setlength{\fboxsep}{1pt}%
\begin{picture}(5100.00,5100.00)%
    \gplgaddtomacro\gplbacktext{%
      \csname LTb\endcsname%
      \put(742,3998){\makebox(0,0)[r]{\strut{}$0.96$}}%
      \csname LTb\endcsname%
      \put(742,4201){\makebox(0,0)[r]{\strut{}$0.98$}}%
      \csname LTb\endcsname%
      \put(742,4405){\makebox(0,0)[r]{\strut{}$1$}}%
      \csname LTb\endcsname%
      \put(742,4608){\makebox(0,0)[r]{\strut{}$1.02$}}%
      \csname LTb\endcsname%
      \put(742,4811){\makebox(0,0)[r]{\strut{}$1.04$}}%
      \csname LTb\endcsname%
      \put(918,3636){\makebox(0,0){\strut{}1.30}}%
      \csname LTb\endcsname%
      \put(1472,3636){\makebox(0,0){\strut{}1.31}}%
      \csname LTb\endcsname%
      \put(2025,3636){\makebox(0,0){\strut{}1.32}}%
      \csname LTb\endcsname%
      \put(2579,3636){\makebox(0,0){\strut{}1.33}}%
      \csname LTb\endcsname%
      \put(3132,3636){\makebox(0,0){\strut{}1.34}}%
      \csname LTb\endcsname%
      \put(3686,3636){\makebox(0,0){\strut{}1.35}}%
      \csname LTb\endcsname%
      \put(4239,3636){\makebox(0,0){\strut{}1.36}}%
      \csname LTb\endcsname%
      \put(4793,3636){\makebox(0,0){\strut{}1.37}}%
    }%
    \gplgaddtomacro\gplfronttext{%
      \csname LTb\endcsname%
      \put(139,4404){\rotatebox{-270}{\makebox(0,0){\strut{}$F$}}}%
    }%
    \gplgaddtomacro\gplbacktext{%
      \csname LTb\endcsname%
      \put(742,2422){\makebox(0,0)[r]{\strut{}$0.96$}}%
      \csname LTb\endcsname%
      \put(742,2625){\makebox(0,0)[r]{\strut{}$0.98$}}%
      \csname LTb\endcsname%
      \put(742,2829){\makebox(0,0)[r]{\strut{}$1$}}%
      \csname LTb\endcsname%
      \put(742,3032){\makebox(0,0)[r]{\strut{}$1.02$}}%
      \csname LTb\endcsname%
      \put(742,3235){\makebox(0,0)[r]{\strut{}$1.04$}}%
      \csname LTb\endcsname%
      \put(918,2060){\makebox(0,0){\strut{}2.29}}%
      \csname LTb\endcsname%
      \put(1472,2060){\makebox(0,0){\strut{}2.30}}%
      \csname LTb\endcsname%
      \put(2025,2060){\makebox(0,0){\strut{}2.31}}%
      \csname LTb\endcsname%
      \put(2579,2060){\makebox(0,0){\strut{}2.32}}%
      \csname LTb\endcsname%
      \put(3132,2060){\makebox(0,0){\strut{}2.33}}%
      \csname LTb\endcsname%
      \put(3686,2060){\makebox(0,0){\strut{}2.34}}%
      \csname LTb\endcsname%
      \put(4239,2060){\makebox(0,0){\strut{}2.35}}%
      \csname LTb\endcsname%
      \put(4793,2060){\makebox(0,0){\strut{}2.36}}%
    }%
    \gplgaddtomacro\gplfronttext{%
      \csname LTb\endcsname%
      \put(139,2828){\rotatebox{-270}{\makebox(0,0){\strut{}$F$}}}%
    }%
    \gplgaddtomacro\gplbacktext{%
      \csname LTb\endcsname%
      \put(742,846){\makebox(0,0)[r]{\strut{}$0.96$}}%
      \csname LTb\endcsname%
      \put(742,1049){\makebox(0,0)[r]{\strut{}$0.98$}}%
      \csname LTb\endcsname%
      \put(742,1253){\makebox(0,0)[r]{\strut{}$1$}}%
      \csname LTb\endcsname%
      \put(742,1456){\makebox(0,0)[r]{\strut{}$1.02$}}%
      \csname LTb\endcsname%
      \put(742,1659){\makebox(0,0)[r]{\strut{}$1.04$}}%
      \csname LTb\endcsname%
      \put(918,484){\makebox(0,0){\strut{}3.29}}%
      \csname LTb\endcsname%
      \put(1472,484){\makebox(0,0){\strut{}3.30}}%
      \csname LTb\endcsname%
      \put(2025,484){\makebox(0,0){\strut{}3.31}}%
      \csname LTb\endcsname%
      \put(2579,484){\makebox(0,0){\strut{}3.32}}%
      \csname LTb\endcsname%
      \put(3132,484){\makebox(0,0){\strut{}3.33}}%
      \csname LTb\endcsname%
      \put(3686,484){\makebox(0,0){\strut{}3.34}}%
      \csname LTb\endcsname%
      \put(4239,484){\makebox(0,0){\strut{}3.35}}%
      \csname LTb\endcsname%
      \put(4793,484){\makebox(0,0){\strut{}3.36}}%
    }%
    \gplgaddtomacro\gplfronttext{%
      \csname LTb\endcsname%
      \put(139,1252){\rotatebox{-270}{\makebox(0,0){\strut{}$F$}}}%
      \csname LTb\endcsname%
      \put(2855,205){\makebox(0,0){\strut{}$\unit[t /]{d}$ since 2009-03-20}}%
    }%
    \gplbacktext
    \put(0,0){\includegraphics[scale=0.5]{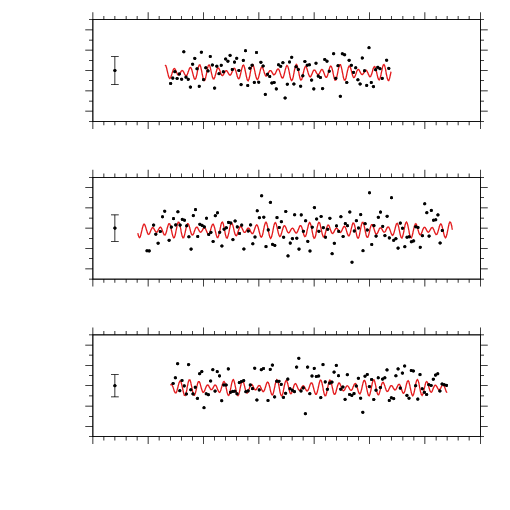}}%
    \gplfronttext
  \end{picture}%
\endgroup


        \end{subfigure}
        \begin{subfigure}[b]{0.49\textwidth}
                \centering
\begingroup
  \makeatletter
  \providecommand\color[2][]{%
    \GenericError{(gnuplot) \space\space\space\@spaces}{%
      Package color not loaded in conjunction with
      terminal option `colourtext'%
    }{See the gnuplot documentation for explanation.%
    }{Either use 'blacktext' in gnuplot or load the package
      color.sty in LaTeX.}%
    \renewcommand\color[2][]{}%
  }%
  \providecommand\includegraphics[2][]{%
    \GenericError{(gnuplot) \space\space\space\@spaces}{%
      Package graphicx or graphics not loaded%
    }{See the gnuplot documentation for explanation.%
    }{The gnuplot epslatex terminal needs graphicx.sty or graphics.sty.}
  }%
  \providecommand\rotatebox[2]{#2}%
  \@ifundefined{ifGPcolor}{%
    \newif\ifGPcolor
    \GPcolortrue
  }{}%
  \@ifundefined{ifGPblacktext}{%
    \newif\ifGPblacktext
    \GPblacktexttrue
  }{}%
  \let\gplgaddtomacro\g@addto@macro
  \gdef\gplbacktext{}%
  \gdef\gplfronttext{}%
  \makeatother
  \ifGPblacktext
    \def\colorrgb#1{}%
    \def\colorgray#1{}%
  \else
    \ifGPcolor
      \def\colorrgb#1{\color[rgb]{#1}}%
      \def\colorgray#1{\color[gray]{#1}}%
      \expandafter\def\csname LTw\endcsname{\color{white}}%
      \expandafter\def\csname LTb\endcsname{\color{black}}%
      \expandafter\def\csname LTa\endcsname{\color{black}}%
      \expandafter\def\csname LT0\endcsname{\color[rgb]{1,0,0}}%
      \expandafter\def\csname LT1\endcsname{\color[rgb]{0,1,0}}%
      \expandafter\def\csname LT2\endcsname{\color[rgb]{0,0,1}}%
      \expandafter\def\csname LT3\endcsname{\color[rgb]{1,0,1}}%
      \expandafter\def\csname LT4\endcsname{\color[rgb]{0,1,1}}%
      \expandafter\def\csname LT5\endcsname{\color[rgb]{1,1,0}}%
      \expandafter\def\csname LT6\endcsname{\color[rgb]{0,0,0}}%
      \expandafter\def\csname LT7\endcsname{\color[rgb]{1,0.3,0}}%
      \expandafter\def\csname LT8\endcsname{\color[rgb]{0.5,0.5,0.5}}%
    \else
      \def\colorrgb#1{\color{black}}%
      \def\colorgray#1{\color[gray]{#1}}%
      \expandafter\def\csname LTw\endcsname{\color{white}}%
      \expandafter\def\csname LTb\endcsname{\color{black}}%
      \expandafter\def\csname LTa\endcsname{\color{black}}%
      \expandafter\def\csname LT0\endcsname{\color{black}}%
      \expandafter\def\csname LT1\endcsname{\color{black}}%
      \expandafter\def\csname LT2\endcsname{\color{black}}%
      \expandafter\def\csname LT3\endcsname{\color{black}}%
      \expandafter\def\csname LT4\endcsname{\color{black}}%
      \expandafter\def\csname LT5\endcsname{\color{black}}%
      \expandafter\def\csname LT6\endcsname{\color{black}}%
      \expandafter\def\csname LT7\endcsname{\color{black}}%
      \expandafter\def\csname LT8\endcsname{\color{black}}%
    \fi
  \fi
    \setlength{\unitlength}{0.0500bp}%
    \ifx\gptboxheight\undefined%
      \newlength{\gptboxheight}%
      \newlength{\gptboxwidth}%
      \newsavebox{\gptboxtext}%
    \fi%
    \setlength{\fboxrule}{0.5pt}%
    \setlength{\fboxsep}{1pt}%
\begin{picture}(5100.00,5100.00)%
    \gplgaddtomacro\gplbacktext{%
      \csname LTb\endcsname%
      \put(640,3214){\makebox(0,0)[r]{\strut{}$0$}}%
      \csname LTb\endcsname%
      \put(640,3656){\makebox(0,0)[r]{\strut{}$0.2$}}%
      \csname LTb\endcsname%
      \put(640,4099){\makebox(0,0)[r]{\strut{}$0.4$}}%
      \csname LTb\endcsname%
      \put(640,4541){\makebox(0,0)[r]{\strut{}$0.6$}}%
      \csname LTb\endcsname%
      \put(1099,4801){\makebox(0,0){\strut{}$5700$}}%
      \csname LTb\endcsname%
      \put(1686,4801){\makebox(0,0){\strut{}$6000$}}%
      \csname LTb\endcsname%
      \put(2274,4801){\makebox(0,0){\strut{}$6300$}}%
      \csname LTb\endcsname%
      \put(2861,4801){\makebox(0,0){\strut{}$6600$}}%
      \csname LTb\endcsname%
      \put(3448,4801){\makebox(0,0){\strut{}$6900$}}%
      \csname LTb\endcsname%
      \put(4036,4801){\makebox(0,0){\strut{}$7200$}}%
      \csname LTb\endcsname%
      \put(4623,4801){\makebox(0,0){\strut{}$7500$}}%
      \csname LTb\endcsname%
      \put(1428,4276){\makebox(0,0)[l]{\strut{}full}}%
    }%
    \gplgaddtomacro\gplfronttext{%
      \csname LTb\endcsname%
      \put(165,3877){\rotatebox{-270}{\makebox(0,0){\strut{}$\unit[A_{full}/]{\%}$}}}%
      \csname LTb\endcsname%
      \put(2855,5079){\makebox(0,0){\strut{}$\unit[f /]{\mu Hz}$}}%
    }%
    \gplgaddtomacro\gplbacktext{%
      \csname LTb\endcsname%
      \put(640,3213){\makebox(0,0)[r]{\strut{}}}%
      \csname LTb\endcsname%
      \put(640,1886){\makebox(0,0)[r]{\strut{}$0$}}%
      \csname LTb\endcsname%
      \put(640,2328){\makebox(0,0)[r]{\strut{}$0.2$}}%
      \csname LTb\endcsname%
      \put(640,2771){\makebox(0,0)[r]{\strut{}$0.4$}}%
      \csname LTb\endcsname%
      \put(1428,2948){\makebox(0,0)[l]{\strut{}epoch}}%
    }%
    \gplgaddtomacro\gplfronttext{%
      \csname LTb\endcsname%
      \put(165,2549){\rotatebox{-270}{\makebox(0,0){\strut{}$\unit[A_{epoch}/]{\%}$}}}%
    }%
    \gplgaddtomacro\gplbacktext{%
      \csname LTb\endcsname%
      \put(640,1885){\makebox(0,0)[r]{\strut{}}}%
      \csname LTb\endcsname%
      \put(640,558){\makebox(0,0)[r]{\strut{}$0$}}%
      \csname LTb\endcsname%
      \put(640,823){\makebox(0,0)[r]{\strut{}$0.2$}}%
      \csname LTb\endcsname%
      \put(640,1089){\makebox(0,0)[r]{\strut{}$0.4$}}%
      \csname LTb\endcsname%
      \put(640,1354){\makebox(0,0)[r]{\strut{}$0.6$}}%
      \csname LTb\endcsname%
      \put(640,1620){\makebox(0,0)[r]{\strut{}$0.8$}}%
      \csname LTb\endcsname%
      \put(816,298){\makebox(0,0){\strut{}$480$}}%
      \csname LTb\endcsname%
      \put(1269,298){\makebox(0,0){\strut{}$500$}}%
      \csname LTb\endcsname%
      \put(1722,298){\makebox(0,0){\strut{}$520$}}%
      \csname LTb\endcsname%
      \put(2176,298){\makebox(0,0){\strut{}$540$}}%
      \csname LTb\endcsname%
      \put(2629,298){\makebox(0,0){\strut{}$560$}}%
      \csname LTb\endcsname%
      \put(3082,298){\makebox(0,0){\strut{}$580$}}%
      \csname LTb\endcsname%
      \put(3535,298){\makebox(0,0){\strut{}$600$}}%
      \csname LTb\endcsname%
      \put(3989,298){\makebox(0,0){\strut{}$620$}}%
      \csname LTb\endcsname%
      \put(4442,298){\makebox(0,0){\strut{}$640$}}%
      \csname LTb\endcsname%
      \put(4895,298){\makebox(0,0){\strut{}$660$}}%
      \csname LTb\endcsname%
      \put(1428,1620){\makebox(0,0)[l]{\strut{}w}}%
    }%
    \gplgaddtomacro\gplfronttext{%
      \csname LTb\endcsname%
      \put(165,1221){\rotatebox{-270}{\makebox(0,0){\strut{}$A_{\text{window}}$}}}%
      \csname LTb\endcsname%
      \put(2855,19){\makebox(0,0){\strut{}$\unit[f /]{d^{-1}}$}}%
    }%
    \gplbacktext
    \put(0,0){\includegraphics[scale=0.5]{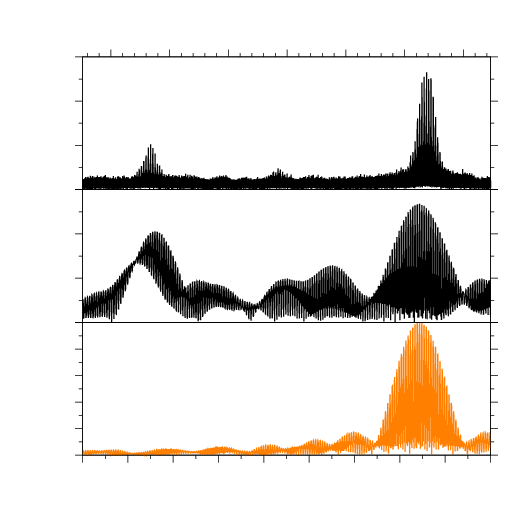}}%
    \gplfronttext
  \end{picture}%
\endgroup


        \end{subfigure} 
        \caption{Example observations of V1636~Ori from March 20, 21, and 22, 2009 at the CAHA (left, from top to bottom), combined used as one $ O-C $ measurement. The error bar on the left of each plot represents the photometric uncertainty. The red line shows the simultaneous two frequency fit to the epoch data. The amplitude spectra on the right hand side show the spectrum of the full data set (top), the spectrum of this epoch (middle) and the respective window function computed at $ f_1 $.}
        \label{fig:v1636-lc-crop}
\end{figure*}

\begin{figure*}[tp]
        \centering
\begingroup
  \makeatletter
  \providecommand\color[2][]{%
    \GenericError{(gnuplot) \space\space\space\@spaces}{%
      Package color not loaded in conjunction with
      terminal option `colourtext'%
    }{See the gnuplot documentation for explanation.%
    }{Either use 'blacktext' in gnuplot or load the package
      color.sty in LaTeX.}%
    \renewcommand\color[2][]{}%
  }%
  \providecommand\includegraphics[2][]{%
    \GenericError{(gnuplot) \space\space\space\@spaces}{%
      Package graphicx or graphics not loaded%
    }{See the gnuplot documentation for explanation.%
    }{The gnuplot epslatex terminal needs graphicx.sty or graphics.sty.}
  }%
  \providecommand\rotatebox[2]{#2}%
  \@ifundefined{ifGPcolor}{%
    \newif\ifGPcolor
    \GPcolortrue
  }{}%
  \@ifundefined{ifGPblacktext}{%
    \newif\ifGPblacktext
    \GPblacktexttrue
  }{}%
  \let\gplgaddtomacro\g@addto@macro
  \gdef\gplbacktext{}%
  \gdef\gplfronttext{}%
  \makeatother
  \ifGPblacktext
    \def\colorrgb#1{}%
    \def\colorgray#1{}%
  \else
    \ifGPcolor
      \def\colorrgb#1{\color[rgb]{#1}}%
      \def\colorgray#1{\color[gray]{#1}}%
      \expandafter\def\csname LTw\endcsname{\color{white}}%
      \expandafter\def\csname LTb\endcsname{\color{black}}%
      \expandafter\def\csname LTa\endcsname{\color{black}}%
      \expandafter\def\csname LT0\endcsname{\color[rgb]{1,0,0}}%
      \expandafter\def\csname LT1\endcsname{\color[rgb]{0,1,0}}%
      \expandafter\def\csname LT2\endcsname{\color[rgb]{0,0,1}}%
      \expandafter\def\csname LT3\endcsname{\color[rgb]{1,0,1}}%
      \expandafter\def\csname LT4\endcsname{\color[rgb]{0,1,1}}%
      \expandafter\def\csname LT5\endcsname{\color[rgb]{1,1,0}}%
      \expandafter\def\csname LT6\endcsname{\color[rgb]{0,0,0}}%
      \expandafter\def\csname LT7\endcsname{\color[rgb]{1,0.3,0}}%
      \expandafter\def\csname LT8\endcsname{\color[rgb]{0.5,0.5,0.5}}%
    \else
      \def\colorrgb#1{\color{black}}%
      \def\colorgray#1{\color[gray]{#1}}%
      \expandafter\def\csname LTw\endcsname{\color{white}}%
      \expandafter\def\csname LTb\endcsname{\color{black}}%
      \expandafter\def\csname LTa\endcsname{\color{black}}%
      \expandafter\def\csname LT0\endcsname{\color{black}}%
      \expandafter\def\csname LT1\endcsname{\color{black}}%
      \expandafter\def\csname LT2\endcsname{\color{black}}%
      \expandafter\def\csname LT3\endcsname{\color{black}}%
      \expandafter\def\csname LT4\endcsname{\color{black}}%
      \expandafter\def\csname LT5\endcsname{\color{black}}%
      \expandafter\def\csname LT6\endcsname{\color{black}}%
      \expandafter\def\csname LT7\endcsname{\color{black}}%
      \expandafter\def\csname LT8\endcsname{\color{black}}%
    \fi
  \fi
    \setlength{\unitlength}{0.0500bp}%
    \ifx\gptboxheight\undefined%
      \newlength{\gptboxheight}%
      \newlength{\gptboxwidth}%
      \newsavebox{\gptboxtext}%
    \fi%
    \setlength{\fboxrule}{0.5pt}%
    \setlength{\fboxsep}{1pt}%
\begin{picture}(10420.00,5100.00)%
    \gplgaddtomacro\gplbacktext{%
      \csname LTb\endcsname%
      \put(742,3998){\makebox(0,0)[r]{\strut{}$0.96$}}%
      \csname LTb\endcsname%
      \put(742,4201){\makebox(0,0)[r]{\strut{}$0.98$}}%
      \csname LTb\endcsname%
      \put(742,4405){\makebox(0,0)[r]{\strut{}$1$}}%
      \csname LTb\endcsname%
      \put(742,4608){\makebox(0,0)[r]{\strut{}$1.02$}}%
      \csname LTb\endcsname%
      \put(742,4811){\makebox(0,0)[r]{\strut{}$1.04$}}%
      \csname LTb\endcsname%
      \put(918,3636){\makebox(0,0){\strut{}8.76}}%
      \csname LTb\endcsname%
      \put(2451,3636){\makebox(0,0){\strut{}8.80}}%
      \csname LTb\endcsname%
      \put(3983,3636){\makebox(0,0){\strut{}8.84}}%
      \csname LTb\endcsname%
      \put(5516,3636){\makebox(0,0){\strut{}8.88}}%
      \csname LTb\endcsname%
      \put(7048,3636){\makebox(0,0){\strut{}8.92}}%
      \csname LTb\endcsname%
      \put(8581,3636){\makebox(0,0){\strut{}8.96}}%
      \csname LTb\endcsname%
      \put(10113,3636){\makebox(0,0){\strut{}9.00}}%
    }%
    \gplgaddtomacro\gplfronttext{%
      \csname LTb\endcsname%
      \put(139,4404){\rotatebox{-270}{\makebox(0,0){\strut{}$F$}}}%
    }%
    \gplgaddtomacro\gplbacktext{%
      \csname LTb\endcsname%
      \put(742,2422){\makebox(0,0)[r]{\strut{}$0.96$}}%
      \csname LTb\endcsname%
      \put(742,2625){\makebox(0,0)[r]{\strut{}$0.98$}}%
      \csname LTb\endcsname%
      \put(742,2829){\makebox(0,0)[r]{\strut{}$1$}}%
      \csname LTb\endcsname%
      \put(742,3032){\makebox(0,0)[r]{\strut{}$1.02$}}%
      \csname LTb\endcsname%
      \put(742,3235){\makebox(0,0)[r]{\strut{}$1.04$}}%
      \csname LTb\endcsname%
      \put(918,2060){\makebox(0,0){\strut{}9.76}}%
      \csname LTb\endcsname%
      \put(2451,2060){\makebox(0,0){\strut{}9.80}}%
      \csname LTb\endcsname%
      \put(3983,2060){\makebox(0,0){\strut{}9.84}}%
      \csname LTb\endcsname%
      \put(5516,2060){\makebox(0,0){\strut{}9.88}}%
      \csname LTb\endcsname%
      \put(7048,2060){\makebox(0,0){\strut{}9.92}}%
      \csname LTb\endcsname%
      \put(8581,2060){\makebox(0,0){\strut{}9.96}}%
      \csname LTb\endcsname%
      \put(10113,2060){\makebox(0,0){\strut{}10.00}}%
    }%
    \gplgaddtomacro\gplfronttext{%
      \csname LTb\endcsname%
      \put(139,2828){\rotatebox{-270}{\makebox(0,0){\strut{}$F$}}}%
    }%
    \gplgaddtomacro\gplbacktext{%
      \csname LTb\endcsname%
      \put(742,846){\makebox(0,0)[r]{\strut{}$0.96$}}%
      \csname LTb\endcsname%
      \put(742,1049){\makebox(0,0)[r]{\strut{}$0.98$}}%
      \csname LTb\endcsname%
      \put(742,1253){\makebox(0,0)[r]{\strut{}$1$}}%
      \csname LTb\endcsname%
      \put(742,1456){\makebox(0,0)[r]{\strut{}$1.02$}}%
      \csname LTb\endcsname%
      \put(742,1659){\makebox(0,0)[r]{\strut{}$1.04$}}%
      \csname LTb\endcsname%
      \put(918,484){\makebox(0,0){\strut{}10.72}}%
      \csname LTb\endcsname%
      \put(2451,484){\makebox(0,0){\strut{}10.76}}%
      \csname LTb\endcsname%
      \put(3983,484){\makebox(0,0){\strut{}10.80}}%
      \csname LTb\endcsname%
      \put(5516,484){\makebox(0,0){\strut{}10.84}}%
      \csname LTb\endcsname%
      \put(7048,484){\makebox(0,0){\strut{}10.88}}%
      \csname LTb\endcsname%
      \put(8581,484){\makebox(0,0){\strut{}10.92}}%
      \csname LTb\endcsname%
      \put(10113,484){\makebox(0,0){\strut{}10.96}}%
    }%
    \gplgaddtomacro\gplfronttext{%
      \csname LTb\endcsname%
      \put(139,1252){\rotatebox{-270}{\makebox(0,0){\strut{}$F$}}}%
      \csname LTb\endcsname%
      \put(5515,205){\makebox(0,0){\strut{}$\unit[t /]{d}$ since 2012-02-13}}%
    }%
    \gplbacktext
    \put(0,0){\includegraphics[scale=0.5]{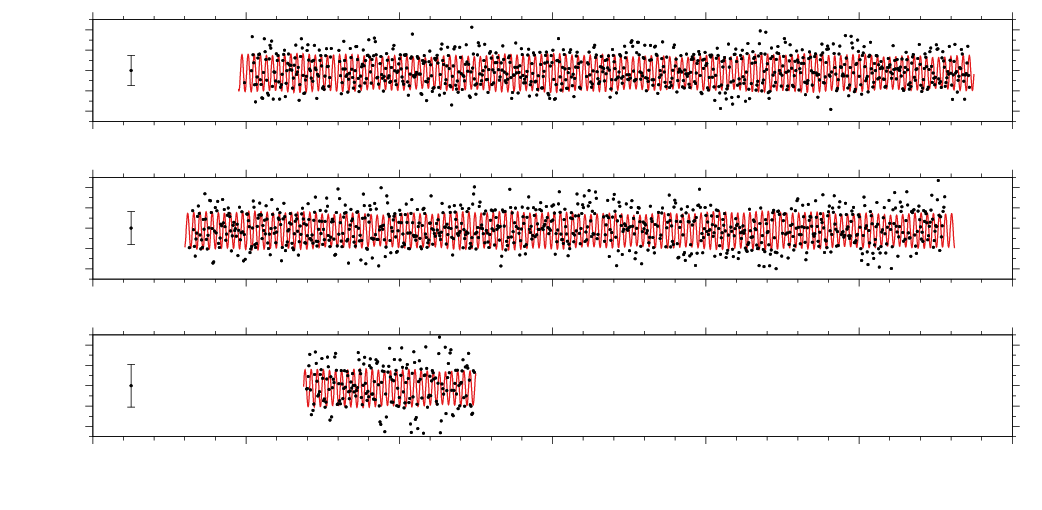}}%
    \gplfronttext
  \end{picture}%
\endgroup


\begingroup
  \makeatletter
  \providecommand\color[2][]{%
    \GenericError{(gnuplot) \space\space\space\@spaces}{%
      Package color not loaded in conjunction with
      terminal option `colourtext'%
    }{See the gnuplot documentation for explanation.%
    }{Either use 'blacktext' in gnuplot or load the package
      color.sty in LaTeX.}%
    \renewcommand\color[2][]{}%
  }%
  \providecommand\includegraphics[2][]{%
    \GenericError{(gnuplot) \space\space\space\@spaces}{%
      Package graphicx or graphics not loaded%
    }{See the gnuplot documentation for explanation.%
    }{The gnuplot epslatex terminal needs graphicx.sty or graphics.sty.}
  }%
  \providecommand\rotatebox[2]{#2}%
  \@ifundefined{ifGPcolor}{%
    \newif\ifGPcolor
    \GPcolortrue
  }{}%
  \@ifundefined{ifGPblacktext}{%
    \newif\ifGPblacktext
    \GPblacktexttrue
  }{}%
  \let\gplgaddtomacro\g@addto@macro
  \gdef\gplbacktext{}%
  \gdef\gplfronttext{}%
  \makeatother
  \ifGPblacktext
    \def\colorrgb#1{}%
    \def\colorgray#1{}%
  \else
    \ifGPcolor
      \def\colorrgb#1{\color[rgb]{#1}}%
      \def\colorgray#1{\color[gray]{#1}}%
      \expandafter\def\csname LTw\endcsname{\color{white}}%
      \expandafter\def\csname LTb\endcsname{\color{black}}%
      \expandafter\def\csname LTa\endcsname{\color{black}}%
      \expandafter\def\csname LT0\endcsname{\color[rgb]{1,0,0}}%
      \expandafter\def\csname LT1\endcsname{\color[rgb]{0,1,0}}%
      \expandafter\def\csname LT2\endcsname{\color[rgb]{0,0,1}}%
      \expandafter\def\csname LT3\endcsname{\color[rgb]{1,0,1}}%
      \expandafter\def\csname LT4\endcsname{\color[rgb]{0,1,1}}%
      \expandafter\def\csname LT5\endcsname{\color[rgb]{1,1,0}}%
      \expandafter\def\csname LT6\endcsname{\color[rgb]{0,0,0}}%
      \expandafter\def\csname LT7\endcsname{\color[rgb]{1,0.3,0}}%
      \expandafter\def\csname LT8\endcsname{\color[rgb]{0.5,0.5,0.5}}%
    \else
      \def\colorrgb#1{\color{black}}%
      \def\colorgray#1{\color[gray]{#1}}%
      \expandafter\def\csname LTw\endcsname{\color{white}}%
      \expandafter\def\csname LTb\endcsname{\color{black}}%
      \expandafter\def\csname LTa\endcsname{\color{black}}%
      \expandafter\def\csname LT0\endcsname{\color{black}}%
      \expandafter\def\csname LT1\endcsname{\color{black}}%
      \expandafter\def\csname LT2\endcsname{\color{black}}%
      \expandafter\def\csname LT3\endcsname{\color{black}}%
      \expandafter\def\csname LT4\endcsname{\color{black}}%
      \expandafter\def\csname LT5\endcsname{\color{black}}%
      \expandafter\def\csname LT6\endcsname{\color{black}}%
      \expandafter\def\csname LT7\endcsname{\color{black}}%
      \expandafter\def\csname LT8\endcsname{\color{black}}%
    \fi
  \fi
    \setlength{\unitlength}{0.0500bp}%
    \ifx\gptboxheight\undefined%
      \newlength{\gptboxheight}%
      \newlength{\gptboxwidth}%
      \newsavebox{\gptboxtext}%
    \fi%
    \setlength{\fboxrule}{0.5pt}%
    \setlength{\fboxsep}{1pt}%
\begin{picture}(10420.00,5100.00)%
    \gplgaddtomacro\gplbacktext{%
      \csname LTb\endcsname%
      \put(742,3896){\makebox(0,0)[r]{\strut{}$0.96$}}%
      \csname LTb\endcsname%
      \put(742,4150){\makebox(0,0)[r]{\strut{}$0.98$}}%
      \csname LTb\endcsname%
      \put(742,4405){\makebox(0,0)[r]{\strut{}$1$}}%
      \csname LTb\endcsname%
      \put(742,4659){\makebox(0,0)[r]{\strut{}$1.02$}}%
      \csname LTb\endcsname%
      \put(742,4913){\makebox(0,0)[r]{\strut{}$1.04$}}%
      \csname LTb\endcsname%
      \put(918,3636){\makebox(0,0){\strut{}2.84}}%
      \csname LTb\endcsname%
      \put(2232,3636){\makebox(0,0){\strut{}2.88}}%
      \csname LTb\endcsname%
      \put(3545,3636){\makebox(0,0){\strut{}2.92}}%
      \csname LTb\endcsname%
      \put(4859,3636){\makebox(0,0){\strut{}2.96}}%
      \csname LTb\endcsname%
      \put(6172,3636){\makebox(0,0){\strut{}3.00}}%
      \csname LTb\endcsname%
      \put(7486,3636){\makebox(0,0){\strut{}3.04}}%
      \csname LTb\endcsname%
      \put(8799,3636){\makebox(0,0){\strut{}3.08}}%
      \csname LTb\endcsname%
      \put(10113,3636){\makebox(0,0){\strut{}3.12}}%
    }%
    \gplgaddtomacro\gplfronttext{%
      \csname LTb\endcsname%
      \put(139,4404){\rotatebox{-270}{\makebox(0,0){\strut{}$F$}}}%
    }%
    \gplgaddtomacro\gplbacktext{%
      \csname LTb\endcsname%
      \put(742,2320){\makebox(0,0)[r]{\strut{}$0.96$}}%
      \csname LTb\endcsname%
      \put(742,2574){\makebox(0,0)[r]{\strut{}$0.98$}}%
      \csname LTb\endcsname%
      \put(742,2829){\makebox(0,0)[r]{\strut{}$1$}}%
      \csname LTb\endcsname%
      \put(742,3083){\makebox(0,0)[r]{\strut{}$1.02$}}%
      \csname LTb\endcsname%
      \put(742,3337){\makebox(0,0)[r]{\strut{}$1.04$}}%
      \csname LTb\endcsname%
      \put(918,2060){\makebox(0,0){\strut{}3.80}}%
      \csname LTb\endcsname%
      \put(2232,2060){\makebox(0,0){\strut{}3.84}}%
      \csname LTb\endcsname%
      \put(3545,2060){\makebox(0,0){\strut{}3.88}}%
      \csname LTb\endcsname%
      \put(4859,2060){\makebox(0,0){\strut{}3.92}}%
      \csname LTb\endcsname%
      \put(6172,2060){\makebox(0,0){\strut{}3.96}}%
      \csname LTb\endcsname%
      \put(7486,2060){\makebox(0,0){\strut{}4.00}}%
      \csname LTb\endcsname%
      \put(8799,2060){\makebox(0,0){\strut{}4.04}}%
      \csname LTb\endcsname%
      \put(10113,2060){\makebox(0,0){\strut{}4.08}}%
    }%
    \gplgaddtomacro\gplfronttext{%
      \csname LTb\endcsname%
      \put(139,2828){\rotatebox{-270}{\makebox(0,0){\strut{}$F$}}}%
    }%
    \gplgaddtomacro\gplbacktext{%
      \csname LTb\endcsname%
      \put(742,744){\makebox(0,0)[r]{\strut{}$0.96$}}%
      \csname LTb\endcsname%
      \put(742,998){\makebox(0,0)[r]{\strut{}$0.98$}}%
      \csname LTb\endcsname%
      \put(742,1253){\makebox(0,0)[r]{\strut{}$1$}}%
      \csname LTb\endcsname%
      \put(742,1507){\makebox(0,0)[r]{\strut{}$1.02$}}%
      \csname LTb\endcsname%
      \put(742,1761){\makebox(0,0)[r]{\strut{}$1.04$}}%
      \csname LTb\endcsname%
      \put(918,484){\makebox(0,0){\strut{}8.88}}%
      \csname LTb\endcsname%
      \put(2232,484){\makebox(0,0){\strut{}8.92}}%
      \csname LTb\endcsname%
      \put(3545,484){\makebox(0,0){\strut{}8.96}}%
      \csname LTb\endcsname%
      \put(4859,484){\makebox(0,0){\strut{}9.00}}%
      \csname LTb\endcsname%
      \put(6172,484){\makebox(0,0){\strut{}9.04}}%
      \csname LTb\endcsname%
      \put(7486,484){\makebox(0,0){\strut{}9.08}}%
      \csname LTb\endcsname%
      \put(8799,484){\makebox(0,0){\strut{}9.12}}%
      \csname LTb\endcsname%
      \put(10113,484){\makebox(0,0){\strut{}9.16}}%
    }%
    \gplgaddtomacro\gplfronttext{%
      \csname LTb\endcsname%
      \put(139,1252){\rotatebox{-270}{\makebox(0,0){\strut{}$F$}}}%
      \csname LTb\endcsname%
      \put(5515,205){\makebox(0,0){\strut{}$\unit[t /]{d}$ since 2008-02-04}}%
    }%
    \gplbacktext
    \put(0,0){\includegraphics[scale=0.5]{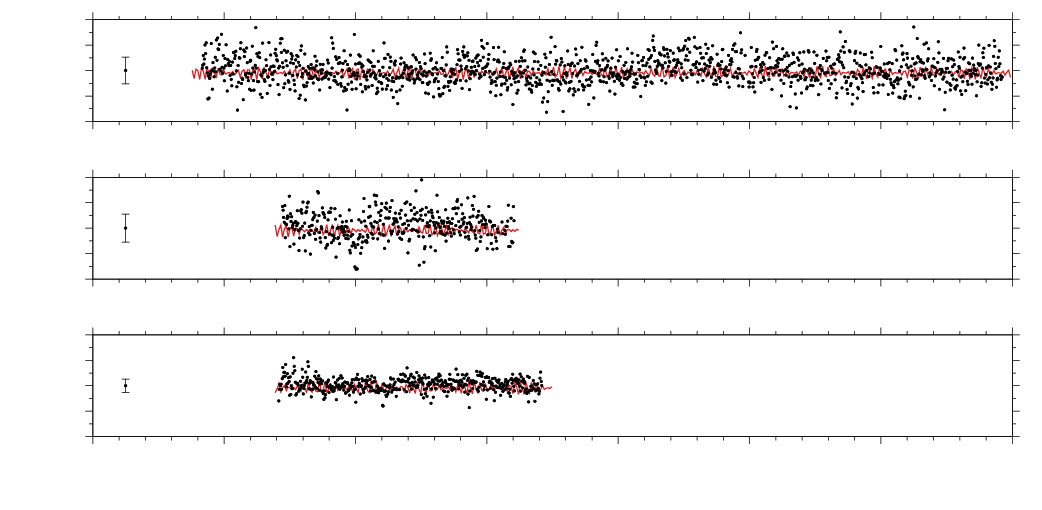}}%
    \gplfronttext
  \end{picture}%
\endgroup


        \caption{Example observations of QQ Vir from February 20, 21, and 22, 2012 at the Monet telescope (top three panels), and V541~Hya from February 6, 7, and 12, 2008 at the SAAO (bottom three panels), combined used as one $ O-C $ measurement each. The error bar on the left of each plot represents the photometric uncertainty. The red line shows the simultaneous three and two frequency fit to the epoch data, repsectively.}
        \label{fig:qqvir-v541hya-lc-crop}
\end{figure*}

\begin{figure*}[tp]
        \centering
        \begin{subfigure}[b]{0.49\textwidth}
                \centering
\begingroup
  \makeatletter
  \providecommand\color[2][]{%
    \GenericError{(gnuplot) \space\space\space\@spaces}{%
      Package color not loaded in conjunction with
      terminal option `colourtext'%
    }{See the gnuplot documentation for explanation.%
    }{Either use 'blacktext' in gnuplot or load the package
      color.sty in LaTeX.}%
    \renewcommand\color[2][]{}%
  }%
  \providecommand\includegraphics[2][]{%
    \GenericError{(gnuplot) \space\space\space\@spaces}{%
      Package graphicx or graphics not loaded%
    }{See the gnuplot documentation for explanation.%
    }{The gnuplot epslatex terminal needs graphicx.sty or graphics.sty.}
  }%
  \providecommand\rotatebox[2]{#2}%
  \@ifundefined{ifGPcolor}{%
    \newif\ifGPcolor
    \GPcolortrue
  }{}%
  \@ifundefined{ifGPblacktext}{%
    \newif\ifGPblacktext
    \GPblacktexttrue
  }{}%
  \let\gplgaddtomacro\g@addto@macro
  \gdef\gplbacktext{}%
  \gdef\gplfronttext{}%
  \makeatother
  \ifGPblacktext
    \def\colorrgb#1{}%
    \def\colorgray#1{}%
  \else
    \ifGPcolor
      \def\colorrgb#1{\color[rgb]{#1}}%
      \def\colorgray#1{\color[gray]{#1}}%
      \expandafter\def\csname LTw\endcsname{\color{white}}%
      \expandafter\def\csname LTb\endcsname{\color{black}}%
      \expandafter\def\csname LTa\endcsname{\color{black}}%
      \expandafter\def\csname LT0\endcsname{\color[rgb]{1,0,0}}%
      \expandafter\def\csname LT1\endcsname{\color[rgb]{0,1,0}}%
      \expandafter\def\csname LT2\endcsname{\color[rgb]{0,0,1}}%
      \expandafter\def\csname LT3\endcsname{\color[rgb]{1,0,1}}%
      \expandafter\def\csname LT4\endcsname{\color[rgb]{0,1,1}}%
      \expandafter\def\csname LT5\endcsname{\color[rgb]{1,1,0}}%
      \expandafter\def\csname LT6\endcsname{\color[rgb]{0,0,0}}%
      \expandafter\def\csname LT7\endcsname{\color[rgb]{1,0.3,0}}%
      \expandafter\def\csname LT8\endcsname{\color[rgb]{0.5,0.5,0.5}}%
    \else
      \def\colorrgb#1{\color{black}}%
      \def\colorgray#1{\color[gray]{#1}}%
      \expandafter\def\csname LTw\endcsname{\color{white}}%
      \expandafter\def\csname LTb\endcsname{\color{black}}%
      \expandafter\def\csname LTa\endcsname{\color{black}}%
      \expandafter\def\csname LT0\endcsname{\color{black}}%
      \expandafter\def\csname LT1\endcsname{\color{black}}%
      \expandafter\def\csname LT2\endcsname{\color{black}}%
      \expandafter\def\csname LT3\endcsname{\color{black}}%
      \expandafter\def\csname LT4\endcsname{\color{black}}%
      \expandafter\def\csname LT5\endcsname{\color{black}}%
      \expandafter\def\csname LT6\endcsname{\color{black}}%
      \expandafter\def\csname LT7\endcsname{\color{black}}%
      \expandafter\def\csname LT8\endcsname{\color{black}}%
    \fi
  \fi
    \setlength{\unitlength}{0.0500bp}%
    \ifx\gptboxheight\undefined%
      \newlength{\gptboxheight}%
      \newlength{\gptboxwidth}%
      \newsavebox{\gptboxtext}%
    \fi%
    \setlength{\fboxrule}{0.5pt}%
    \setlength{\fboxsep}{1pt}%
\begin{picture}(5100.00,5100.00)%
    \gplgaddtomacro\gplbacktext{%
      \csname LTb\endcsname%
      \put(640,3214){\makebox(0,0)[r]{\strut{}$0$}}%
      \csname LTb\endcsname%
      \put(640,3479){\makebox(0,0)[r]{\strut{}$0.5$}}%
      \csname LTb\endcsname%
      \put(640,3745){\makebox(0,0)[r]{\strut{}$1$}}%
      \csname LTb\endcsname%
      \put(640,4010){\makebox(0,0)[r]{\strut{}$1.5$}}%
      \csname LTb\endcsname%
      \put(640,4276){\makebox(0,0)[r]{\strut{}$2$}}%
      \csname LTb\endcsname%
      \put(640,4541){\makebox(0,0)[r]{\strut{}$2.5$}}%
      \csname LTb\endcsname%
      \put(970,4801){\makebox(0,0){\strut{}$6200$}}%
      \csname LTb\endcsname%
      \put(1440,4801){\makebox(0,0){\strut{}$6400$}}%
      \csname LTb\endcsname%
      \put(1910,4801){\makebox(0,0){\strut{}$6600$}}%
      \csname LTb\endcsname%
      \put(2380,4801){\makebox(0,0){\strut{}$6800$}}%
      \csname LTb\endcsname%
      \put(2850,4801){\makebox(0,0){\strut{}$7000$}}%
      \csname LTb\endcsname%
      \put(3320,4801){\makebox(0,0){\strut{}$7200$}}%
      \csname LTb\endcsname%
      \put(3790,4801){\makebox(0,0){\strut{}$7400$}}%
      \csname LTb\endcsname%
      \put(4260,4801){\makebox(0,0){\strut{}$7600$}}%
      \csname LTb\endcsname%
      \put(4730,4801){\makebox(0,0){\strut{}$7800$}}%
      \csname LTb\endcsname%
      \put(1428,4276){\makebox(0,0)[l]{\strut{}full}}%
    }%
    \gplgaddtomacro\gplfronttext{%
      \csname LTb\endcsname%
      \put(165,3877){\rotatebox{-270}{\makebox(0,0){\strut{}$\unit[A_{full}/]{\%}$}}}%
      \csname LTb\endcsname%
      \put(2855,5079){\makebox(0,0){\strut{}$\unit[f /]{\mu Hz}$}}%
    }%
    \gplgaddtomacro\gplbacktext{%
      \csname LTb\endcsname%
      \put(640,1886){\makebox(0,0)[r]{\strut{}$0$}}%
      \csname LTb\endcsname%
      \put(640,2265){\makebox(0,0)[r]{\strut{}$0.5$}}%
      \csname LTb\endcsname%
      \put(640,2644){\makebox(0,0)[r]{\strut{}$1$}}%
      \csname LTb\endcsname%
      \put(640,3023){\makebox(0,0)[r]{\strut{}$1.5$}}%
      \csname LTb\endcsname%
      \put(1428,2948){\makebox(0,0)[l]{\strut{}epoch}}%
    }%
    \gplgaddtomacro\gplfronttext{%
      \csname LTb\endcsname%
      \put(165,2549){\rotatebox{-270}{\makebox(0,0){\strut{}$\unit[A_{epoch}/]{\%}$}}}%
    }%
    \gplgaddtomacro\gplbacktext{%
      \csname LTb\endcsname%
      \put(640,1885){\makebox(0,0)[r]{\strut{}}}%
      \csname LTb\endcsname%
      \put(640,558){\makebox(0,0)[r]{\strut{}$0$}}%
      \csname LTb\endcsname%
      \put(640,823){\makebox(0,0)[r]{\strut{}$0.2$}}%
      \csname LTb\endcsname%
      \put(640,1089){\makebox(0,0)[r]{\strut{}$0.4$}}%
      \csname LTb\endcsname%
      \put(640,1354){\makebox(0,0)[r]{\strut{}$0.6$}}%
      \csname LTb\endcsname%
      \put(640,1620){\makebox(0,0)[r]{\strut{}$0.8$}}%
      \csname LTb\endcsname%
      \put(1088,298){\makebox(0,0){\strut{}$540$}}%
      \csname LTb\endcsname%
      \put(1632,298){\makebox(0,0){\strut{}$560$}}%
      \csname LTb\endcsname%
      \put(2176,298){\makebox(0,0){\strut{}$580$}}%
      \csname LTb\endcsname%
      \put(2720,298){\makebox(0,0){\strut{}$600$}}%
      \csname LTb\endcsname%
      \put(3263,298){\makebox(0,0){\strut{}$620$}}%
      \csname LTb\endcsname%
      \put(3807,298){\makebox(0,0){\strut{}$640$}}%
      \csname LTb\endcsname%
      \put(4351,298){\makebox(0,0){\strut{}$660$}}%
      \csname LTb\endcsname%
      \put(4895,298){\makebox(0,0){\strut{}$680$}}%
      \csname LTb\endcsname%
      \put(1428,1620){\makebox(0,0)[l]{\strut{}w}}%
    }%
    \gplgaddtomacro\gplfronttext{%
      \csname LTb\endcsname%
      \put(165,1221){\rotatebox{-270}{\makebox(0,0){\strut{}$A_{\text{window}}$}}}%
      \csname LTb\endcsname%
      \put(2855,19){\makebox(0,0){\strut{}$\unit[f /]{d^{-1}}$}}%
    }%
    \gplbacktext
    \put(0,0){\includegraphics[scale=0.5]{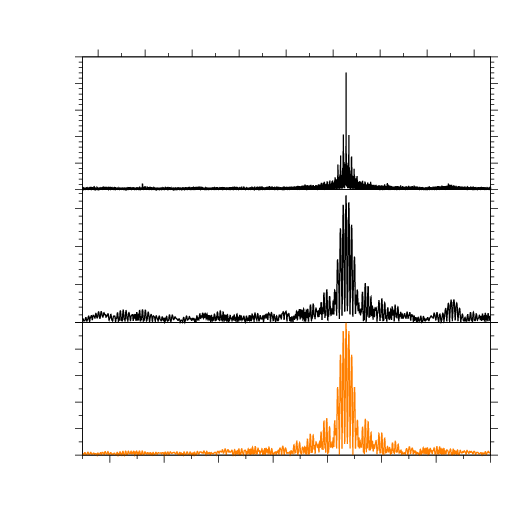}}%
    \gplfronttext
  \end{picture}%
\endgroup


        \end{subfigure}
        \begin{subfigure}[b]{0.49\textwidth}
                \centering
\begingroup
  \makeatletter
  \providecommand\color[2][]{%
    \GenericError{(gnuplot) \space\space\space\@spaces}{%
      Package color not loaded in conjunction with
      terminal option `colourtext'%
    }{See the gnuplot documentation for explanation.%
    }{Either use 'blacktext' in gnuplot or load the package
      color.sty in LaTeX.}%
    \renewcommand\color[2][]{}%
  }%
  \providecommand\includegraphics[2][]{%
    \GenericError{(gnuplot) \space\space\space\@spaces}{%
      Package graphicx or graphics not loaded%
    }{See the gnuplot documentation for explanation.%
    }{The gnuplot epslatex terminal needs graphicx.sty or graphics.sty.}
  }%
  \providecommand\rotatebox[2]{#2}%
  \@ifundefined{ifGPcolor}{%
    \newif\ifGPcolor
    \GPcolortrue
  }{}%
  \@ifundefined{ifGPblacktext}{%
    \newif\ifGPblacktext
    \GPblacktexttrue
  }{}%
  \let\gplgaddtomacro\g@addto@macro
  \gdef\gplbacktext{}%
  \gdef\gplfronttext{}%
  \makeatother
  \ifGPblacktext
    \def\colorrgb#1{}%
    \def\colorgray#1{}%
  \else
    \ifGPcolor
      \def\colorrgb#1{\color[rgb]{#1}}%
      \def\colorgray#1{\color[gray]{#1}}%
      \expandafter\def\csname LTw\endcsname{\color{white}}%
      \expandafter\def\csname LTb\endcsname{\color{black}}%
      \expandafter\def\csname LTa\endcsname{\color{black}}%
      \expandafter\def\csname LT0\endcsname{\color[rgb]{1,0,0}}%
      \expandafter\def\csname LT1\endcsname{\color[rgb]{0,1,0}}%
      \expandafter\def\csname LT2\endcsname{\color[rgb]{0,0,1}}%
      \expandafter\def\csname LT3\endcsname{\color[rgb]{1,0,1}}%
      \expandafter\def\csname LT4\endcsname{\color[rgb]{0,1,1}}%
      \expandafter\def\csname LT5\endcsname{\color[rgb]{1,1,0}}%
      \expandafter\def\csname LT6\endcsname{\color[rgb]{0,0,0}}%
      \expandafter\def\csname LT7\endcsname{\color[rgb]{1,0.3,0}}%
      \expandafter\def\csname LT8\endcsname{\color[rgb]{0.5,0.5,0.5}}%
    \else
      \def\colorrgb#1{\color{black}}%
      \def\colorgray#1{\color[gray]{#1}}%
      \expandafter\def\csname LTw\endcsname{\color{white}}%
      \expandafter\def\csname LTb\endcsname{\color{black}}%
      \expandafter\def\csname LTa\endcsname{\color{black}}%
      \expandafter\def\csname LT0\endcsname{\color{black}}%
      \expandafter\def\csname LT1\endcsname{\color{black}}%
      \expandafter\def\csname LT2\endcsname{\color{black}}%
      \expandafter\def\csname LT3\endcsname{\color{black}}%
      \expandafter\def\csname LT4\endcsname{\color{black}}%
      \expandafter\def\csname LT5\endcsname{\color{black}}%
      \expandafter\def\csname LT6\endcsname{\color{black}}%
      \expandafter\def\csname LT7\endcsname{\color{black}}%
      \expandafter\def\csname LT8\endcsname{\color{black}}%
    \fi
  \fi
    \setlength{\unitlength}{0.0500bp}%
    \ifx\gptboxheight\undefined%
      \newlength{\gptboxheight}%
      \newlength{\gptboxwidth}%
      \newsavebox{\gptboxtext}%
    \fi%
    \setlength{\fboxrule}{0.5pt}%
    \setlength{\fboxsep}{1pt}%
\begin{picture}(5100.00,5100.00)%
    \gplgaddtomacro\gplbacktext{%
      \csname LTb\endcsname%
      \put(640,3214){\makebox(0,0)[r]{\strut{}$0$}}%
      \csname LTb\endcsname%
      \put(640,3656){\makebox(0,0)[r]{\strut{}$0.1$}}%
      \csname LTb\endcsname%
      \put(640,4099){\makebox(0,0)[r]{\strut{}$0.2$}}%
      \csname LTb\endcsname%
      \put(640,4541){\makebox(0,0)[r]{\strut{}$0.3$}}%
      \csname LTb\endcsname%
      \put(1194,4801){\makebox(0,0){\strut{}$6400$}}%
      \csname LTb\endcsname%
      \put(1697,4801){\makebox(0,0){\strut{}$6600$}}%
      \csname LTb\endcsname%
      \put(2201,4801){\makebox(0,0){\strut{}$6800$}}%
      \csname LTb\endcsname%
      \put(2704,4801){\makebox(0,0){\strut{}$7000$}}%
      \csname LTb\endcsname%
      \put(3207,4801){\makebox(0,0){\strut{}$7200$}}%
      \csname LTb\endcsname%
      \put(3711,4801){\makebox(0,0){\strut{}$7400$}}%
      \csname LTb\endcsname%
      \put(4214,4801){\makebox(0,0){\strut{}$7600$}}%
      \csname LTb\endcsname%
      \put(4718,4801){\makebox(0,0){\strut{}$7800$}}%
      \csname LTb\endcsname%
      \put(1428,4276){\makebox(0,0)[l]{\strut{}full}}%
    }%
    \gplgaddtomacro\gplfronttext{%
      \csname LTb\endcsname%
      \put(165,3877){\rotatebox{-270}{\makebox(0,0){\strut{}$\unit[A_{full}/]{\%}$}}}%
      \csname LTb\endcsname%
      \put(2855,5079){\makebox(0,0){\strut{}$\unit[f /]{\mu Hz}$}}%
    }%
    \gplgaddtomacro\gplbacktext{%
      \csname LTb\endcsname%
      \put(640,1886){\makebox(0,0)[r]{\strut{}$0$}}%
      \csname LTb\endcsname%
      \put(640,2417){\makebox(0,0)[r]{\strut{}$0.2$}}%
      \csname LTb\endcsname%
      \put(640,2948){\makebox(0,0)[r]{\strut{}$0.4$}}%
      \csname LTb\endcsname%
      \put(1428,2948){\makebox(0,0)[l]{\strut{}epoch}}%
    }%
    \gplgaddtomacro\gplfronttext{%
      \csname LTb\endcsname%
      \put(165,2549){\rotatebox{-270}{\makebox(0,0){\strut{}$\unit[A_{epoch}/]{\%}$}}}%
    }%
    \gplgaddtomacro\gplbacktext{%
      \csname LTb\endcsname%
      \put(640,1885){\makebox(0,0)[r]{\strut{}}}%
      \csname LTb\endcsname%
      \put(640,558){\makebox(0,0)[r]{\strut{}$0$}}%
      \csname LTb\endcsname%
      \put(640,823){\makebox(0,0)[r]{\strut{}$0.2$}}%
      \csname LTb\endcsname%
      \put(640,1089){\makebox(0,0)[r]{\strut{}$0.4$}}%
      \csname LTb\endcsname%
      \put(640,1354){\makebox(0,0)[r]{\strut{}$0.6$}}%
      \csname LTb\endcsname%
      \put(640,1620){\makebox(0,0)[r]{\strut{}$0.8$}}%
      \csname LTb\endcsname%
      \put(816,298){\makebox(0,0){\strut{}$540$}}%
      \csname LTb\endcsname%
      \put(1399,298){\makebox(0,0){\strut{}$560$}}%
      \csname LTb\endcsname%
      \put(1981,298){\makebox(0,0){\strut{}$580$}}%
      \csname LTb\endcsname%
      \put(2564,298){\makebox(0,0){\strut{}$600$}}%
      \csname LTb\endcsname%
      \put(3147,298){\makebox(0,0){\strut{}$620$}}%
      \csname LTb\endcsname%
      \put(3730,298){\makebox(0,0){\strut{}$640$}}%
      \csname LTb\endcsname%
      \put(4312,298){\makebox(0,0){\strut{}$660$}}%
      \csname LTb\endcsname%
      \put(4895,298){\makebox(0,0){\strut{}$680$}}%
      \csname LTb\endcsname%
      \put(1428,1620){\makebox(0,0)[l]{\strut{}w}}%
    }%
    \gplgaddtomacro\gplfronttext{%
      \csname LTb\endcsname%
      \put(165,1221){\rotatebox{-270}{\makebox(0,0){\strut{}$A_{\text{window}}$}}}%
      \csname LTb\endcsname%
      \put(2855,19){\makebox(0,0){\strut{}$\unit[f /]{d^{-1}}$}}%
    }%
    \gplbacktext
    \put(0,0){\includegraphics[scale=0.5]{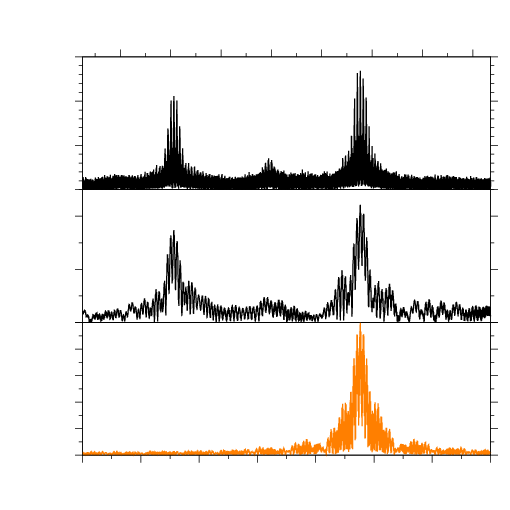}}%
    \gplfronttext
  \end{picture}%
\endgroup


        \end{subfigure}
        \caption{Amplitude spectra for epoch data in Fig.~\ref{fig:qqvir-v541hya-lc-crop} of QQ~Vir (left) and V541~Hya (right) show the spectrum of the full data set (top), the spectrum of this epoch (middle), and the respective window function computed at $ f_1 $.}
        \label{fig:qqvir-v541hya-epoch-power}
\end{figure*}

The results of the simultaneous fit for each target in this paper are summarised in Table~\ref{tab:fit}. We list pulsation modes not used for the timing analysis in Table~\ref{tab:add-freq}. Fig.~\ref{fig:dwlyn-lc-crop}, \ref{fig:v1636-lc-crop}, \ref{fig:qqvir-v541hya-lc-crop}, and \ref{fig:qqvir-v541hya-epoch-power} show example light curves of the targets for one epoch each, including their multi frequency fit and the respective amplitude spectrum.

\begin{table*}[t]
\centering
        \caption{Parameters of the simultaneously fitted pulsations per target over their full observational time span and the pulsation period $ P $. The phase $ \phi $ refers to the time corresponding to the first zero-crossing of the function after the first measurement $ t_0 $ in MBJD.}
        \label{tab:fit}
        \begin{tabular}{llr@{.}lr@{.}lr@{.}lr@{.}l}
                \toprule
                Target & & \multicolumn{2}{c}{$ \unit[f/]{d^{-1}} $} & \multicolumn{2}{c}{$ \unit[P/]{s} $} & \multicolumn{2}{c}{$ \unit[A/]{\%} $} & \multicolumn{2}{c}{$ \unit[\phi/]{d} $} \\
                \midrule
                DW Lyn& $ f_1 $ & 237 & 941160\,(8) & 363 & 114982(12) & 2 & 19\,(9) & 54394 & 741\,(5) \\
                & $ f_2 $ & 225 & 15898\,(5) & 383 & 72887(9) & 0 & 35\,(9) & 54394 & 74\,(3) \\
                V1636 Ori & $ f_{1} $ & 631 & 7346\,(2) & 136 & 76629(5) & 0 & 54\,(3) & 54698 & 72\,(3) \\
                & $ f_2 $ & 509 & 9780\,(3) & 169 & 4191(1) & 0 & 24\,(3) & 54698 & 72\,(7) \\
                QQ Vir & $ f_{1} $ & 626 & 877627\,(3) & 137 & 8259429(7) & 2 & 6\,(1) & 52117 & 924\,(6) \\
                 & $ f_{2} $ & 552 & 00714\,(9) & 156 & 51971(3) & 0 & 10\,(9) & 52117 & 9\,(1) \\
                 & $ f_{3} $ & 642 & 0515\,(1) & 134 & 56864(3) & 0 & 07\,(1) & 52117 & 9\,(2) \\
                V541 Hya & $ f_{1} $ & 635 & 32218\,(5) & 135 & 993993(11) & 0 & 31\,(8) & 53413 & 88\,(3) \\
                & $ f_2 $ & 571 & 28556\,(3) & 151 & 237850(8) & 0 & 21\,(7) &  53413 & 88\,(3) \\
                \bottomrule
        \end{tabular}
\end{table*}

\section{Results and discussion} \label{sec:results}
In the following, we discuss the implications of the obtained amplitude spectra and $ O-C $ measurements on the evolutionary state and presence of sub-stellar companions to the targets.

\subsection{DW Lyn} \label{sec:results-dwlyn}
\begin{figure}[tp]
        \centering
\begingroup
  \makeatletter
  \providecommand\color[2][]{%
    \GenericError{(gnuplot) \space\space\space\@spaces}{%
      Package color not loaded in conjunction with
      terminal option `colourtext'%
    }{See the gnuplot documentation for explanation.%
    }{Either use 'blacktext' in gnuplot or load the package
      color.sty in LaTeX.}%
    \renewcommand\color[2][]{}%
  }%
  \providecommand\includegraphics[2][]{%
    \GenericError{(gnuplot) \space\space\space\@spaces}{%
      Package graphicx or graphics not loaded%
    }{See the gnuplot documentation for explanation.%
    }{The gnuplot epslatex terminal needs graphicx.sty or graphics.sty.}
  }%
  \providecommand\rotatebox[2]{#2}%
  \@ifundefined{ifGPcolor}{%
    \newif\ifGPcolor
    \GPcolortrue
  }{}%
  \@ifundefined{ifGPblacktext}{%
    \newif\ifGPblacktext
    \GPblacktexttrue
  }{}%
  \let\gplgaddtomacro\g@addto@macro
  \gdef\gplbacktext{}%
  \gdef\gplfronttext{}%
  \makeatother
  \ifGPblacktext
    \def\colorrgb#1{}%
    \def\colorgray#1{}%
  \else
    \ifGPcolor
      \def\colorrgb#1{\color[rgb]{#1}}%
      \def\colorgray#1{\color[gray]{#1}}%
      \expandafter\def\csname LTw\endcsname{\color{white}}%
      \expandafter\def\csname LTb\endcsname{\color{black}}%
      \expandafter\def\csname LTa\endcsname{\color{black}}%
      \expandafter\def\csname LT0\endcsname{\color[rgb]{1,0,0}}%
      \expandafter\def\csname LT1\endcsname{\color[rgb]{0,1,0}}%
      \expandafter\def\csname LT2\endcsname{\color[rgb]{0,0,1}}%
      \expandafter\def\csname LT3\endcsname{\color[rgb]{1,0,1}}%
      \expandafter\def\csname LT4\endcsname{\color[rgb]{0,1,1}}%
      \expandafter\def\csname LT5\endcsname{\color[rgb]{1,1,0}}%
      \expandafter\def\csname LT6\endcsname{\color[rgb]{0,0,0}}%
      \expandafter\def\csname LT7\endcsname{\color[rgb]{1,0.3,0}}%
      \expandafter\def\csname LT8\endcsname{\color[rgb]{0.5,0.5,0.5}}%
    \else
      \def\colorrgb#1{\color{black}}%
      \def\colorgray#1{\color[gray]{#1}}%
      \expandafter\def\csname LTw\endcsname{\color{white}}%
      \expandafter\def\csname LTb\endcsname{\color{black}}%
      \expandafter\def\csname LTa\endcsname{\color{black}}%
      \expandafter\def\csname LT0\endcsname{\color{black}}%
      \expandafter\def\csname LT1\endcsname{\color{black}}%
      \expandafter\def\csname LT2\endcsname{\color{black}}%
      \expandafter\def\csname LT3\endcsname{\color{black}}%
      \expandafter\def\csname LT4\endcsname{\color{black}}%
      \expandafter\def\csname LT5\endcsname{\color{black}}%
      \expandafter\def\csname LT6\endcsname{\color{black}}%
      \expandafter\def\csname LT7\endcsname{\color{black}}%
      \expandafter\def\csname LT8\endcsname{\color{black}}%
    \fi
  \fi
    \setlength{\unitlength}{0.0500bp}%
    \ifx\gptboxheight\undefined%
      \newlength{\gptboxheight}%
      \newlength{\gptboxwidth}%
      \newsavebox{\gptboxtext}%
    \fi%
    \setlength{\fboxrule}{0.5pt}%
    \setlength{\fboxsep}{1pt}%
\begin{picture}(5100.00,5580.00)%
    \gplgaddtomacro\gplbacktext{%
      \csname LTb\endcsname%
      \put(640,3905){\makebox(0,0)[r]{\strut{}$0$}}%
      \csname LTb\endcsname%
      \put(640,4128){\makebox(0,0)[r]{\strut{}$0.5$}}%
      \csname LTb\endcsname%
      \put(640,4351){\makebox(0,0)[r]{\strut{}$1$}}%
      \csname LTb\endcsname%
      \put(640,4575){\makebox(0,0)[r]{\strut{}$1.5$}}%
      \csname LTb\endcsname%
      \put(640,4798){\makebox(0,0)[r]{\strut{}$2$}}%
      \csname LTb\endcsname%
      \put(640,5021){\makebox(0,0)[r]{\strut{}$2.5$}}%
      \csname LTb\endcsname%
      \put(1093,5281){\makebox(0,0){\strut{}$-0.2$}}%
      \csname LTb\endcsname%
      \put(1534,5281){\makebox(0,0){\strut{}$-0.15$}}%
      \csname LTb\endcsname%
      \put(1974,5281){\makebox(0,0){\strut{}$-0.1$}}%
      \csname LTb\endcsname%
      \put(2415,5281){\makebox(0,0){\strut{}$-0.05$}}%
      \csname LTb\endcsname%
      \put(2856,5281){\makebox(0,0){\strut{}$0$}}%
      \csname LTb\endcsname%
      \put(3296,5281){\makebox(0,0){\strut{}$0.05$}}%
      \csname LTb\endcsname%
      \put(3737,5281){\makebox(0,0){\strut{}$0.1$}}%
      \csname LTb\endcsname%
      \put(4177,5281){\makebox(0,0){\strut{}$0.15$}}%
      \csname LTb\endcsname%
      \put(4618,5281){\makebox(0,0){\strut{}$0.2$}}%
      \csname LTb\endcsname%
      \put(4487,4798){\makebox(0,0)[l]{\strut{}$f_1$}}%
    }%
    \gplgaddtomacro\gplfronttext{%
      \csname LTb\endcsname%
      \put(165,4463){\rotatebox{-270}{\makebox(0,0){\strut{}$\unit[A_{f_1}/]{\%}$}}}%
      \csname LTb\endcsname%
      \put(2855,5559){\makebox(0,0){\strut{}$\unit[f /]{\mu Hz}$}}%
    }%
    \gplgaddtomacro\gplbacktext{%
      \csname LTb\endcsname%
      \put(640,3348){\makebox(0,0)[r]{\strut{}$0$}}%
      \csname LTb\endcsname%
      \put(640,3571){\makebox(0,0)[r]{\strut{}$0.5$}}%
      \csname LTb\endcsname%
      \put(640,3794){\makebox(0,0)[r]{\strut{}$1$}}%
    }%
    \gplgaddtomacro\gplfronttext{%
      \csname LTb\endcsname%
      \put(165,3626){\rotatebox{-270}{\makebox(0,0){\strut{}$\unit[A_{1,res}/]{\%}$}}}%
    }%
    \gplgaddtomacro\gplbacktext{%
      \csname LTb\endcsname%
      \put(640,3347){\makebox(0,0)[r]{\strut{}}}%
      \csname LTb\endcsname%
      \put(640,2232){\makebox(0,0)[r]{\strut{}$0$}}%
      \csname LTb\endcsname%
      \put(640,2511){\makebox(0,0)[r]{\strut{}$0.1$}}%
      \csname LTb\endcsname%
      \put(640,2790){\makebox(0,0)[r]{\strut{}$0.2$}}%
      \csname LTb\endcsname%
      \put(640,3068){\makebox(0,0)[r]{\strut{}$0.3$}}%
      \csname LTb\endcsname%
      \put(4487,3124){\makebox(0,0)[l]{\strut{}$f_2$}}%
    }%
    \gplgaddtomacro\gplfronttext{%
      \csname LTb\endcsname%
      \put(165,2789){\rotatebox{-270}{\makebox(0,0){\strut{}$\unit[A_{f_2}/]{\%}$}}}%
    }%
    \gplgaddtomacro\gplbacktext{%
      \csname LTb\endcsname%
      \put(640,2231){\makebox(0,0)[r]{\strut{}}}%
      \csname LTb\endcsname%
      \put(640,1674){\makebox(0,0)[r]{\strut{}$0$}}%
      \csname LTb\endcsname%
      \put(640,1953){\makebox(0,0)[r]{\strut{}$0.1$}}%
    }%
    \gplgaddtomacro\gplfronttext{%
      \csname LTb\endcsname%
      \put(165,1952){\rotatebox{-270}{\makebox(0,0){\strut{}$\unit[A_{2,res}/]{\%}$}}}%
    }%
    \gplgaddtomacro\gplbacktext{%
      \csname LTb\endcsname%
      \put(640,1673){\makebox(0,0)[r]{\strut{}}}%
      \csname LTb\endcsname%
      \put(640,558){\makebox(0,0)[r]{\strut{}$0$}}%
      \csname LTb\endcsname%
      \put(640,781){\makebox(0,0)[r]{\strut{}$0.2$}}%
      \csname LTb\endcsname%
      \put(640,1004){\makebox(0,0)[r]{\strut{}$0.4$}}%
      \csname LTb\endcsname%
      \put(640,1227){\makebox(0,0)[r]{\strut{}$0.6$}}%
      \csname LTb\endcsname%
      \put(640,1450){\makebox(0,0)[r]{\strut{}$0.8$}}%
      \csname LTb\endcsname%
      \put(816,298){\makebox(0,0){\strut{}$-0.02$}}%
      \csname LTb\endcsname%
      \put(1326,298){\makebox(0,0){\strut{}$-0.015$}}%
      \csname LTb\endcsname%
      \put(1836,298){\makebox(0,0){\strut{}$-0.01$}}%
      \csname LTb\endcsname%
      \put(2346,298){\makebox(0,0){\strut{}$-0.005$}}%
      \csname LTb\endcsname%
      \put(2856,298){\makebox(0,0){\strut{}$0$}}%
      \csname LTb\endcsname%
      \put(3365,298){\makebox(0,0){\strut{}$0.005$}}%
      \csname LTb\endcsname%
      \put(3875,298){\makebox(0,0){\strut{}$0.01$}}%
      \csname LTb\endcsname%
      \put(4385,298){\makebox(0,0){\strut{}$0.015$}}%
      \csname LTb\endcsname%
      \put(4895,298){\makebox(0,0){\strut{}$0.02$}}%
      \csname LTb\endcsname%
      \put(4487,1450){\makebox(0,0)[l]{\strut{}w}}%
    }%
    \gplgaddtomacro\gplfronttext{%
      \csname LTb\endcsname%
      \put(165,1115){\rotatebox{-270}{\makebox(0,0){\strut{}$A_{\text{window}}$}}}%
      \csname LTb\endcsname%
      \put(2855,19){\makebox(0,0){\strut{}$\unit[f /]{d^{-1}}$}}%
    }%
    \gplbacktext
    \put(0,0){\includegraphics[scale=0.5]{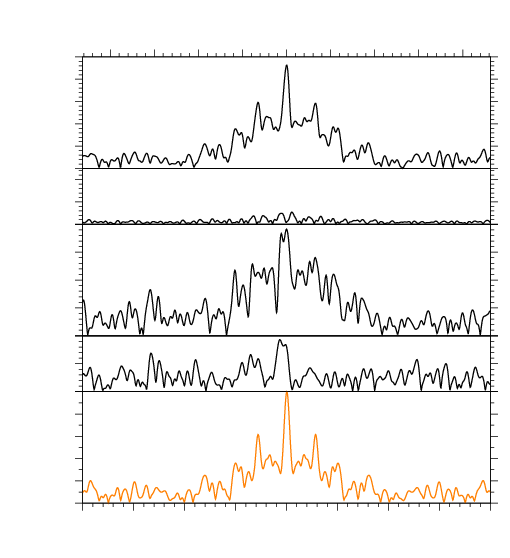}}%
    \gplfronttext
  \end{picture}%
\endgroup


        \caption{Amplitude spectrum of DW Lyn of the main pulsation frequency $ f_1 =  \unit[237.941160]{d^{-1}} $ (top), $ f_2 =  \unit[225.15898]{d^{-1}} $ (middle) with the respective residuals after the pre-whitening below, and the normalised window function (bottom).}
        \label{fig:dwlyn-window}
\end{figure}

The amplitude spectrum of DW Lyn in Fig.~\ref{fig:dwlyn-power} reveals two strong pulsation modes at $ f_1 = \unit[237.941160]{d^{-1}} $ and $ f_2 = \unit[225.15898]{d^{-1}} $. A closer look to the amplitude spectrum in Fig.~\ref{fig:dwlyn-window} reveals small asymmetries compared to the window function. The pre-whitening of both frequencies leaves residuals well above noise level in the amplitude spectrum, indicating unresolved multiplets or mode splitting, especially for $ f_2 $. 

\begin{figure*}[tp]
        \centering
\begingroup
  \makeatletter
  \providecommand\color[2][]{%
    \GenericError{(gnuplot) \space\space\space\@spaces}{%
      Package color not loaded in conjunction with
      terminal option `colourtext'%
    }{See the gnuplot documentation for explanation.%
    }{Either use 'blacktext' in gnuplot or load the package
      color.sty in LaTeX.}%
    \renewcommand\color[2][]{}%
  }%
  \providecommand\includegraphics[2][]{%
    \GenericError{(gnuplot) \space\space\space\@spaces}{%
      Package graphicx or graphics not loaded%
    }{See the gnuplot documentation for explanation.%
    }{The gnuplot epslatex terminal needs graphicx.sty or graphics.sty.}
  }%
  \providecommand\rotatebox[2]{#2}%
  \@ifundefined{ifGPcolor}{%
    \newif\ifGPcolor
    \GPcolortrue
  }{}%
  \@ifundefined{ifGPblacktext}{%
    \newif\ifGPblacktext
    \GPblacktexttrue
  }{}%
  \let\gplgaddtomacro\g@addto@macro
  \gdef\gplbacktext{}%
  \gdef\gplfronttext{}%
  \makeatother
  \ifGPblacktext
    \def\colorrgb#1{}%
    \def\colorgray#1{}%
  \else
    \ifGPcolor
      \def\colorrgb#1{\color[rgb]{#1}}%
      \def\colorgray#1{\color[gray]{#1}}%
      \expandafter\def\csname LTw\endcsname{\color{white}}%
      \expandafter\def\csname LTb\endcsname{\color{black}}%
      \expandafter\def\csname LTa\endcsname{\color{black}}%
      \expandafter\def\csname LT0\endcsname{\color[rgb]{1,0,0}}%
      \expandafter\def\csname LT1\endcsname{\color[rgb]{0,1,0}}%
      \expandafter\def\csname LT2\endcsname{\color[rgb]{0,0,1}}%
      \expandafter\def\csname LT3\endcsname{\color[rgb]{1,0,1}}%
      \expandafter\def\csname LT4\endcsname{\color[rgb]{0,1,1}}%
      \expandafter\def\csname LT5\endcsname{\color[rgb]{1,1,0}}%
      \expandafter\def\csname LT6\endcsname{\color[rgb]{0,0,0}}%
      \expandafter\def\csname LT7\endcsname{\color[rgb]{1,0.3,0}}%
      \expandafter\def\csname LT8\endcsname{\color[rgb]{0.5,0.5,0.5}}%
    \else
      \def\colorrgb#1{\color{black}}%
      \def\colorgray#1{\color[gray]{#1}}%
      \expandafter\def\csname LTw\endcsname{\color{white}}%
      \expandafter\def\csname LTb\endcsname{\color{black}}%
      \expandafter\def\csname LTa\endcsname{\color{black}}%
      \expandafter\def\csname LT0\endcsname{\color{black}}%
      \expandafter\def\csname LT1\endcsname{\color{black}}%
      \expandafter\def\csname LT2\endcsname{\color{black}}%
      \expandafter\def\csname LT3\endcsname{\color{black}}%
      \expandafter\def\csname LT4\endcsname{\color{black}}%
      \expandafter\def\csname LT5\endcsname{\color{black}}%
      \expandafter\def\csname LT6\endcsname{\color{black}}%
      \expandafter\def\csname LT7\endcsname{\color{black}}%
      \expandafter\def\csname LT8\endcsname{\color{black}}%
    \fi
  \fi
    \setlength{\unitlength}{0.0500bp}%
    \ifx\gptboxheight\undefined%
      \newlength{\gptboxheight}%
      \newlength{\gptboxwidth}%
      \newsavebox{\gptboxtext}%
    \fi%
    \setlength{\fboxrule}{0.5pt}%
    \setlength{\fboxsep}{1pt}%
\begin{picture}(10420.00,5800.00)%
    \gplgaddtomacro\gplbacktext{%
      \csname LTb\endcsname%
      \put(844,3866){\makebox(0,0)[r]{\strut{}$1.4$}}%
      \csname LTb\endcsname%
      \put(844,4061){\makebox(0,0)[r]{\strut{}$1.6$}}%
      \csname LTb\endcsname%
      \put(844,4256){\makebox(0,0)[r]{\strut{}$1.8$}}%
      \csname LTb\endcsname%
      \put(844,4451){\makebox(0,0)[r]{\strut{}$2$}}%
      \csname LTb\endcsname%
      \put(844,4647){\makebox(0,0)[r]{\strut{}$2.2$}}%
      \csname LTb\endcsname%
      \put(844,4842){\makebox(0,0)[r]{\strut{}$2.4$}}%
      \csname LTb\endcsname%
      \put(844,5037){\makebox(0,0)[r]{\strut{}$2.6$}}%
      \csname LTb\endcsname%
      \put(844,5232){\makebox(0,0)[r]{\strut{}$2.8$}}%
      \csname LTb\endcsname%
      \put(844,5427){\makebox(0,0)[r]{\strut{}$3$}}%
      \csname LTb\endcsname%
      \put(9779,3866){\makebox(0,0)[l]{\strut{}$0.1$}}%
      \csname LTb\endcsname%
      \put(9779,4061){\makebox(0,0)[l]{\strut{}$0.2$}}%
      \csname LTb\endcsname%
      \put(9779,4256){\makebox(0,0)[l]{\strut{}$0.3$}}%
      \csname LTb\endcsname%
      \put(9779,4451){\makebox(0,0)[l]{\strut{}$0.4$}}%
      \csname LTb\endcsname%
      \put(9779,4647){\makebox(0,0)[l]{\strut{}$0.5$}}%
      \csname LTb\endcsname%
      \put(9779,4842){\makebox(0,0)[l]{\strut{}$0.6$}}%
      \csname LTb\endcsname%
      \put(9779,5037){\makebox(0,0)[l]{\strut{}$0.7$}}%
      \csname LTb\endcsname%
      \put(9779,5232){\makebox(0,0)[l]{\strut{}$0.8$}}%
      \csname LTb\endcsname%
      \put(9779,5427){\makebox(0,0)[l]{\strut{}$0.9$}}%
      \csname LTb\endcsname%
      \put(1170,5687){\makebox(0,0){\strut{}2007}}%
      \csname LTb\endcsname%
      \put(2741,5687){\makebox(0,0){\strut{}2008}}%
      \csname LTb\endcsname%
      \put(4307,5687){\makebox(0,0){\strut{}2009}}%
      \csname LTb\endcsname%
      \put(5874,5687){\makebox(0,0){\strut{}2010}}%
      \csname LTb\endcsname%
      \put(7440,5687){\makebox(0,0){\strut{}2011}}%
      \csname LTb\endcsname%
      \put(9011,5687){\makebox(0,0){\strut{}2012}}%
    }%
    \gplgaddtomacro\gplfronttext{%
      \csname LTb\endcsname%
      \put(343,4646){\rotatebox{-270}{\makebox(0,0){\strut{}$\unit[A(f_1)/]{\%}$}}}%
      \csname LTb\endcsname%
      \put(10279,4646){\rotatebox{-270}{\makebox(0,0){\strut{}$\unit[A(f_2)/]{\%}$}}}%
      \csname LTb\endcsname%
      \put(8815,5260){\makebox(0,0)[r]{\strut{}$f_1$}}%
      \csname LTb\endcsname%
      \put(8815,5074){\makebox(0,0)[r]{\strut{}$f_2$}}%
    }%
    \gplgaddtomacro\gplbacktext{%
      \csname LTb\endcsname%
      \put(844,2305){\makebox(0,0)[r]{\strut{}$-500$}}%
      \csname LTb\endcsname%
      \put(844,2589){\makebox(0,0)[r]{\strut{}$-400$}}%
      \csname LTb\endcsname%
      \put(844,2873){\makebox(0,0)[r]{\strut{}$-300$}}%
      \csname LTb\endcsname%
      \put(844,3156){\makebox(0,0)[r]{\strut{}$-200$}}%
      \csname LTb\endcsname%
      \put(844,3440){\makebox(0,0)[r]{\strut{}$-100$}}%
      \csname LTb\endcsname%
      \put(844,3724){\makebox(0,0)[r]{\strut{}$0$}}%
      \csname LTb\endcsname%
      \put(9779,2305){\makebox(0,0)[l]{\strut{}}}%
      \csname LTb\endcsname%
      \put(9779,2589){\makebox(0,0)[l]{\strut{}}}%
      \csname LTb\endcsname%
      \put(9779,2873){\makebox(0,0)[l]{\strut{}}}%
      \csname LTb\endcsname%
      \put(9779,3156){\makebox(0,0)[l]{\strut{}}}%
      \csname LTb\endcsname%
      \put(9779,3440){\makebox(0,0)[l]{\strut{}}}%
      \csname LTb\endcsname%
      \put(9779,3724){\makebox(0,0)[l]{\strut{}}}%
    }%
    \gplgaddtomacro\gplfronttext{%
      \csname LTb\endcsname%
      \put(241,3085){\rotatebox{-270}{\makebox(0,0){\strut{}$\unit[(O-C)/]{s}$}}}%
    }%
    \gplgaddtomacro\gplbacktext{%
      \csname LTb\endcsname%
      \put(844,842){\makebox(0,0)[r]{\strut{}$-100$}}%
      \csname LTb\endcsname%
      \put(844,1037){\makebox(0,0)[r]{\strut{}$-80$}}%
      \csname LTb\endcsname%
      \put(844,1232){\makebox(0,0)[r]{\strut{}$-60$}}%
      \csname LTb\endcsname%
      \put(844,1427){\makebox(0,0)[r]{\strut{}$-40$}}%
      \csname LTb\endcsname%
      \put(844,1622){\makebox(0,0)[r]{\strut{}$-20$}}%
      \csname LTb\endcsname%
      \put(844,1817){\makebox(0,0)[r]{\strut{}$0$}}%
      \csname LTb\endcsname%
      \put(844,2012){\makebox(0,0)[r]{\strut{}$20$}}%
      \csname LTb\endcsname%
      \put(844,2207){\makebox(0,0)[r]{\strut{}$40$}}%
      \csname LTb\endcsname%
      \put(1020,484){\makebox(0,0){\strut{}$54400$}}%
      \csname LTb\endcsname%
      \put(1878,484){\makebox(0,0){\strut{}$54600$}}%
      \csname LTb\endcsname%
      \put(2737,484){\makebox(0,0){\strut{}$54800$}}%
      \csname LTb\endcsname%
      \put(3595,484){\makebox(0,0){\strut{}$55000$}}%
      \csname LTb\endcsname%
      \put(4453,484){\makebox(0,0){\strut{}$55200$}}%
      \csname LTb\endcsname%
      \put(5312,484){\makebox(0,0){\strut{}$55400$}}%
      \csname LTb\endcsname%
      \put(6170,484){\makebox(0,0){\strut{}$55600$}}%
      \csname LTb\endcsname%
      \put(7028,484){\makebox(0,0){\strut{}$55800$}}%
      \csname LTb\endcsname%
      \put(7886,484){\makebox(0,0){\strut{}$56000$}}%
      \csname LTb\endcsname%
      \put(8745,484){\makebox(0,0){\strut{}$56200$}}%
      \csname LTb\endcsname%
      \put(9603,484){\makebox(0,0){\strut{}$56400$}}%
      \csname LTb\endcsname%
      \put(9779,842){\makebox(0,0)[l]{\strut{}}}%
      \csname LTb\endcsname%
      \put(9779,1037){\makebox(0,0)[l]{\strut{}}}%
      \csname LTb\endcsname%
      \put(9779,1232){\makebox(0,0)[l]{\strut{}}}%
      \csname LTb\endcsname%
      \put(9779,1427){\makebox(0,0)[l]{\strut{}}}%
      \csname LTb\endcsname%
      \put(9779,1622){\makebox(0,0)[l]{\strut{}}}%
      \csname LTb\endcsname%
      \put(9779,1817){\makebox(0,0)[l]{\strut{}}}%
      \csname LTb\endcsname%
      \put(9779,2012){\makebox(0,0)[l]{\strut{}}}%
      \csname LTb\endcsname%
      \put(9779,2207){\makebox(0,0)[l]{\strut{}}}%
    }%
    \gplgaddtomacro\gplfronttext{%
      \csname LTb\endcsname%
      \put(241,1524){\rotatebox{-270}{\makebox(0,0){\strut{}residuals$\unit[/]{s}$}}}%
      \csname LTb\endcsname%
      \put(5311,205){\makebox(0,0){\strut{}$\unit[t/]{MBJD_{TDB}}$}}%
    }%
    \gplbacktext
    \put(0,0){\includegraphics[scale=0.5]{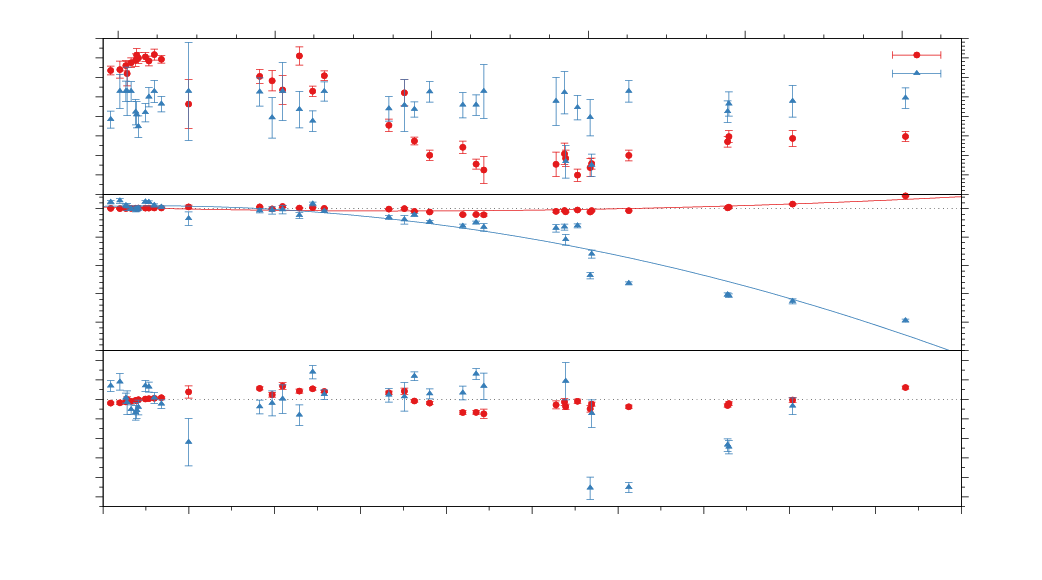}}%
    \gplfronttext
  \end{picture}%
\endgroup


        \caption{Results for the two main pulsations of DW Lyn. \textit{Top panel:} Amplitudes. \textit{Middle panel:} Fits of the $ O-C $ data with second order polynomials in time. \textit{Lower panel:} Residuals.}
        \label{fig:dwlyn-oc}
\end{figure*}

The S/N of modes at higher frequencies, for example, at about $ \unit[320]{d^{-1}} $ and $ \unit[480]{d^{-1}} $, are too small for a stable $ O-C $ analysis (see Table~\ref{tab:add-freq}). 
Therefore, the $ O-C $ diagram in Fig.~\ref{fig:dwlyn-oc} shows the analysis of the two main pulsation modes, with the time-dependent variation of the pulsation amplitudes. 

In order to determine evolutionary timescales of the pulsations, we investigated the long-term evolution in the $ O-C $ data. A constant change in period results in a second-order term as a function of time \citep{sterken_o-c_2005}, which allows us to derive a value for the secular change of the period $ \dot{P,} $ and hence the evolutionary timescale. Results of the fits of the second order polynomial are included in Fig.~\ref{fig:dwlyn-oc}, which are $ \dot{P}/P_{f1} = \left( 5.8 \pm 0.2 \right) \times \unit[10^{-5}]{d^{-1}} $ and $ \dot{P}/P_{f2} = \left( -29.3 \pm 0.8 \right) \times \unit[10^{-5}]{d^{-1}} $. 
Assuming $ \dot{P} $ is based on stellar evolution, stellar model calculations show that the sign of the rate of period change indicates the phase of the sdB after the zero-age extreme horizontal branch (ZAEHB) \citep{charpinet_adiabatic_2002}. For p-modes, a positive $ \dot{P} $ relates to the first evolutionary phase of the ZAEHB, in which the surface gravity decreases due to \element{He} burning in the core. A negative $ \dot{P} $ would correspond to the second evolutionary phase, in which the sdB contracts because the depletion of \element{He} in its core, and this happens before the post-EHB evolution. The turning point between these two states occurs between 87 and 91 Myr after the ZAEHB. According to our measurement of a positive $ \dot{P} $ for $ f_1 $, DW Lyn would still be in its first evolutionary phase. With the lack of a mode identification from an asteroseismic model for DW~Lyn, we can not directly compare the measured $ \dot{P} $ with theoretical predictions from \citet{charpinet_adiabatic_2002}. 
However, stellar models with pulsation periods of around $ \unit[360]{s} $ show values for $ \dot{P} $ with a comparable order of magnitude to our measurement $ \dot{P} = \left( 4.3 \pm 0.15 \right) \times \unit[10^{-1}]{s \, Myr^{-1}} $, for example, $ \dot{P} = 1.62 \unit{s \, Myr^{-1}} $ for a model with a mode of $ l=0,k=0 $ at the age of $ \unit[67.83]{Myr} $ \citep[appendix C]{charpinet_adiabatic_2002}.
The large $ \dot{P} $ of $ f_2 $ is consistent with the apparent mode splitting seen in the amplitude spectra in Fig.~\ref{fig:dwlyn-window}, and thus does not reflect the evolutionary phase of DW Lyn.

After subtracting the long-term trend,
small timescale features are evident. For example, the $ O-C $ data for $ f_2 $ show an oscillating behaviour with a significance of $ \unit[3]{\sigma} $ within the first 200 days, while the arrival times for $ f_1 $ remain constant during the same period of time. In later epochs, the $ O-C $ data for both frequencies agree mostly within $ \unit[2]{\sigma} $. During the second half of the observations, the phase of $ f_2 $ jumps by about $ \unit[100]{s} $. This behavior lacks an explanation.

Additionally, the evolution of the pulsation-amplitudes in Fig.~\ref{fig:dwlyn-oc} shows a comparable oscillating behaviour for $ f_2 $ within the first epochs similar to the change in arrival times. Although the periodic variations in amplitude are not as significant as for the phase, the occurrence of simultaneous phase- and amplitude-modulations indicate a mode beating of two close, unresolved frequencies. The residuals in the amplitude spectrum support this explanation. In later observations, the amplitude remains almost constant within the uncertainties. The beating mode might lose energy or shift frequency over time.
The amplitude for the $ f_1 $ pulsation drops by about 1 per cent (amplitude), or about 35 per cent (relative) to the second half of the observation campaign with a similar quasi-periodic variation as the phase. The residuals in the amplitude spectrum show no indication of an unresolved frequency leading to mode-beating. Besides stochastically driven pulsation modes, \citet{kilkenny_amplitude_2010} suggested energy transfer between modes as possible explanation for amplitude variations. For both frequencies, a possible interaction between amplitude and phase of pulsations is not well understood.

\subsection{V1636 Ori} 
\begin{figure}[tp]
	\centering
\begingroup
  \makeatletter
  \providecommand\color[2][]{%
    \GenericError{(gnuplot) \space\space\space\@spaces}{%
      Package color not loaded in conjunction with
      terminal option `colourtext'%
    }{See the gnuplot documentation for explanation.%
    }{Either use 'blacktext' in gnuplot or load the package
      color.sty in LaTeX.}%
    \renewcommand\color[2][]{}%
  }%
  \providecommand\includegraphics[2][]{%
    \GenericError{(gnuplot) \space\space\space\@spaces}{%
      Package graphicx or graphics not loaded%
    }{See the gnuplot documentation for explanation.%
    }{The gnuplot epslatex terminal needs graphicx.sty or graphics.sty.}
  }%
  \providecommand\rotatebox[2]{#2}%
  \@ifundefined{ifGPcolor}{%
    \newif\ifGPcolor
    \GPcolortrue
  }{}%
  \@ifundefined{ifGPblacktext}{%
    \newif\ifGPblacktext
    \GPblacktexttrue
  }{}%
  \let\gplgaddtomacro\g@addto@macro
  \gdef\gplbacktext{}%
  \gdef\gplfronttext{}%
  \makeatother
  \ifGPblacktext
    \def\colorrgb#1{}%
    \def\colorgray#1{}%
  \else
    \ifGPcolor
      \def\colorrgb#1{\color[rgb]{#1}}%
      \def\colorgray#1{\color[gray]{#1}}%
      \expandafter\def\csname LTw\endcsname{\color{white}}%
      \expandafter\def\csname LTb\endcsname{\color{black}}%
      \expandafter\def\csname LTa\endcsname{\color{black}}%
      \expandafter\def\csname LT0\endcsname{\color[rgb]{1,0,0}}%
      \expandafter\def\csname LT1\endcsname{\color[rgb]{0,1,0}}%
      \expandafter\def\csname LT2\endcsname{\color[rgb]{0,0,1}}%
      \expandafter\def\csname LT3\endcsname{\color[rgb]{1,0,1}}%
      \expandafter\def\csname LT4\endcsname{\color[rgb]{0,1,1}}%
      \expandafter\def\csname LT5\endcsname{\color[rgb]{1,1,0}}%
      \expandafter\def\csname LT6\endcsname{\color[rgb]{0,0,0}}%
      \expandafter\def\csname LT7\endcsname{\color[rgb]{1,0.3,0}}%
      \expandafter\def\csname LT8\endcsname{\color[rgb]{0.5,0.5,0.5}}%
    \else
      \def\colorrgb#1{\color{black}}%
      \def\colorgray#1{\color[gray]{#1}}%
      \expandafter\def\csname LTw\endcsname{\color{white}}%
      \expandafter\def\csname LTb\endcsname{\color{black}}%
      \expandafter\def\csname LTa\endcsname{\color{black}}%
      \expandafter\def\csname LT0\endcsname{\color{black}}%
      \expandafter\def\csname LT1\endcsname{\color{black}}%
      \expandafter\def\csname LT2\endcsname{\color{black}}%
      \expandafter\def\csname LT3\endcsname{\color{black}}%
      \expandafter\def\csname LT4\endcsname{\color{black}}%
      \expandafter\def\csname LT5\endcsname{\color{black}}%
      \expandafter\def\csname LT6\endcsname{\color{black}}%
      \expandafter\def\csname LT7\endcsname{\color{black}}%
      \expandafter\def\csname LT8\endcsname{\color{black}}%
    \fi
  \fi
    \setlength{\unitlength}{0.0500bp}%
    \ifx\gptboxheight\undefined%
      \newlength{\gptboxheight}%
      \newlength{\gptboxwidth}%
      \newsavebox{\gptboxtext}%
    \fi%
    \setlength{\fboxrule}{0.5pt}%
    \setlength{\fboxsep}{1pt}%
\begin{picture}(5100.00,5580.00)%
    \gplgaddtomacro\gplbacktext{%
      \csname LTb\endcsname%
      \put(640,3905){\makebox(0,0)[r]{\strut{}$0$}}%
      \csname LTb\endcsname%
      \put(640,4184){\makebox(0,0)[r]{\strut{}$0.2$}}%
      \csname LTb\endcsname%
      \put(640,4463){\makebox(0,0)[r]{\strut{}$0.4$}}%
      \csname LTb\endcsname%
      \put(640,4742){\makebox(0,0)[r]{\strut{}$0.6$}}%
      \csname LTb\endcsname%
      \put(640,5021){\makebox(0,0)[r]{\strut{}$0.8$}}%
      \csname LTb\endcsname%
      \put(1093,5281){\makebox(0,0){\strut{}$-0.2$}}%
      \csname LTb\endcsname%
      \put(1534,5281){\makebox(0,0){\strut{}$-0.15$}}%
      \csname LTb\endcsname%
      \put(1974,5281){\makebox(0,0){\strut{}$-0.1$}}%
      \csname LTb\endcsname%
      \put(2415,5281){\makebox(0,0){\strut{}$-0.05$}}%
      \csname LTb\endcsname%
      \put(2856,5281){\makebox(0,0){\strut{}$0$}}%
      \csname LTb\endcsname%
      \put(3296,5281){\makebox(0,0){\strut{}$0.05$}}%
      \csname LTb\endcsname%
      \put(3737,5281){\makebox(0,0){\strut{}$0.1$}}%
      \csname LTb\endcsname%
      \put(4177,5281){\makebox(0,0){\strut{}$0.15$}}%
      \csname LTb\endcsname%
      \put(4618,5281){\makebox(0,0){\strut{}$0.2$}}%
      \csname LTb\endcsname%
      \put(4487,4798){\makebox(0,0)[l]{\strut{}$f_1$}}%
    }%
    \gplgaddtomacro\gplfronttext{%
      \csname LTb\endcsname%
      \put(165,4463){\rotatebox{-270}{\makebox(0,0){\strut{}$\unit[A_{f_1}/]{\%}$}}}%
      \csname LTb\endcsname%
      \put(2855,5559){\makebox(0,0){\strut{}$\unit[f /]{\mu Hz}$}}%
    }%
    \gplgaddtomacro\gplbacktext{%
      \csname LTb\endcsname%
      \put(640,3905){\makebox(0,0)[r]{\strut{}}}%
      \csname LTb\endcsname%
      \put(640,3348){\makebox(0,0)[r]{\strut{}$0$}}%
      \csname LTb\endcsname%
      \put(640,3627){\makebox(0,0)[r]{\strut{}$0.2$}}%
    }%
    \gplgaddtomacro\gplfronttext{%
      \csname LTb\endcsname%
      \put(165,3626){\rotatebox{-270}{\makebox(0,0){\strut{}$\unit[A_{1,res}/]{\%}$}}}%
    }%
    \gplgaddtomacro\gplbacktext{%
      \csname LTb\endcsname%
      \put(640,2232){\makebox(0,0)[r]{\strut{}$0$}}%
      \csname LTb\endcsname%
      \put(640,2630){\makebox(0,0)[r]{\strut{}$0.1$}}%
      \csname LTb\endcsname%
      \put(640,3028){\makebox(0,0)[r]{\strut{}$0.2$}}%
      \csname LTb\endcsname%
      \put(4487,3124){\makebox(0,0)[l]{\strut{}$f_2$}}%
    }%
    \gplgaddtomacro\gplfronttext{%
      \csname LTb\endcsname%
      \put(165,2789){\rotatebox{-270}{\makebox(0,0){\strut{}$\unit[A_{f_2}/]{\%}$}}}%
    }%
    \gplgaddtomacro\gplbacktext{%
      \csname LTb\endcsname%
      \put(640,1674){\makebox(0,0)[r]{\strut{}$0$}}%
      \csname LTb\endcsname%
      \put(640,2072){\makebox(0,0)[r]{\strut{}$0.1$}}%
    }%
    \gplgaddtomacro\gplfronttext{%
      \csname LTb\endcsname%
      \put(165,1952){\rotatebox{-270}{\makebox(0,0){\strut{}$\unit[A_{2,res}/]{\%}$}}}%
    }%
    \gplgaddtomacro\gplbacktext{%
      \csname LTb\endcsname%
      \put(640,1673){\makebox(0,0)[r]{\strut{}}}%
      \csname LTb\endcsname%
      \put(640,558){\makebox(0,0)[r]{\strut{}$0$}}%
      \csname LTb\endcsname%
      \put(640,781){\makebox(0,0)[r]{\strut{}$0.2$}}%
      \csname LTb\endcsname%
      \put(640,1004){\makebox(0,0)[r]{\strut{}$0.4$}}%
      \csname LTb\endcsname%
      \put(640,1227){\makebox(0,0)[r]{\strut{}$0.6$}}%
      \csname LTb\endcsname%
      \put(640,1450){\makebox(0,0)[r]{\strut{}$0.8$}}%
      \csname LTb\endcsname%
      \put(816,298){\makebox(0,0){\strut{}$-0.02$}}%
      \csname LTb\endcsname%
      \put(1326,298){\makebox(0,0){\strut{}$-0.015$}}%
      \csname LTb\endcsname%
      \put(1836,298){\makebox(0,0){\strut{}$-0.01$}}%
      \csname LTb\endcsname%
      \put(2346,298){\makebox(0,0){\strut{}$-0.005$}}%
      \csname LTb\endcsname%
      \put(2856,298){\makebox(0,0){\strut{}$0$}}%
      \csname LTb\endcsname%
      \put(3365,298){\makebox(0,0){\strut{}$0.005$}}%
      \csname LTb\endcsname%
      \put(3875,298){\makebox(0,0){\strut{}$0.01$}}%
      \csname LTb\endcsname%
      \put(4385,298){\makebox(0,0){\strut{}$0.015$}}%
      \csname LTb\endcsname%
      \put(4895,298){\makebox(0,0){\strut{}$0.02$}}%
      \csname LTb\endcsname%
      \put(4487,1450){\makebox(0,0)[l]{\strut{}w}}%
    }%
    \gplgaddtomacro\gplfronttext{%
      \csname LTb\endcsname%
      \put(165,1115){\rotatebox{-270}{\makebox(0,0){\strut{}$A_{\text{window}}$}}}%
      \csname LTb\endcsname%
      \put(2855,19){\makebox(0,0){\strut{}$\unit[f /]{d^{-1}}$}}%
    }%
    \gplbacktext
    \put(0,0){\includegraphics[scale=0.5]{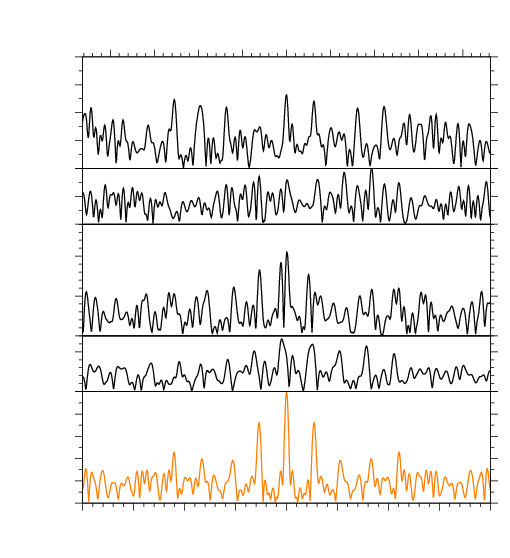}}%
    \gplfronttext
  \end{picture}%
\endgroup


	\caption{Amplitude spectrum of V1636 Ori of the main pulsation frequency $ f_1 =  \unit[631.7346]{d^{-1}} $ (top), $ f_2 =  \unit[509.9780]{d^{-1}} $ (middle) with the respective residuals after the pre-whitening below and the normalised window-function (bottom).}
	\label{fig:v1636ori-window}
\end{figure}

The amplitude spectrum of V1636 Ori in Fig.~\ref{fig:v1636ori-power} shows two main pulsation modes with frequencies at $ f_1 = \unit[631.7346]{d^{-1}} $ and $ f_2 =  \unit[509.9780]{d^{-1}} $. The S/N is not sufficient to use a third pulsation mode at $ \unit[566.2]{d^{-1}}$ ($ \unit[6553]{\mu Hz} $, \citealt{reed_resolving_2007}). The amplitude spectrum of TESS data in Fig.~\ref{fig:tess-power} shows no evidence for g-mode pulsations with amplitudes greater than 0.4 per cent.
A detailed look at the spectra of the two main frequencies in Fig.~\ref{fig:v1636ori-window} shows mode splitting, likely due to a change in frequency over the long observation time.

\begin{figure*}[tp]
        \centering
\begingroup
  \makeatletter
  \providecommand\color[2][]{%
    \GenericError{(gnuplot) \space\space\space\@spaces}{%
      Package color not loaded in conjunction with
      terminal option `colourtext'%
    }{See the gnuplot documentation for explanation.%
    }{Either use 'blacktext' in gnuplot or load the package
      color.sty in LaTeX.}%
    \renewcommand\color[2][]{}%
  }%
  \providecommand\includegraphics[2][]{%
    \GenericError{(gnuplot) \space\space\space\@spaces}{%
      Package graphicx or graphics not loaded%
    }{See the gnuplot documentation for explanation.%
    }{The gnuplot epslatex terminal needs graphicx.sty or graphics.sty.}
  }%
  \providecommand\rotatebox[2]{#2}%
  \@ifundefined{ifGPcolor}{%
    \newif\ifGPcolor
    \GPcolortrue
  }{}%
  \@ifundefined{ifGPblacktext}{%
    \newif\ifGPblacktext
    \GPblacktexttrue
  }{}%
  \let\gplgaddtomacro\g@addto@macro
  \gdef\gplbacktext{}%
  \gdef\gplfronttext{}%
  \makeatother
  \ifGPblacktext
    \def\colorrgb#1{}%
    \def\colorgray#1{}%
  \else
    \ifGPcolor
      \def\colorrgb#1{\color[rgb]{#1}}%
      \def\colorgray#1{\color[gray]{#1}}%
      \expandafter\def\csname LTw\endcsname{\color{white}}%
      \expandafter\def\csname LTb\endcsname{\color{black}}%
      \expandafter\def\csname LTa\endcsname{\color{black}}%
      \expandafter\def\csname LT0\endcsname{\color[rgb]{1,0,0}}%
      \expandafter\def\csname LT1\endcsname{\color[rgb]{0,1,0}}%
      \expandafter\def\csname LT2\endcsname{\color[rgb]{0,0,1}}%
      \expandafter\def\csname LT3\endcsname{\color[rgb]{1,0,1}}%
      \expandafter\def\csname LT4\endcsname{\color[rgb]{0,1,1}}%
      \expandafter\def\csname LT5\endcsname{\color[rgb]{1,1,0}}%
      \expandafter\def\csname LT6\endcsname{\color[rgb]{0,0,0}}%
      \expandafter\def\csname LT7\endcsname{\color[rgb]{1,0.3,0}}%
      \expandafter\def\csname LT8\endcsname{\color[rgb]{0.5,0.5,0.5}}%
    \else
      \def\colorrgb#1{\color{black}}%
      \def\colorgray#1{\color[gray]{#1}}%
      \expandafter\def\csname LTw\endcsname{\color{white}}%
      \expandafter\def\csname LTb\endcsname{\color{black}}%
      \expandafter\def\csname LTa\endcsname{\color{black}}%
      \expandafter\def\csname LT0\endcsname{\color{black}}%
      \expandafter\def\csname LT1\endcsname{\color{black}}%
      \expandafter\def\csname LT2\endcsname{\color{black}}%
      \expandafter\def\csname LT3\endcsname{\color{black}}%
      \expandafter\def\csname LT4\endcsname{\color{black}}%
      \expandafter\def\csname LT5\endcsname{\color{black}}%
      \expandafter\def\csname LT6\endcsname{\color{black}}%
      \expandafter\def\csname LT7\endcsname{\color{black}}%
      \expandafter\def\csname LT8\endcsname{\color{black}}%
    \fi
  \fi
    \setlength{\unitlength}{0.0500bp}%
    \ifx\gptboxheight\undefined%
      \newlength{\gptboxheight}%
      \newlength{\gptboxwidth}%
      \newsavebox{\gptboxtext}%
    \fi%
    \setlength{\fboxrule}{0.5pt}%
    \setlength{\fboxsep}{1pt}%
\begin{picture}(10420.00,5800.00)%
    \gplgaddtomacro\gplbacktext{%
      \csname LTb\endcsname%
      \put(844,3866){\makebox(0,0)[r]{\strut{}$0$}}%
      \csname LTb\endcsname%
      \put(844,4022){\makebox(0,0)[r]{\strut{}$0.1$}}%
      \csname LTb\endcsname%
      \put(844,4178){\makebox(0,0)[r]{\strut{}$0.2$}}%
      \csname LTb\endcsname%
      \put(844,4334){\makebox(0,0)[r]{\strut{}$0.3$}}%
      \csname LTb\endcsname%
      \put(844,4490){\makebox(0,0)[r]{\strut{}$0.4$}}%
      \csname LTb\endcsname%
      \put(844,4647){\makebox(0,0)[r]{\strut{}$0.5$}}%
      \csname LTb\endcsname%
      \put(844,4803){\makebox(0,0)[r]{\strut{}$0.6$}}%
      \csname LTb\endcsname%
      \put(844,4959){\makebox(0,0)[r]{\strut{}$0.7$}}%
      \csname LTb\endcsname%
      \put(844,5115){\makebox(0,0)[r]{\strut{}$0.8$}}%
      \csname LTb\endcsname%
      \put(844,5271){\makebox(0,0)[r]{\strut{}$0.9$}}%
      \csname LTb\endcsname%
      \put(844,5427){\makebox(0,0)[r]{\strut{}$1$}}%
      \csname LTb\endcsname%
      \put(9779,3866){\makebox(0,0)[l]{\strut{}$-0.1$}}%
      \csname LTb\endcsname%
      \put(9779,4089){\makebox(0,0)[l]{\strut{}$0$}}%
      \csname LTb\endcsname%
      \put(9779,4312){\makebox(0,0)[l]{\strut{}$0.1$}}%
      \csname LTb\endcsname%
      \put(9779,4535){\makebox(0,0)[l]{\strut{}$0.2$}}%
      \csname LTb\endcsname%
      \put(9779,4758){\makebox(0,0)[l]{\strut{}$0.3$}}%
      \csname LTb\endcsname%
      \put(9779,4981){\makebox(0,0)[l]{\strut{}$0.4$}}%
      \csname LTb\endcsname%
      \put(9779,5204){\makebox(0,0)[l]{\strut{}$0.5$}}%
      \csname LTb\endcsname%
      \put(9779,5427){\makebox(0,0)[l]{\strut{}$0.6$}}%
      \csname LTb\endcsname%
      \put(1684,5687){\makebox(0,0){\strut{}2008}}%
      \csname LTb\endcsname%
      \put(2888,5687){\makebox(0,0){\strut{}2009}}%
      \csname LTb\endcsname%
      \put(4093,5687){\makebox(0,0){\strut{}2010}}%
      \csname LTb\endcsname%
      \put(5298,5687){\makebox(0,0){\strut{}2011}}%
      \csname LTb\endcsname%
      \put(6507,5687){\makebox(0,0){\strut{}2012}}%
      \csname LTb\endcsname%
      \put(7711,5687){\makebox(0,0){\strut{}2013}}%
      \csname LTb\endcsname%
      \put(8916,5687){\makebox(0,0){\strut{}2014}}%
    }%
    \gplgaddtomacro\gplfronttext{%
      \csname LTb\endcsname%
      \put(343,4646){\rotatebox{-270}{\makebox(0,0){\strut{}$\unit[A(f_1)/]{\%}$}}}%
      \csname LTb\endcsname%
      \put(10381,4646){\rotatebox{-270}{\makebox(0,0){\strut{}$\unit[A(f_2)/]{\%}$}}}%
      \csname LTb\endcsname%
      \put(8815,4219){\makebox(0,0)[r]{\strut{}$f_1$}}%
      \csname LTb\endcsname%
      \put(8815,4033){\makebox(0,0)[r]{\strut{}$f_2$}}%
    }%
    \gplgaddtomacro\gplbacktext{%
      \csname LTb\endcsname%
      \put(844,2305){\makebox(0,0)[r]{\strut{}$-50$}}%
      \csname LTb\endcsname%
      \put(844,2660){\makebox(0,0)[r]{\strut{}$0$}}%
      \csname LTb\endcsname%
      \put(844,3015){\makebox(0,0)[r]{\strut{}$50$}}%
      \csname LTb\endcsname%
      \put(844,3369){\makebox(0,0)[r]{\strut{}$100$}}%
      \csname LTb\endcsname%
      \put(844,3724){\makebox(0,0)[r]{\strut{}$150$}}%
      \csname LTb\endcsname%
      \put(9779,2305){\makebox(0,0)[l]{\strut{}}}%
      \csname LTb\endcsname%
      \put(9779,2660){\makebox(0,0)[l]{\strut{}}}%
      \csname LTb\endcsname%
      \put(9779,3015){\makebox(0,0)[l]{\strut{}}}%
      \csname LTb\endcsname%
      \put(9779,3369){\makebox(0,0)[l]{\strut{}}}%
      \csname LTb\endcsname%
      \put(9779,3724){\makebox(0,0)[l]{\strut{}}}%
    }%
    \gplgaddtomacro\gplfronttext{%
      \csname LTb\endcsname%
      \put(343,3085){\rotatebox{-270}{\makebox(0,0){\strut{}$\unit[(O-C)/]{s}$}}}%
    }%
    \gplgaddtomacro\gplbacktext{%
      \csname LTb\endcsname%
      \put(844,744){\makebox(0,0)[r]{\strut{}$-100$}}%
      \csname LTb\endcsname%
      \put(844,1099){\makebox(0,0)[r]{\strut{}$-50$}}%
      \csname LTb\endcsname%
      \put(844,1453){\makebox(0,0)[r]{\strut{}$0$}}%
      \csname LTb\endcsname%
      \put(844,1808){\makebox(0,0)[r]{\strut{}$50$}}%
      \csname LTb\endcsname%
      \put(844,2162){\makebox(0,0)[r]{\strut{}$100$}}%
      \csname LTb\endcsname%
      \put(1020,484){\makebox(0,0){\strut{}$54600$}}%
      \csname LTb\endcsname%
      \put(1680,484){\makebox(0,0){\strut{}$54800$}}%
      \csname LTb\endcsname%
      \put(2340,484){\makebox(0,0){\strut{}$55000$}}%
      \csname LTb\endcsname%
      \put(3001,484){\makebox(0,0){\strut{}$55200$}}%
      \csname LTb\endcsname%
      \put(3661,484){\makebox(0,0){\strut{}$55400$}}%
      \csname LTb\endcsname%
      \put(4321,484){\makebox(0,0){\strut{}$55600$}}%
      \csname LTb\endcsname%
      \put(4981,484){\makebox(0,0){\strut{}$55800$}}%
      \csname LTb\endcsname%
      \put(5642,484){\makebox(0,0){\strut{}$56000$}}%
      \csname LTb\endcsname%
      \put(6302,484){\makebox(0,0){\strut{}$56200$}}%
      \csname LTb\endcsname%
      \put(6962,484){\makebox(0,0){\strut{}$56400$}}%
      \csname LTb\endcsname%
      \put(7622,484){\makebox(0,0){\strut{}$56600$}}%
      \csname LTb\endcsname%
      \put(8283,484){\makebox(0,0){\strut{}$56800$}}%
      \csname LTb\endcsname%
      \put(8943,484){\makebox(0,0){\strut{}$57000$}}%
      \csname LTb\endcsname%
      \put(9603,484){\makebox(0,0){\strut{}$57200$}}%
      \csname LTb\endcsname%
      \put(9779,744){\makebox(0,0)[l]{\strut{}}}%
      \csname LTb\endcsname%
      \put(9779,1099){\makebox(0,0)[l]{\strut{}}}%
      \csname LTb\endcsname%
      \put(9779,1453){\makebox(0,0)[l]{\strut{}}}%
      \csname LTb\endcsname%
      \put(9779,1808){\makebox(0,0)[l]{\strut{}}}%
      \csname LTb\endcsname%
      \put(9779,2162){\makebox(0,0)[l]{\strut{}}}%
    }%
    \gplgaddtomacro\gplfronttext{%
      \csname LTb\endcsname%
      \put(241,1524){\rotatebox{-270}{\makebox(0,0){\strut{}residuals$\unit[/]{s}$}}}%
      \csname LTb\endcsname%
      \put(5311,205){\makebox(0,0){\strut{}$\unit[t/]{MBJD_{TDB}}$}}%
    }%
    \gplbacktext
    \put(0,0){\includegraphics[scale=0.5]{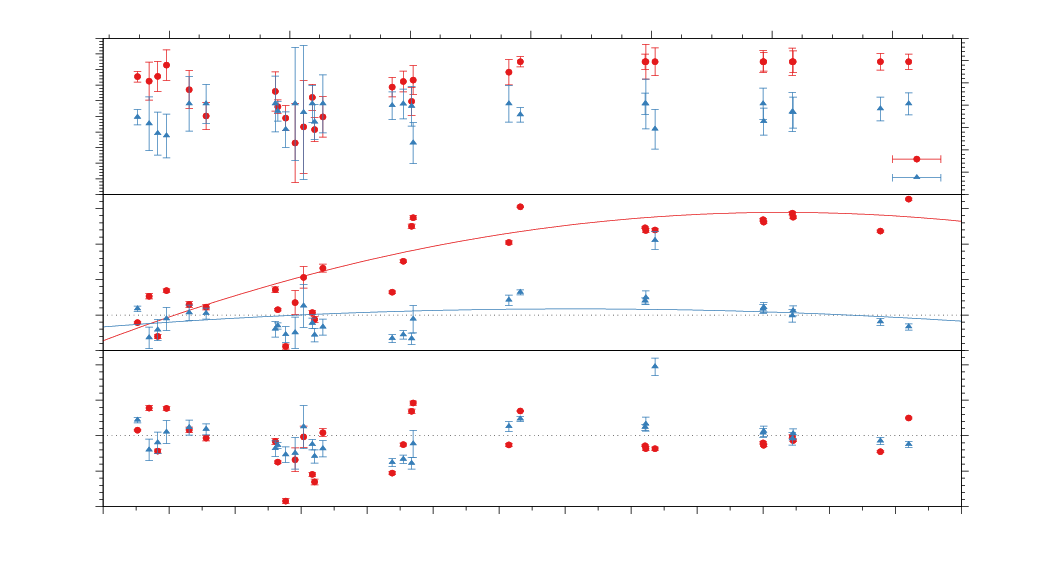}}%
    \gplfronttext
  \end{picture}%
\endgroup


        \caption{Results for the two main pulsations of V1636 Ori. \textit{Top panel:} Amplitudes. \textit{Middle panel:} Fits of the $ O-C $ data with second order polynomials in time. \textit{Lower panel:} Residuals.}
        \label{fig:v1636ori-oc}
\end{figure*}

The $ O-C $ diagram in Fig.~\ref{fig:v1636ori-oc} shows the two main pulsation modes and the variation of the pulsation amplitudes. 


From the second order fit in time, we derive the changes in period $ \dot{P}/P_{f1} = \left( -8.54 \pm 0.14 \right)\times \unit[10^{-5}]{d^{-1}} $ and $ \dot{P}/P_{f2} = \left( -2.5 \pm 0.5 \right)\times \unit[10^{-5}]{d^{-1}} $. We caution the interpretation of these values as evolutionary timescales since the  apparent mode splitting seen in Fig.~\ref{fig:v1636ori-window} could explain these trends as well.

The residuals after subtracting the long term trend show a large variation. They change by up to about $ \unit[\pm50]{s} $ for $ f_1 $ ($ \sim \unit[14]{\sigma} $ significance) and up to about $ \unit[\pm30]{s} $ for $ f_2 $ ($ \sim \unit[3]{\sigma} $ significance). The amplitude for $ f_1 $ drops by about 0.25 per cent (amplitude) or about 33 per cent (relative)  in the time between $ MBJD = \unit[55100]{d} \text{ and } \unit[55300]{d}$, and returns to its previous level afterwards, while the amplitude for $ f_2 $ remains constant within the uncertainties. This decrease in amplitude coincides with earlier arrival times in the $ O-C $ diagram. As already discussed in the previous section, a possible amplitude- and phase-interaction is not well understood. The $ f_1 $ pulsation mode may not be coherent on such long timescales but of a short-term stochastic nature not resolvable by our data set \citep[e.g.  KIC~2991276,][]{ostensen_stochastic_2014}.


\subsection{QQ Vir}
\begin{figure}[tp]
        \centering
\begingroup
  \makeatletter
  \providecommand\color[2][]{%
    \GenericError{(gnuplot) \space\space\space\@spaces}{%
      Package color not loaded in conjunction with
      terminal option `colourtext'%
    }{See the gnuplot documentation for explanation.%
    }{Either use 'blacktext' in gnuplot or load the package
      color.sty in LaTeX.}%
    \renewcommand\color[2][]{}%
  }%
  \providecommand\includegraphics[2][]{%
    \GenericError{(gnuplot) \space\space\space\@spaces}{%
      Package graphicx or graphics not loaded%
    }{See the gnuplot documentation for explanation.%
    }{The gnuplot epslatex terminal needs graphicx.sty or graphics.sty.}
  }%
  \providecommand\rotatebox[2]{#2}%
  \@ifundefined{ifGPcolor}{%
    \newif\ifGPcolor
    \GPcolortrue
  }{}%
  \@ifundefined{ifGPblacktext}{%
    \newif\ifGPblacktext
    \GPblacktexttrue
  }{}%
  \let\gplgaddtomacro\g@addto@macro
  \gdef\gplbacktext{}%
  \gdef\gplfronttext{}%
  \makeatother
  \ifGPblacktext
    \def\colorrgb#1{}%
    \def\colorgray#1{}%
  \else
    \ifGPcolor
      \def\colorrgb#1{\color[rgb]{#1}}%
      \def\colorgray#1{\color[gray]{#1}}%
      \expandafter\def\csname LTw\endcsname{\color{white}}%
      \expandafter\def\csname LTb\endcsname{\color{black}}%
      \expandafter\def\csname LTa\endcsname{\color{black}}%
      \expandafter\def\csname LT0\endcsname{\color[rgb]{1,0,0}}%
      \expandafter\def\csname LT1\endcsname{\color[rgb]{0,1,0}}%
      \expandafter\def\csname LT2\endcsname{\color[rgb]{0,0,1}}%
      \expandafter\def\csname LT3\endcsname{\color[rgb]{1,0,1}}%
      \expandafter\def\csname LT4\endcsname{\color[rgb]{0,1,1}}%
      \expandafter\def\csname LT5\endcsname{\color[rgb]{1,1,0}}%
      \expandafter\def\csname LT6\endcsname{\color[rgb]{0,0,0}}%
      \expandafter\def\csname LT7\endcsname{\color[rgb]{1,0.3,0}}%
      \expandafter\def\csname LT8\endcsname{\color[rgb]{0.5,0.5,0.5}}%
    \else
      \def\colorrgb#1{\color{black}}%
      \def\colorgray#1{\color[gray]{#1}}%
      \expandafter\def\csname LTw\endcsname{\color{white}}%
      \expandafter\def\csname LTb\endcsname{\color{black}}%
      \expandafter\def\csname LTa\endcsname{\color{black}}%
      \expandafter\def\csname LT0\endcsname{\color{black}}%
      \expandafter\def\csname LT1\endcsname{\color{black}}%
      \expandafter\def\csname LT2\endcsname{\color{black}}%
      \expandafter\def\csname LT3\endcsname{\color{black}}%
      \expandafter\def\csname LT4\endcsname{\color{black}}%
      \expandafter\def\csname LT5\endcsname{\color{black}}%
      \expandafter\def\csname LT6\endcsname{\color{black}}%
      \expandafter\def\csname LT7\endcsname{\color{black}}%
      \expandafter\def\csname LT8\endcsname{\color{black}}%
    \fi
  \fi
    \setlength{\unitlength}{0.0500bp}%
    \ifx\gptboxheight\undefined%
      \newlength{\gptboxheight}%
      \newlength{\gptboxwidth}%
      \newsavebox{\gptboxtext}%
    \fi%
    \setlength{\fboxrule}{0.5pt}%
    \setlength{\fboxsep}{1pt}%
\begin{picture}(5100.00,7520.00)%
    \gplgaddtomacro\gplbacktext{%
      \csname LTb\endcsname%
      \put(640,5640){\makebox(0,0)[r]{\strut{}$0$}}%
      \csname LTb\endcsname%
      \put(640,5865){\makebox(0,0)[r]{\strut{}$0.5$}}%
      \csname LTb\endcsname%
      \put(640,6091){\makebox(0,0)[r]{\strut{}$1$}}%
      \csname LTb\endcsname%
      \put(640,6316){\makebox(0,0)[r]{\strut{}$1.5$}}%
      \csname LTb\endcsname%
      \put(640,6542){\makebox(0,0)[r]{\strut{}$2$}}%
      \csname LTb\endcsname%
      \put(640,6767){\makebox(0,0)[r]{\strut{}$2.5$}}%
      \csname LTb\endcsname%
      \put(1093,7027){\makebox(0,0){\strut{}$-0.2$}}%
      \csname LTb\endcsname%
      \put(1974,7027){\makebox(0,0){\strut{}$-0.1$}}%
      \csname LTb\endcsname%
      \put(2856,7027){\makebox(0,0){\strut{}$0$}}%
      \csname LTb\endcsname%
      \put(3737,7027){\makebox(0,0){\strut{}$0.1$}}%
      \csname LTb\endcsname%
      \put(4618,7027){\makebox(0,0){\strut{}$0.2$}}%
      \csname LTb\endcsname%
      \put(4487,6542){\makebox(0,0)[l]{\strut{}$f_1$}}%
    }%
    \gplgaddtomacro\gplfronttext{%
      \csname LTb\endcsname%
      \put(165,6203){\rotatebox{-270}{\makebox(0,0){\strut{}$\unit[A_{f_1}/]{\%}$}}}%
      \csname LTb\endcsname%
      \put(2855,7305){\makebox(0,0){\strut{}$\unit[f /]{\mu Hz}$}}%
    }%
    \gplgaddtomacro\gplbacktext{%
      \csname LTb\endcsname%
      \put(640,5076){\makebox(0,0)[r]{\strut{}$0$}}%
      \csname LTb\endcsname%
      \put(640,5301){\makebox(0,0)[r]{\strut{}$0.5$}}%
      \csname LTb\endcsname%
      \put(640,5526){\makebox(0,0)[r]{\strut{}$1$}}%
    }%
    \gplgaddtomacro\gplfronttext{%
      \csname LTb\endcsname%
      \put(165,5357){\rotatebox{-270}{\makebox(0,0){\strut{}$\unit[A_{1,res}/]{\%}$}}}%
    }%
    \gplgaddtomacro\gplbacktext{%
      \csname LTb\endcsname%
      \put(640,3948){\makebox(0,0)[r]{\strut{}$0$}}%
      \csname LTb\endcsname%
      \put(640,4699){\makebox(0,0)[r]{\strut{}$0.1$}}%
      \csname LTb\endcsname%
      \put(4487,4850){\makebox(0,0)[l]{\strut{}$f_2$}}%
    }%
    \gplgaddtomacro\gplfronttext{%
      \csname LTb\endcsname%
      \put(165,4511){\rotatebox{-270}{\makebox(0,0){\strut{}$\unit[A_{f_2}/]{\%}$}}}%
    }%
    \gplgaddtomacro\gplbacktext{%
      \csname LTb\endcsname%
      \put(640,3384){\makebox(0,0)[r]{\strut{}$0$}}%
      \csname LTb\endcsname%
      \put(640,3759){\makebox(0,0)[r]{\strut{}$0.05$}}%
    }%
    \gplgaddtomacro\gplfronttext{%
      \csname LTb\endcsname%
      \put(63,3665){\rotatebox{-270}{\makebox(0,0){\strut{}$\unit[A_{2,res}/]{\%}$}}}%
    }%
    \gplgaddtomacro\gplbacktext{%
      \csname LTb\endcsname%
      \put(640,2256){\makebox(0,0)[r]{\strut{}$0$}}%
      \csname LTb\endcsname%
      \put(640,3007){\makebox(0,0)[r]{\strut{}$0.1$}}%
      \csname LTb\endcsname%
      \put(4487,3158){\makebox(0,0)[l]{\strut{}$f_3$}}%
    }%
    \gplgaddtomacro\gplfronttext{%
      \csname LTb\endcsname%
      \put(165,2819){\rotatebox{-270}{\makebox(0,0){\strut{}$\unit[A_{f_3}/]{\%}$}}}%
    }%
    \gplgaddtomacro\gplbacktext{%
      \csname LTb\endcsname%
      \put(640,1692){\makebox(0,0)[r]{\strut{}$0$}}%
      \csname LTb\endcsname%
      \put(640,2067){\makebox(0,0)[r]{\strut{}$0.1$}}%
    }%
    \gplgaddtomacro\gplfronttext{%
      \csname LTb\endcsname%
      \put(165,1973){\rotatebox{-270}{\makebox(0,0){\strut{}$\unit[A_{3,res}/]{\%}$}}}%
    }%
    \gplgaddtomacro\gplbacktext{%
      \csname LTb\endcsname%
      \put(640,1691){\makebox(0,0)[r]{\strut{}}}%
      \csname LTb\endcsname%
      \put(640,564){\makebox(0,0)[r]{\strut{}$0$}}%
      \csname LTb\endcsname%
      \put(640,789){\makebox(0,0)[r]{\strut{}$0.2$}}%
      \csname LTb\endcsname%
      \put(640,1015){\makebox(0,0)[r]{\strut{}$0.4$}}%
      \csname LTb\endcsname%
      \put(640,1240){\makebox(0,0)[r]{\strut{}$0.6$}}%
      \csname LTb\endcsname%
      \put(640,1466){\makebox(0,0)[r]{\strut{}$0.8$}}%
      \csname LTb\endcsname%
      \put(816,304){\makebox(0,0){\strut{}$-0.02$}}%
      \csname LTb\endcsname%
      \put(1836,304){\makebox(0,0){\strut{}$-0.01$}}%
      \csname LTb\endcsname%
      \put(2856,304){\makebox(0,0){\strut{}$0$}}%
      \csname LTb\endcsname%
      \put(3875,304){\makebox(0,0){\strut{}$0.01$}}%
      \csname LTb\endcsname%
      \put(4895,304){\makebox(0,0){\strut{}$0.02$}}%
      \csname LTb\endcsname%
      \put(4487,1466){\makebox(0,0)[l]{\strut{}w}}%
    }%
    \gplgaddtomacro\gplfronttext{%
      \csname LTb\endcsname%
      \put(165,1127){\rotatebox{-270}{\makebox(0,0){\strut{}$A_{\text{window}}$}}}%
      \csname LTb\endcsname%
      \put(2855,25){\makebox(0,0){\strut{}$\unit[f /]{d^{-1}}$}}%
    }%
    \gplbacktext
    \put(0,0){\includegraphics[scale=0.5]{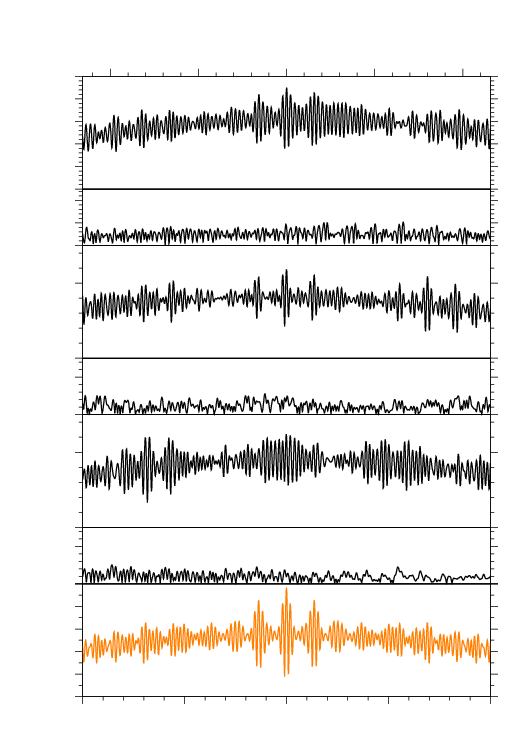}}%
    \gplfronttext
  \end{picture}%
\endgroup


        \caption{Amplitude spectrum of QQ Vir of the main pulsation frequency $ f_1 =  \unit[626.877628]{d^{-1}} $ (top), $ f_2 =  \unit[552.00713]{d^{-1}} $ (top middle),  $ f_3 =  \unit[642.0516]{d^{-1}} $ (bottom middle) with the respective residuals after the pre-whitening below and the normalised window-function (bottom).}
        \label{fig:qqvir-window}
\end{figure}

Fig.~\ref{fig:qqvir-power} shows the amplitude spectrum for the QQ Vir observations. The main frequency at about $ f_1 = \unit[626.877628]{d^{-1}} $ is presented in Fig.~\ref{fig:qqvir-window} in detail and shows asymmetries compared to the window function. After the pre-whitening process, a close frequency at about $ \unit[626.881270]{d^{-1}} $ remains but attempts to model this pulsation fail with uncertainties too large for the timing analysis. There appear two more frequencies suitable for our study. The amplitude spectra around $ f_2 =  \unit[552.00713]{d^{-1}} $  and $ f_3 = \unit[642.0516]{d^{-1}} $ are presented next to $ f_1 $ in Fig.~\ref{fig:qqvir-window}. Another peak at about $ \unit[665]{d^{-1}} $ consists of at least two frequencies at $ \unit[664.488549]{d^{-1}} $ and $ \unit[665.478133]{d^{-1}} $, but they are not sufficiently resolvable within the individual epochs, and lead to uncertainties in the $ O-C $ analysis that are too large.

\begin{figure*}[tp]
        \centering
\begingroup
  \makeatletter
  \providecommand\color[2][]{%
    \GenericError{(gnuplot) \space\space\space\@spaces}{%
      Package color not loaded in conjunction with
      terminal option `colourtext'%
    }{See the gnuplot documentation for explanation.%
    }{Either use 'blacktext' in gnuplot or load the package
      color.sty in LaTeX.}%
    \renewcommand\color[2][]{}%
  }%
  \providecommand\includegraphics[2][]{%
    \GenericError{(gnuplot) \space\space\space\@spaces}{%
      Package graphicx or graphics not loaded%
    }{See the gnuplot documentation for explanation.%
    }{The gnuplot epslatex terminal needs graphicx.sty or graphics.sty.}
  }%
  \providecommand\rotatebox[2]{#2}%
  \@ifundefined{ifGPcolor}{%
    \newif\ifGPcolor
    \GPcolortrue
  }{}%
  \@ifundefined{ifGPblacktext}{%
    \newif\ifGPblacktext
    \GPblacktexttrue
  }{}%
  \let\gplgaddtomacro\g@addto@macro
  \gdef\gplbacktext{}%
  \gdef\gplfronttext{}%
  \makeatother
  \ifGPblacktext
    \def\colorrgb#1{}%
    \def\colorgray#1{}%
  \else
    \ifGPcolor
      \def\colorrgb#1{\color[rgb]{#1}}%
      \def\colorgray#1{\color[gray]{#1}}%
      \expandafter\def\csname LTw\endcsname{\color{white}}%
      \expandafter\def\csname LTb\endcsname{\color{black}}%
      \expandafter\def\csname LTa\endcsname{\color{black}}%
      \expandafter\def\csname LT0\endcsname{\color[rgb]{1,0,0}}%
      \expandafter\def\csname LT1\endcsname{\color[rgb]{0,1,0}}%
      \expandafter\def\csname LT2\endcsname{\color[rgb]{0,0,1}}%
      \expandafter\def\csname LT3\endcsname{\color[rgb]{1,0,1}}%
      \expandafter\def\csname LT4\endcsname{\color[rgb]{0,1,1}}%
      \expandafter\def\csname LT5\endcsname{\color[rgb]{1,1,0}}%
      \expandafter\def\csname LT6\endcsname{\color[rgb]{0,0,0}}%
      \expandafter\def\csname LT7\endcsname{\color[rgb]{1,0.3,0}}%
      \expandafter\def\csname LT8\endcsname{\color[rgb]{0.5,0.5,0.5}}%
    \else
      \def\colorrgb#1{\color{black}}%
      \def\colorgray#1{\color[gray]{#1}}%
      \expandafter\def\csname LTw\endcsname{\color{white}}%
      \expandafter\def\csname LTb\endcsname{\color{black}}%
      \expandafter\def\csname LTa\endcsname{\color{black}}%
      \expandafter\def\csname LT0\endcsname{\color{black}}%
      \expandafter\def\csname LT1\endcsname{\color{black}}%
      \expandafter\def\csname LT2\endcsname{\color{black}}%
      \expandafter\def\csname LT3\endcsname{\color{black}}%
      \expandafter\def\csname LT4\endcsname{\color{black}}%
      \expandafter\def\csname LT5\endcsname{\color{black}}%
      \expandafter\def\csname LT6\endcsname{\color{black}}%
      \expandafter\def\csname LT7\endcsname{\color{black}}%
      \expandafter\def\csname LT8\endcsname{\color{black}}%
    \fi
  \fi
    \setlength{\unitlength}{0.0500bp}%
    \ifx\gptboxheight\undefined%
      \newlength{\gptboxheight}%
      \newlength{\gptboxwidth}%
      \newsavebox{\gptboxtext}%
    \fi%
    \setlength{\fboxrule}{0.5pt}%
    \setlength{\fboxsep}{1pt}%
\begin{picture}(10420.00,5800.00)%
    \gplgaddtomacro\gplbacktext{%
      \csname LTb\endcsname%
      \put(844,3866){\makebox(0,0)[r]{\strut{}$0.5$}}%
      \csname LTb\endcsname%
      \put(844,4089){\makebox(0,0)[r]{\strut{}$1$}}%
      \csname LTb\endcsname%
      \put(844,4312){\makebox(0,0)[r]{\strut{}$1.5$}}%
      \csname LTb\endcsname%
      \put(844,4535){\makebox(0,0)[r]{\strut{}$2$}}%
      \csname LTb\endcsname%
      \put(844,4758){\makebox(0,0)[r]{\strut{}$2.5$}}%
      \csname LTb\endcsname%
      \put(844,4981){\makebox(0,0)[r]{\strut{}$3$}}%
      \csname LTb\endcsname%
      \put(844,5204){\makebox(0,0)[r]{\strut{}$3.5$}}%
      \csname LTb\endcsname%
      \put(844,5427){\makebox(0,0)[r]{\strut{}$4$}}%
      \csname LTb\endcsname%
      \put(9779,3866){\makebox(0,0)[l]{\strut{}$-1.5$}}%
      \csname LTb\endcsname%
      \put(9779,4089){\makebox(0,0)[l]{\strut{}$-1$}}%
      \csname LTb\endcsname%
      \put(9779,4312){\makebox(0,0)[l]{\strut{}$-0.5$}}%
      \csname LTb\endcsname%
      \put(9779,4535){\makebox(0,0)[l]{\strut{}$0$}}%
      \csname LTb\endcsname%
      \put(9779,4758){\makebox(0,0)[l]{\strut{}$0.5$}}%
      \csname LTb\endcsname%
      \put(9779,4981){\makebox(0,0)[l]{\strut{}$1$}}%
      \csname LTb\endcsname%
      \put(9779,5204){\makebox(0,0)[l]{\strut{}$1.5$}}%
      \csname LTb\endcsname%
      \put(9779,5427){\makebox(0,0)[l]{\strut{}$2$}}%
      \csname LTb\endcsname%
      \put(1456,5687){\makebox(0,0){\strut{}2001}}%
      \csname LTb\endcsname%
      \put(2109,5687){\makebox(0,0){\strut{}2002}}%
      \csname LTb\endcsname%
      \put(2762,5687){\makebox(0,0){\strut{}2003}}%
      \csname LTb\endcsname%
      \put(3416,5687){\makebox(0,0){\strut{}2004}}%
      \csname LTb\endcsname%
      \put(4069,5687){\makebox(0,0){\strut{}2005}}%
      \csname LTb\endcsname%
      \put(4721,5687){\makebox(0,0){\strut{}2006}}%
      \csname LTb\endcsname%
      \put(5374,5687){\makebox(0,0){\strut{}2007}}%
      \csname LTb\endcsname%
      \put(6029,5687){\makebox(0,0){\strut{}2008}}%
      \csname LTb\endcsname%
      \put(6681,5687){\makebox(0,0){\strut{}2009}}%
      \csname LTb\endcsname%
      \put(7334,5687){\makebox(0,0){\strut{}2010}}%
      \csname LTb\endcsname%
      \put(7987,5687){\makebox(0,0){\strut{}2011}}%
      \csname LTb\endcsname%
      \put(8641,5687){\makebox(0,0){\strut{}2012}}%
      \csname LTb\endcsname%
      \put(9294,5687){\makebox(0,0){\strut{}2013}}%
    }%
    \gplgaddtomacro\gplfronttext{%
      \csname LTb\endcsname%
      \put(343,4646){\rotatebox{-270}{\makebox(0,0){\strut{}$\unit[A(f_1)/]{\%}$}}}%
      \csname LTb\endcsname%
      \put(10381,4646){\rotatebox{-270}{\makebox(0,0){\strut{}$\unit[A(f_2,f_3)/]{\%}$}}}%
      \csname LTb\endcsname%
      \put(8815,5260){\makebox(0,0)[r]{\strut{}$f_1$}}%
      \csname LTb\endcsname%
      \put(8815,5074){\makebox(0,0)[r]{\strut{}$f_2$}}%
      \csname LTb\endcsname%
      \put(8815,4888){\makebox(0,0)[r]{\strut{}$f_3$}}%
    }%
    \gplgaddtomacro\gplbacktext{%
      \csname LTb\endcsname%
      \put(844,3866){\makebox(0,0)[r]{\strut{}}}%
      \csname LTb\endcsname%
      \put(844,2447){\makebox(0,0)[r]{\strut{}$-400$}}%
      \csname LTb\endcsname%
      \put(844,2731){\makebox(0,0)[r]{\strut{}$-200$}}%
      \csname LTb\endcsname%
      \put(844,3015){\makebox(0,0)[r]{\strut{}$0$}}%
      \csname LTb\endcsname%
      \put(844,3298){\makebox(0,0)[r]{\strut{}$200$}}%
      \csname LTb\endcsname%
      \put(844,3582){\makebox(0,0)[r]{\strut{}$400$}}%
      \csname LTb\endcsname%
      \put(9779,2447){\makebox(0,0)[l]{\strut{}}}%
      \csname LTb\endcsname%
      \put(9779,2731){\makebox(0,0)[l]{\strut{}}}%
      \csname LTb\endcsname%
      \put(9779,3015){\makebox(0,0)[l]{\strut{}}}%
      \csname LTb\endcsname%
      \put(9779,3298){\makebox(0,0)[l]{\strut{}}}%
      \csname LTb\endcsname%
      \put(9779,3582){\makebox(0,0)[l]{\strut{}}}%
      \csname LTb\endcsname%
      \put(9779,3866){\makebox(0,0)[l]{\strut{}}}%
    }%
    \gplgaddtomacro\gplfronttext{%
      \csname LTb\endcsname%
      \put(241,3085){\rotatebox{-270}{\makebox(0,0){\strut{}$\unit[(O-C)/]{s}$}}}%
    }%
    \gplgaddtomacro\gplbacktext{%
      \csname LTb\endcsname%
      \put(844,855){\makebox(0,0)[r]{\strut{}$-300$}}%
      \csname LTb\endcsname%
      \put(844,1078){\makebox(0,0)[r]{\strut{}$-200$}}%
      \csname LTb\endcsname%
      \put(844,1301){\makebox(0,0)[r]{\strut{}$-100$}}%
      \csname LTb\endcsname%
      \put(844,1524){\makebox(0,0)[r]{\strut{}$0$}}%
      \csname LTb\endcsname%
      \put(844,1747){\makebox(0,0)[r]{\strut{}$100$}}%
      \csname LTb\endcsname%
      \put(844,1970){\makebox(0,0)[r]{\strut{}$200$}}%
      \csname LTb\endcsname%
      \put(844,2193){\makebox(0,0)[r]{\strut{}$300$}}%
      \csname LTb\endcsname%
      \put(1020,484){\makebox(0,0){\strut{}$52000$}}%
      \csname LTb\endcsname%
      \put(2808,484){\makebox(0,0){\strut{}$53000$}}%
      \csname LTb\endcsname%
      \put(4596,484){\makebox(0,0){\strut{}$54000$}}%
      \csname LTb\endcsname%
      \put(6384,484){\makebox(0,0){\strut{}$55000$}}%
      \csname LTb\endcsname%
      \put(8173,484){\makebox(0,0){\strut{}$56000$}}%
      \csname LTb\endcsname%
      \put(9779,855){\makebox(0,0)[l]{\strut{}}}%
      \csname LTb\endcsname%
      \put(9779,1078){\makebox(0,0)[l]{\strut{}}}%
      \csname LTb\endcsname%
      \put(9779,1301){\makebox(0,0)[l]{\strut{}}}%
      \csname LTb\endcsname%
      \put(9779,1524){\makebox(0,0)[l]{\strut{}}}%
      \csname LTb\endcsname%
      \put(9779,1747){\makebox(0,0)[l]{\strut{}}}%
      \csname LTb\endcsname%
      \put(9779,1970){\makebox(0,0)[l]{\strut{}}}%
      \csname LTb\endcsname%
      \put(9779,2193){\makebox(0,0)[l]{\strut{}}}%
    }%
    \gplgaddtomacro\gplfronttext{%
      \csname LTb\endcsname%
      \put(241,1524){\rotatebox{-270}{\makebox(0,0){\strut{}residuals$\unit[/]{s}$}}}%
      \csname LTb\endcsname%
      \put(5311,205){\makebox(0,0){\strut{}$\unit[t/]{MBJD_{TDB}}$}}%
    }%
    \gplbacktext
    \put(0,0){\includegraphics[scale=0.5]{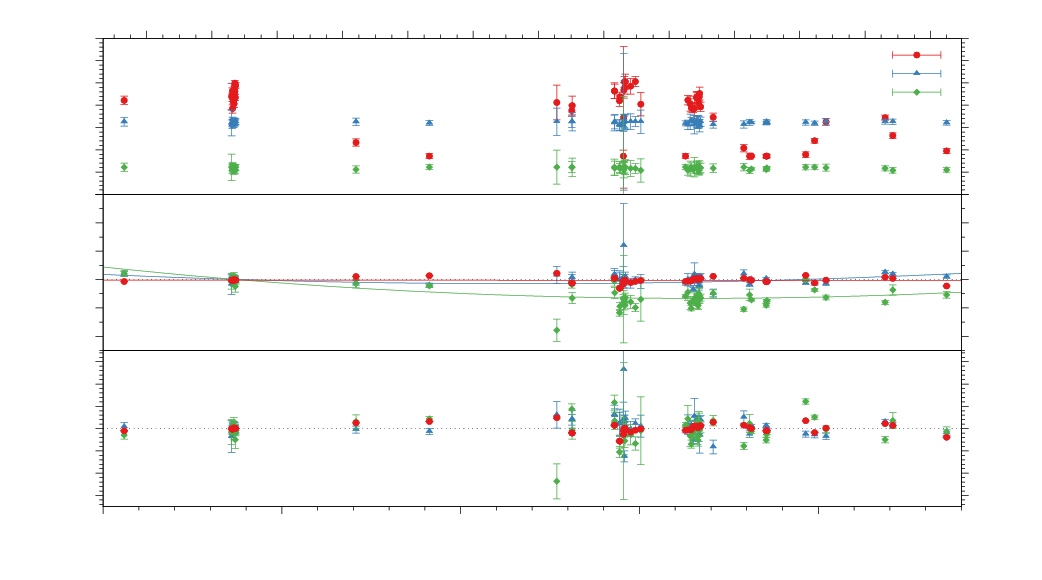}}%
    \gplfronttext
  \end{picture}%
\endgroup


        \caption{Results for the three main pulsations of QQ Vir. \textit{Top panel:} Amplitudes. $ f_3 $ has a vertical offset of -1 for clarity. \textit{Middle panel:} Fits of the $ O-C $ data with second order polynomials in time. \textit{Lower panel:} Residuals.}
        \label{fig:qqvir-oc}
\end{figure*}

Figure~\ref{fig:qqvir-oc} shows the resulting $ O-C $ diagram and the amplitudes at different epochs. Due to the large observational gap from 2003 to 2008 with only one block of observations in between, we had difficulties avoiding errors in cycle count. In order to avoid a phase jump, we increased the averaging window for initial phase values to $ q = 6 $. With this set up, the changes in pulsation frequencies read as follows: $ \dot{P}/P_{f1} = \left( 1.7 \pm 1.6 \right)\times \unit[10^{-7}]{d^{-1}} $, $ \dot{P}/P_{f2} = \left( 2.4 \pm 0.4 \right)\times \unit[10^{-5}]{d^{-1}} $ and $ \dot{P}/P_{f3} = \left( 4.0 \pm 0.5 \right)\times \unit[10^{-6}]{d^{-1}} $. While $ f_2 $ and $ f_3 $ show no significant variation of pulsation amplitude, $ f_1 $ varies by 1.5 per cent (amplitude) or 50 per cent (relative). Thus, the corresponding phase changes should be interpreted with caution. \citet{charpinet_rapidly_2006} identified the radial order $ k $ and degree $ l $ from asteroseismic modelling to be $ f_1 $: $ l=2,k=2 $; $ f_2 $: $ l=4,k=1 $; $ f_3 $: $ l=3,k=2 $. These combinations do not allow a direct comparison of our $ \dot{P} $ measurements to the model calculations from \citet{charpinet_adiabatic_2002}, but the sign of $ \dot{P} $ indicates QQ~Vir to be in the stage of \element{He} burning.

\subsection{V541 Hya}
\begin{figure}[tp]
        \centering
\begingroup
  \makeatletter
  \providecommand\color[2][]{%
    \GenericError{(gnuplot) \space\space\space\@spaces}{%
      Package color not loaded in conjunction with
      terminal option `colourtext'%
    }{See the gnuplot documentation for explanation.%
    }{Either use 'blacktext' in gnuplot or load the package
      color.sty in LaTeX.}%
    \renewcommand\color[2][]{}%
  }%
  \providecommand\includegraphics[2][]{%
    \GenericError{(gnuplot) \space\space\space\@spaces}{%
      Package graphicx or graphics not loaded%
    }{See the gnuplot documentation for explanation.%
    }{The gnuplot epslatex terminal needs graphicx.sty or graphics.sty.}
  }%
  \providecommand\rotatebox[2]{#2}%
  \@ifundefined{ifGPcolor}{%
    \newif\ifGPcolor
    \GPcolortrue
  }{}%
  \@ifundefined{ifGPblacktext}{%
    \newif\ifGPblacktext
    \GPblacktexttrue
  }{}%
  \let\gplgaddtomacro\g@addto@macro
  \gdef\gplbacktext{}%
  \gdef\gplfronttext{}%
  \makeatother
  \ifGPblacktext
    \def\colorrgb#1{}%
    \def\colorgray#1{}%
  \else
    \ifGPcolor
      \def\colorrgb#1{\color[rgb]{#1}}%
      \def\colorgray#1{\color[gray]{#1}}%
      \expandafter\def\csname LTw\endcsname{\color{white}}%
      \expandafter\def\csname LTb\endcsname{\color{black}}%
      \expandafter\def\csname LTa\endcsname{\color{black}}%
      \expandafter\def\csname LT0\endcsname{\color[rgb]{1,0,0}}%
      \expandafter\def\csname LT1\endcsname{\color[rgb]{0,1,0}}%
      \expandafter\def\csname LT2\endcsname{\color[rgb]{0,0,1}}%
      \expandafter\def\csname LT3\endcsname{\color[rgb]{1,0,1}}%
      \expandafter\def\csname LT4\endcsname{\color[rgb]{0,1,1}}%
      \expandafter\def\csname LT5\endcsname{\color[rgb]{1,1,0}}%
      \expandafter\def\csname LT6\endcsname{\color[rgb]{0,0,0}}%
      \expandafter\def\csname LT7\endcsname{\color[rgb]{1,0.3,0}}%
      \expandafter\def\csname LT8\endcsname{\color[rgb]{0.5,0.5,0.5}}%
    \else
      \def\colorrgb#1{\color{black}}%
      \def\colorgray#1{\color[gray]{#1}}%
      \expandafter\def\csname LTw\endcsname{\color{white}}%
      \expandafter\def\csname LTb\endcsname{\color{black}}%
      \expandafter\def\csname LTa\endcsname{\color{black}}%
      \expandafter\def\csname LT0\endcsname{\color{black}}%
      \expandafter\def\csname LT1\endcsname{\color{black}}%
      \expandafter\def\csname LT2\endcsname{\color{black}}%
      \expandafter\def\csname LT3\endcsname{\color{black}}%
      \expandafter\def\csname LT4\endcsname{\color{black}}%
      \expandafter\def\csname LT5\endcsname{\color{black}}%
      \expandafter\def\csname LT6\endcsname{\color{black}}%
      \expandafter\def\csname LT7\endcsname{\color{black}}%
      \expandafter\def\csname LT8\endcsname{\color{black}}%
    \fi
  \fi
    \setlength{\unitlength}{0.0500bp}%
    \ifx\gptboxheight\undefined%
      \newlength{\gptboxheight}%
      \newlength{\gptboxwidth}%
      \newsavebox{\gptboxtext}%
    \fi%
    \setlength{\fboxrule}{0.5pt}%
    \setlength{\fboxsep}{1pt}%
\begin{picture}(5100.00,5580.00)%
    \gplgaddtomacro\gplbacktext{%
      \csname LTb\endcsname%
      \put(640,3905){\makebox(0,0)[r]{\strut{}$0$}}%
      \csname LTb\endcsname%
      \put(640,4224){\makebox(0,0)[r]{\strut{}$0.1$}}%
      \csname LTb\endcsname%
      \put(640,4543){\makebox(0,0)[r]{\strut{}$0.2$}}%
      \csname LTb\endcsname%
      \put(640,4862){\makebox(0,0)[r]{\strut{}$0.3$}}%
      \csname LTb\endcsname%
      \put(1093,5281){\makebox(0,0){\strut{}$-0.2$}}%
      \csname LTb\endcsname%
      \put(1534,5281){\makebox(0,0){\strut{}$-0.15$}}%
      \csname LTb\endcsname%
      \put(1974,5281){\makebox(0,0){\strut{}$-0.1$}}%
      \csname LTb\endcsname%
      \put(2415,5281){\makebox(0,0){\strut{}$-0.05$}}%
      \csname LTb\endcsname%
      \put(2856,5281){\makebox(0,0){\strut{}$0$}}%
      \csname LTb\endcsname%
      \put(3296,5281){\makebox(0,0){\strut{}$0.05$}}%
      \csname LTb\endcsname%
      \put(3737,5281){\makebox(0,0){\strut{}$0.1$}}%
      \csname LTb\endcsname%
      \put(4177,5281){\makebox(0,0){\strut{}$0.15$}}%
      \csname LTb\endcsname%
      \put(4618,5281){\makebox(0,0){\strut{}$0.2$}}%
      \csname LTb\endcsname%
      \put(4487,4798){\makebox(0,0)[l]{\strut{}$f_1$}}%
    }%
    \gplgaddtomacro\gplfronttext{%
      \csname LTb\endcsname%
      \put(165,4463){\rotatebox{-270}{\makebox(0,0){\strut{}$\unit[A_{f_1}/]{\%}$}}}%
      \csname LTb\endcsname%
      \put(2855,5559){\makebox(0,0){\strut{}$\unit[f /]{\mu Hz}$}}%
    }%
    \gplgaddtomacro\gplbacktext{%
      \csname LTb\endcsname%
      \put(640,3348){\makebox(0,0)[r]{\strut{}$0$}}%
      \csname LTb\endcsname%
      \put(640,3666){\makebox(0,0)[r]{\strut{}$0.1$}}%
    }%
    \gplgaddtomacro\gplfronttext{%
      \csname LTb\endcsname%
      \put(165,3626){\rotatebox{-270}{\makebox(0,0){\strut{}$\unit[A_{1,res}/]{\%}$}}}%
    }%
    \gplgaddtomacro\gplbacktext{%
      \csname LTb\endcsname%
      \put(640,2232){\makebox(0,0)[r]{\strut{}$0$}}%
      \csname LTb\endcsname%
      \put(640,2678){\makebox(0,0)[r]{\strut{}$0.1$}}%
      \csname LTb\endcsname%
      \put(640,3124){\makebox(0,0)[r]{\strut{}$0.2$}}%
      \csname LTb\endcsname%
      \put(4487,3124){\makebox(0,0)[l]{\strut{}$f_2$}}%
    }%
    \gplgaddtomacro\gplfronttext{%
      \csname LTb\endcsname%
      \put(165,2789){\rotatebox{-270}{\makebox(0,0){\strut{}$\unit[A_{f_2}/]{\%}$}}}%
    }%
    \gplgaddtomacro\gplbacktext{%
      \csname LTb\endcsname%
      \put(640,1674){\makebox(0,0)[r]{\strut{}$0$}}%
      \csname LTb\endcsname%
      \put(640,2120){\makebox(0,0)[r]{\strut{}$0.1$}}%
    }%
    \gplgaddtomacro\gplfronttext{%
      \csname LTb\endcsname%
      \put(165,1952){\rotatebox{-270}{\makebox(0,0){\strut{}$\unit[A_{2,res}/]{\%}$}}}%
    }%
    \gplgaddtomacro\gplbacktext{%
      \csname LTb\endcsname%
      \put(640,1673){\makebox(0,0)[r]{\strut{}}}%
      \csname LTb\endcsname%
      \put(640,558){\makebox(0,0)[r]{\strut{}$0$}}%
      \csname LTb\endcsname%
      \put(640,781){\makebox(0,0)[r]{\strut{}$0.2$}}%
      \csname LTb\endcsname%
      \put(640,1004){\makebox(0,0)[r]{\strut{}$0.4$}}%
      \csname LTb\endcsname%
      \put(640,1227){\makebox(0,0)[r]{\strut{}$0.6$}}%
      \csname LTb\endcsname%
      \put(640,1450){\makebox(0,0)[r]{\strut{}$0.8$}}%
      \csname LTb\endcsname%
      \put(816,298){\makebox(0,0){\strut{}$-0.02$}}%
      \csname LTb\endcsname%
      \put(1326,298){\makebox(0,0){\strut{}$-0.015$}}%
      \csname LTb\endcsname%
      \put(1836,298){\makebox(0,0){\strut{}$-0.01$}}%
      \csname LTb\endcsname%
      \put(2346,298){\makebox(0,0){\strut{}$-0.005$}}%
      \csname LTb\endcsname%
      \put(2856,298){\makebox(0,0){\strut{}$0$}}%
      \csname LTb\endcsname%
      \put(3365,298){\makebox(0,0){\strut{}$0.005$}}%
      \csname LTb\endcsname%
      \put(3875,298){\makebox(0,0){\strut{}$0.01$}}%
      \csname LTb\endcsname%
      \put(4385,298){\makebox(0,0){\strut{}$0.015$}}%
      \csname LTb\endcsname%
      \put(4895,298){\makebox(0,0){\strut{}$0.02$}}%
      \csname LTb\endcsname%
      \put(4487,1450){\makebox(0,0)[l]{\strut{}w}}%
    }%
    \gplgaddtomacro\gplfronttext{%
      \csname LTb\endcsname%
      \put(165,1115){\rotatebox{-270}{\makebox(0,0){\strut{}$A_{\text{window}}$}}}%
      \csname LTb\endcsname%
      \put(2855,19){\makebox(0,0){\strut{}$\unit[f /]{d^{-1}}$}}%
    }%
    \gplbacktext
    \put(0,0){\includegraphics[scale=0.5]{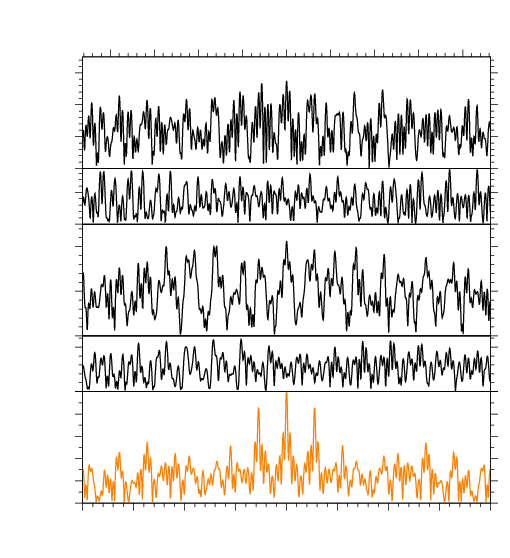}}%
    \gplfronttext
  \end{picture}%
\endgroup


        \caption{Amplitude spectrum of V541 Hya of the main pulsation frequency $ f_1 =  \unit[635.32218]{d^{-1}} $ (top), $ f_2 =  \unit[571.28556]{d^{-1}} $ (middle) with the respective residuals after the pre-whitening below and the normalised window-function (bottom).}
        \label{fig:v541hya-window}
\end{figure}

The amplitude spectrum in Fig.~\ref{fig:v541hya-power} shows two pulsation modes with frequencies at $ f_1 = \unit[635.32218]{d^{-1}} $ and at $ f_2 = \unit[571.28556]{d^{-1}} $.
Both of them show a complex behaviour (Fig.~\ref{fig:v541hya-window}), indicating unresolved multiplets and/or frequency changes that we see also in the O-C diagrams (Fig.~\ref{fig:v541hya-oc}).
The S/N for a third frequency at $ \unit[603.88741]{d^{-1}} $ is not sufficient for the $ O-C $ analysis. Similar to V1636 Ori, the amplitude spectrum obtained from the TESS light curve in Fig.~\ref{fig:tess-power} shows no evidence for g-mode pulsations with amplitudes greater than 0.4 per cent.

\begin{figure}[tp]
        \centering
\begingroup
  \makeatletter
  \providecommand\color[2][]{%
    \GenericError{(gnuplot) \space\space\space\@spaces}{%
      Package color not loaded in conjunction with
      terminal option `colourtext'%
    }{See the gnuplot documentation for explanation.%
    }{Either use 'blacktext' in gnuplot or load the package
      color.sty in LaTeX.}%
    \renewcommand\color[2][]{}%
  }%
  \providecommand\includegraphics[2][]{%
    \GenericError{(gnuplot) \space\space\space\@spaces}{%
      Package graphicx or graphics not loaded%
    }{See the gnuplot documentation for explanation.%
    }{The gnuplot epslatex terminal needs graphicx.sty or graphics.sty.}
  }%
  \providecommand\rotatebox[2]{#2}%
  \@ifundefined{ifGPcolor}{%
    \newif\ifGPcolor
    \GPcolortrue
  }{}%
  \@ifundefined{ifGPblacktext}{%
    \newif\ifGPblacktext
    \GPblacktexttrue
  }{}%
  \let\gplgaddtomacro\g@addto@macro
  \gdef\gplbacktext{}%
  \gdef\gplfronttext{}%
  \makeatother
  \ifGPblacktext
    \def\colorrgb#1{}%
    \def\colorgray#1{}%
  \else
    \ifGPcolor
      \def\colorrgb#1{\color[rgb]{#1}}%
      \def\colorgray#1{\color[gray]{#1}}%
      \expandafter\def\csname LTw\endcsname{\color{white}}%
      \expandafter\def\csname LTb\endcsname{\color{black}}%
      \expandafter\def\csname LTa\endcsname{\color{black}}%
      \expandafter\def\csname LT0\endcsname{\color[rgb]{1,0,0}}%
      \expandafter\def\csname LT1\endcsname{\color[rgb]{0,1,0}}%
      \expandafter\def\csname LT2\endcsname{\color[rgb]{0,0,1}}%
      \expandafter\def\csname LT3\endcsname{\color[rgb]{1,0,1}}%
      \expandafter\def\csname LT4\endcsname{\color[rgb]{0,1,1}}%
      \expandafter\def\csname LT5\endcsname{\color[rgb]{1,1,0}}%
      \expandafter\def\csname LT6\endcsname{\color[rgb]{0,0,0}}%
      \expandafter\def\csname LT7\endcsname{\color[rgb]{1,0.3,0}}%
      \expandafter\def\csname LT8\endcsname{\color[rgb]{0.5,0.5,0.5}}%
    \else
      \def\colorrgb#1{\color{black}}%
      \def\colorgray#1{\color[gray]{#1}}%
      \expandafter\def\csname LTw\endcsname{\color{white}}%
      \expandafter\def\csname LTb\endcsname{\color{black}}%
      \expandafter\def\csname LTa\endcsname{\color{black}}%
      \expandafter\def\csname LT0\endcsname{\color{black}}%
      \expandafter\def\csname LT1\endcsname{\color{black}}%
      \expandafter\def\csname LT2\endcsname{\color{black}}%
      \expandafter\def\csname LT3\endcsname{\color{black}}%
      \expandafter\def\csname LT4\endcsname{\color{black}}%
      \expandafter\def\csname LT5\endcsname{\color{black}}%
      \expandafter\def\csname LT6\endcsname{\color{black}}%
      \expandafter\def\csname LT7\endcsname{\color{black}}%
      \expandafter\def\csname LT8\endcsname{\color{black}}%
    \fi
  \fi
    \setlength{\unitlength}{0.0500bp}%
    \ifx\gptboxheight\undefined%
      \newlength{\gptboxheight}%
      \newlength{\gptboxwidth}%
      \newsavebox{\gptboxtext}%
    \fi%
    \setlength{\fboxrule}{0.5pt}%
    \setlength{\fboxsep}{1pt}%
\begin{picture}(5100.00,3140.00)%
    \gplgaddtomacro\gplbacktext{%
      \csname LTb\endcsname%
      \put(640,1570){\makebox(0,0)[r]{\strut{}$0$}}%
      \csname LTb\endcsname%
      \put(640,1823){\makebox(0,0)[r]{\strut{}$0.02$}}%
      \csname LTb\endcsname%
      \put(640,2076){\makebox(0,0)[r]{\strut{}$0.04$}}%
      \csname LTb\endcsname%
      \put(640,2328){\makebox(0,0)[r]{\strut{}$0.06$}}%
      \csname LTb\endcsname%
      \put(640,2581){\makebox(0,0)[r]{\strut{}$0.08$}}%
      \csname LTb\endcsname%
      \put(1093,2841){\makebox(0,0){\strut{}$-6$}}%
      \csname LTb\endcsname%
      \put(1387,2841){\makebox(0,0){\strut{}$-5$}}%
      \csname LTb\endcsname%
      \put(1681,2841){\makebox(0,0){\strut{}$-4$}}%
      \csname LTb\endcsname%
      \put(1974,2841){\makebox(0,0){\strut{}$-3$}}%
      \csname LTb\endcsname%
      \put(2268,2841){\makebox(0,0){\strut{}$-2$}}%
      \csname LTb\endcsname%
      \put(2562,2841){\makebox(0,0){\strut{}$-1$}}%
      \csname LTb\endcsname%
      \put(2856,2841){\makebox(0,0){\strut{}$0$}}%
      \csname LTb\endcsname%
      \put(3149,2841){\makebox(0,0){\strut{}$1$}}%
      \csname LTb\endcsname%
      \put(3443,2841){\makebox(0,0){\strut{}$2$}}%
      \csname LTb\endcsname%
      \put(3737,2841){\makebox(0,0){\strut{}$3$}}%
      \csname LTb\endcsname%
      \put(4030,2841){\makebox(0,0){\strut{}$4$}}%
      \csname LTb\endcsname%
      \put(4324,2841){\makebox(0,0){\strut{}$5$}}%
      \csname LTb\endcsname%
      \put(4618,2841){\makebox(0,0){\strut{}$6$}}%
      \csname LTb\endcsname%
      \put(4487,2379){\makebox(0,0)[l]{\strut{}$f_3$}}%
    }%
    \gplgaddtomacro\gplfronttext{%
      \csname LTb\endcsname%
      \put(63,2075){\rotatebox{-270}{\makebox(0,0){\strut{}$\unit[A_{obs}/]{\%}$}}}%
      \csname LTb\endcsname%
      \put(2855,3119){\makebox(0,0){\strut{}$\unit[f /]{\mu Hz}$}}%
    }%
    \gplgaddtomacro\gplbacktext{%
      \csname LTb\endcsname%
      \put(640,1569){\makebox(0,0)[r]{\strut{}}}%
      \csname LTb\endcsname%
      \put(640,558){\makebox(0,0)[r]{\strut{}$0$}}%
      \csname LTb\endcsname%
      \put(640,760){\makebox(0,0)[r]{\strut{}$0.2$}}%
      \csname LTb\endcsname%
      \put(640,962){\makebox(0,0)[r]{\strut{}$0.4$}}%
      \csname LTb\endcsname%
      \put(640,1165){\makebox(0,0)[r]{\strut{}$0.6$}}%
      \csname LTb\endcsname%
      \put(640,1367){\makebox(0,0)[r]{\strut{}$0.8$}}%
      \csname LTb\endcsname%
      \put(816,298){\makebox(0,0){\strut{}$-0.6$}}%
      \csname LTb\endcsname%
      \put(1496,298){\makebox(0,0){\strut{}$-0.4$}}%
      \csname LTb\endcsname%
      \put(2176,298){\makebox(0,0){\strut{}$-0.2$}}%
      \csname LTb\endcsname%
      \put(2856,298){\makebox(0,0){\strut{}$0$}}%
      \csname LTb\endcsname%
      \put(3535,298){\makebox(0,0){\strut{}$0.2$}}%
      \csname LTb\endcsname%
      \put(4215,298){\makebox(0,0){\strut{}$0.4$}}%
      \csname LTb\endcsname%
      \put(4895,298){\makebox(0,0){\strut{}$0.6$}}%
      \csname LTb\endcsname%
      \put(4487,1367){\makebox(0,0)[l]{\strut{}w}}%
    }%
    \gplgaddtomacro\gplfronttext{%
      \csname LTb\endcsname%
      \put(165,1063){\rotatebox{-270}{\makebox(0,0){\strut{}$A_{window}$}}}%
      \csname LTb\endcsname%
      \put(2855,19){\makebox(0,0){\strut{}$\unit[f /]{d^{-1}}$}}%
    }%
    \gplbacktext
    \put(0,0){\includegraphics[scale=0.5]{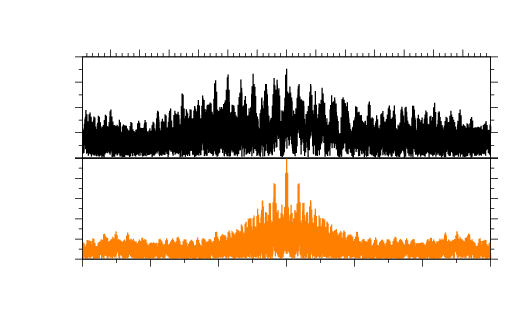}}%
    \gplfronttext
  \end{picture}%
\endgroup


        \caption{Amplitude $ A $ spectrum with respect to the pulsation frequency $ f_3 =  \unit[603.88741]{d^{-1}} $ of V541 Hya (top) and the normalised window function (bottom).}
        \label{fig:v541hya-windowf3}
\end{figure}

\begin{figure*}[tp]
        \centering
\begingroup
  \makeatletter
  \providecommand\color[2][]{%
    \GenericError{(gnuplot) \space\space\space\@spaces}{%
      Package color not loaded in conjunction with
      terminal option `colourtext'%
    }{See the gnuplot documentation for explanation.%
    }{Either use 'blacktext' in gnuplot or load the package
      color.sty in LaTeX.}%
    \renewcommand\color[2][]{}%
  }%
  \providecommand\includegraphics[2][]{%
    \GenericError{(gnuplot) \space\space\space\@spaces}{%
      Package graphicx or graphics not loaded%
    }{See the gnuplot documentation for explanation.%
    }{The gnuplot epslatex terminal needs graphicx.sty or graphics.sty.}
  }%
  \providecommand\rotatebox[2]{#2}%
  \@ifundefined{ifGPcolor}{%
    \newif\ifGPcolor
    \GPcolortrue
  }{}%
  \@ifundefined{ifGPblacktext}{%
    \newif\ifGPblacktext
    \GPblacktexttrue
  }{}%
  \let\gplgaddtomacro\g@addto@macro
  \gdef\gplbacktext{}%
  \gdef\gplfronttext{}%
  \makeatother
  \ifGPblacktext
    \def\colorrgb#1{}%
    \def\colorgray#1{}%
  \else
    \ifGPcolor
      \def\colorrgb#1{\color[rgb]{#1}}%
      \def\colorgray#1{\color[gray]{#1}}%
      \expandafter\def\csname LTw\endcsname{\color{white}}%
      \expandafter\def\csname LTb\endcsname{\color{black}}%
      \expandafter\def\csname LTa\endcsname{\color{black}}%
      \expandafter\def\csname LT0\endcsname{\color[rgb]{1,0,0}}%
      \expandafter\def\csname LT1\endcsname{\color[rgb]{0,1,0}}%
      \expandafter\def\csname LT2\endcsname{\color[rgb]{0,0,1}}%
      \expandafter\def\csname LT3\endcsname{\color[rgb]{1,0,1}}%
      \expandafter\def\csname LT4\endcsname{\color[rgb]{0,1,1}}%
      \expandafter\def\csname LT5\endcsname{\color[rgb]{1,1,0}}%
      \expandafter\def\csname LT6\endcsname{\color[rgb]{0,0,0}}%
      \expandafter\def\csname LT7\endcsname{\color[rgb]{1,0.3,0}}%
      \expandafter\def\csname LT8\endcsname{\color[rgb]{0.5,0.5,0.5}}%
    \else
      \def\colorrgb#1{\color{black}}%
      \def\colorgray#1{\color[gray]{#1}}%
      \expandafter\def\csname LTw\endcsname{\color{white}}%
      \expandafter\def\csname LTb\endcsname{\color{black}}%
      \expandafter\def\csname LTa\endcsname{\color{black}}%
      \expandafter\def\csname LT0\endcsname{\color{black}}%
      \expandafter\def\csname LT1\endcsname{\color{black}}%
      \expandafter\def\csname LT2\endcsname{\color{black}}%
      \expandafter\def\csname LT3\endcsname{\color{black}}%
      \expandafter\def\csname LT4\endcsname{\color{black}}%
      \expandafter\def\csname LT5\endcsname{\color{black}}%
      \expandafter\def\csname LT6\endcsname{\color{black}}%
      \expandafter\def\csname LT7\endcsname{\color{black}}%
      \expandafter\def\csname LT8\endcsname{\color{black}}%
    \fi
  \fi
    \setlength{\unitlength}{0.0500bp}%
    \ifx\gptboxheight\undefined%
      \newlength{\gptboxheight}%
      \newlength{\gptboxwidth}%
      \newsavebox{\gptboxtext}%
    \fi%
    \setlength{\fboxrule}{0.5pt}%
    \setlength{\fboxsep}{1pt}%
\begin{picture}(10420.00,5800.00)%
    \gplgaddtomacro\gplbacktext{%
      \csname LTb\endcsname%
      \put(844,3866){\makebox(0,0)[r]{\strut{}$0$}}%
      \csname LTb\endcsname%
      \put(844,4126){\makebox(0,0)[r]{\strut{}$0.1$}}%
      \csname LTb\endcsname%
      \put(844,4386){\makebox(0,0)[r]{\strut{}$0.2$}}%
      \csname LTb\endcsname%
      \put(844,4647){\makebox(0,0)[r]{\strut{}$0.3$}}%
      \csname LTb\endcsname%
      \put(844,4907){\makebox(0,0)[r]{\strut{}$0.4$}}%
      \csname LTb\endcsname%
      \put(844,5167){\makebox(0,0)[r]{\strut{}$0.5$}}%
      \csname LTb\endcsname%
      \put(844,5427){\makebox(0,0)[r]{\strut{}$0.6$}}%
      \csname LTb\endcsname%
      \put(9779,3866){\makebox(0,0)[l]{\strut{}$-0.1$}}%
      \csname LTb\endcsname%
      \put(9779,4126){\makebox(0,0)[l]{\strut{}$0$}}%
      \csname LTb\endcsname%
      \put(9779,4386){\makebox(0,0)[l]{\strut{}$0.1$}}%
      \csname LTb\endcsname%
      \put(9779,4647){\makebox(0,0)[l]{\strut{}$0.2$}}%
      \csname LTb\endcsname%
      \put(9779,4907){\makebox(0,0)[l]{\strut{}$0.3$}}%
      \csname LTb\endcsname%
      \put(9779,5167){\makebox(0,0)[l]{\strut{}$0.4$}}%
      \csname LTb\endcsname%
      \put(9779,5427){\makebox(0,0)[l]{\strut{}$0.5$}}%
      \csname LTb\endcsname%
      \put(1668,5687){\makebox(0,0){\strut{}2004}}%
      \csname LTb\endcsname%
      \put(2365,5687){\makebox(0,0){\strut{}2005}}%
      \csname LTb\endcsname%
      \put(3061,5687){\makebox(0,0){\strut{}2006}}%
      \csname LTb\endcsname%
      \put(3757,5687){\makebox(0,0){\strut{}2007}}%
      \csname LTb\endcsname%
      \put(4455,5687){\makebox(0,0){\strut{}2008}}%
      \csname LTb\endcsname%
      \put(5151,5687){\makebox(0,0){\strut{}2009}}%
      \csname LTb\endcsname%
      \put(5847,5687){\makebox(0,0){\strut{}2010}}%
      \csname LTb\endcsname%
      \put(6544,5687){\makebox(0,0){\strut{}2011}}%
      \csname LTb\endcsname%
      \put(7242,5687){\makebox(0,0){\strut{}2012}}%
      \csname LTb\endcsname%
      \put(7938,5687){\makebox(0,0){\strut{}2013}}%
      \csname LTb\endcsname%
      \put(8634,5687){\makebox(0,0){\strut{}2014}}%
      \csname LTb\endcsname%
      \put(9330,5687){\makebox(0,0){\strut{}2015}}%
    }%
    \gplgaddtomacro\gplfronttext{%
      \csname LTb\endcsname%
      \put(343,4646){\rotatebox{-270}{\makebox(0,0){\strut{}$\unit[A(f_1)/]{\%}$}}}%
      \csname LTb\endcsname%
      \put(10381,4646){\rotatebox{-270}{\makebox(0,0){\strut{}$\unit[A(f_2)/]{\%}$}}}%
      \csname LTb\endcsname%
      \put(8815,4219){\makebox(0,0)[r]{\strut{}$f_1$}}%
      \csname LTb\endcsname%
      \put(8815,4033){\makebox(0,0)[r]{\strut{}$f_2$}}%
    }%
    \gplgaddtomacro\gplbacktext{%
      \csname LTb\endcsname%
      \put(844,3866){\makebox(0,0)[r]{\strut{}}}%
      \csname LTb\endcsname%
      \put(844,2425){\makebox(0,0)[r]{\strut{}$-40$}}%
      \csname LTb\endcsname%
      \put(844,2665){\makebox(0,0)[r]{\strut{}$-20$}}%
      \csname LTb\endcsname%
      \put(844,2905){\makebox(0,0)[r]{\strut{}$0$}}%
      \csname LTb\endcsname%
      \put(844,3146){\makebox(0,0)[r]{\strut{}$20$}}%
      \csname LTb\endcsname%
      \put(844,3386){\makebox(0,0)[r]{\strut{}$40$}}%
      \csname LTb\endcsname%
      \put(844,3626){\makebox(0,0)[r]{\strut{}$60$}}%
      \csname LTb\endcsname%
      \put(9779,2425){\makebox(0,0)[l]{\strut{}}}%
      \csname LTb\endcsname%
      \put(9779,2665){\makebox(0,0)[l]{\strut{}}}%
      \csname LTb\endcsname%
      \put(9779,2905){\makebox(0,0)[l]{\strut{}}}%
      \csname LTb\endcsname%
      \put(9779,3146){\makebox(0,0)[l]{\strut{}}}%
      \csname LTb\endcsname%
      \put(9779,3386){\makebox(0,0)[l]{\strut{}}}%
      \csname LTb\endcsname%
      \put(9779,3626){\makebox(0,0)[l]{\strut{}}}%
      \csname LTb\endcsname%
      \put(9779,3866){\makebox(0,0)[l]{\strut{}}}%
    }%
    \gplgaddtomacro\gplfronttext{%
      \csname LTb\endcsname%
      \put(343,3085){\rotatebox{-270}{\makebox(0,0){\strut{}$\unit[(O-C)/]{s}$}}}%
    }%
    \gplgaddtomacro\gplbacktext{%
      \csname LTb\endcsname%
      \put(844,864){\makebox(0,0)[r]{\strut{}$-60$}}%
      \csname LTb\endcsname%
      \put(844,1104){\makebox(0,0)[r]{\strut{}$-40$}}%
      \csname LTb\endcsname%
      \put(844,1344){\makebox(0,0)[r]{\strut{}$-20$}}%
      \csname LTb\endcsname%
      \put(844,1584){\makebox(0,0)[r]{\strut{}$0$}}%
      \csname LTb\endcsname%
      \put(844,1824){\makebox(0,0)[r]{\strut{}$20$}}%
      \csname LTb\endcsname%
      \put(844,2064){\makebox(0,0)[r]{\strut{}$40$}}%
      \csname LTb\endcsname%
      \put(844,2304){\makebox(0,0)[r]{\strut{}$60$}}%
      \csname LTb\endcsname%
      \put(1020,484){\makebox(0,0){\strut{}$53000$}}%
      \csname LTb\endcsname%
      \put(2927,484){\makebox(0,0){\strut{}$54000$}}%
      \csname LTb\endcsname%
      \put(4835,484){\makebox(0,0){\strut{}$55000$}}%
      \csname LTb\endcsname%
      \put(6742,484){\makebox(0,0){\strut{}$56000$}}%
      \csname LTb\endcsname%
      \put(8649,484){\makebox(0,0){\strut{}$57000$}}%
      \csname LTb\endcsname%
      \put(9779,864){\makebox(0,0)[l]{\strut{}}}%
      \csname LTb\endcsname%
      \put(9779,1104){\makebox(0,0)[l]{\strut{}}}%
      \csname LTb\endcsname%
      \put(9779,1344){\makebox(0,0)[l]{\strut{}}}%
      \csname LTb\endcsname%
      \put(9779,1584){\makebox(0,0)[l]{\strut{}}}%
      \csname LTb\endcsname%
      \put(9779,1824){\makebox(0,0)[l]{\strut{}}}%
      \csname LTb\endcsname%
      \put(9779,2064){\makebox(0,0)[l]{\strut{}}}%
      \csname LTb\endcsname%
      \put(9779,2304){\makebox(0,0)[l]{\strut{}}}%
    }%
    \gplgaddtomacro\gplfronttext{%
      \csname LTb\endcsname%
      \put(343,1524){\rotatebox{-270}{\makebox(0,0){\strut{}residuals$\unit[/]{s}$}}}%
      \csname LTb\endcsname%
      \put(5311,205){\makebox(0,0){\strut{}$\unit[t/]{MBJD_{TDB}}$}}%
    }%
    \gplbacktext
    \put(0,0){\includegraphics[scale=0.5]{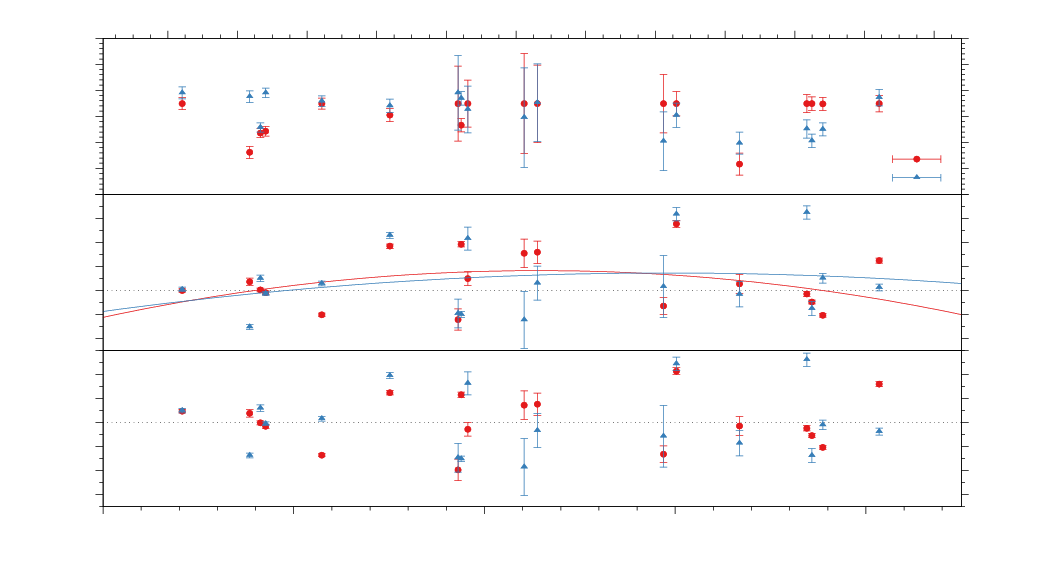}}%
    \gplfronttext
  \end{picture}%
\endgroup


        \caption{Results for the two main pulsations of V541 Hya. \textit{Top panel:} Amplitudes. \textit{Middle panel:} Fits of the $ O-C $ data with second order polynomials in time. \textit{Lower panel:} Residuals.}
        \label{fig:v541hya-oc}
\end{figure*}

\citet{randall_observations_2009} speculated about rotational mode splitting for $ f_3 $ with $ \Delta f_{3,-}  = \unit[5.12]{\mu Hz} $ and $ \Delta f_{3,+}  = \unit[3.68]{\mu Hz} $.
The asteroseismic modelling associates $ f_1 $ with a $ l=0 $ mode and $ f_2 $ with $ l=0\text{ or }1 $ mode (depending on the favoured model). $ f_3 $ corresponds to a $ l=2 $ mode. They caution this interpretation due to their limited resolution in frequency space, the mode splitting could be an unresolved quintuplet. Our data set shows no clear evidence for a mode splitting with $ \Delta f_{3,-}  = \unit[5.12]{\mu Hz} $ or $ \Delta f_{3,+}  = \unit[3.68]{\mu Hz} $ (see Fig.~\ref{fig:v541hya-windowf3}) but rather a mode splitting for $ f_1 $ and $ f_2 $ with about $ \Delta f = \unit[0.08]{\mu Hz} $ (Fig.~\ref{fig:v541hya-window}). Assuming these modes are of degree $ l=1 $, this could be interpreted as a triplet. But \citet{randall_observations_2009} model these modes with a degree of $ l=0, $ which does not support a mode splitting into triplets.

The $ O-C $ diagram in Fig.~\ref{fig:v541hya-oc} shows the analysis of the two main pulsation modes and the variation of the pulsation amplitudes. The second order fits in time correspond to changes in period of $ \dot{P}/P_{f1} = \left( -1.49 \pm 0.11 \right)\times \unit[10^{-5}]{d^{-1}} $ and $ \dot{P}/P_{f2} = \left( -0.7 \pm 1.5 \right)\times \unit[10^{-5}]{d^{-1}} $. For $ f_2, $ the change in period does not significantly differ from the null hypothesis. Assuming these changes origin from stellar evolution, V541~Hya might just have passed the point of sign change in $ \dot{P} $ and at the beginning of the contraction phase. While the arrival times scatter widely, the amplitudes of both pulsations remain almost constant within the uncertainties. If V541~Hya is in its evolution close to starting the contraction phase, as indicated by a $ \dot{P} $ close to zero, the changes in stellar structure may cancel the strict phase coherence.

\subsection{Testing the sub-stellar companion hypothesis}

In order to set upper limits to the mass of a companion, we computed a series of synthetic $ O-C $ curves for different orbital periods and companion masses, assuming circular orbits, and compared these curves with the $ O-C $ measurements after subtracting the long-term variations. 

For each synthetic $ O-C $ curve, we selected the phase that gives the best fit to  the data using a weighted least squares algorithm. For each observational point, we computed the difference, in absolute value and in $\sigma$ units (where $\sigma$ is the $ O-C $ error), between $ O-C $ and the synthetic value. The greyscale in Figs.~\ref{fig:rv-planet} and \ref{fig:rv-planet2} corresponds to the mean value of this difference in $\sigma$ units, which means that the presence of a companion is indicated by a minimum (bright areas) of this parameter. We see that in V1636~Ori, QQ~Vir, and V541~Hya, the mean difference for $ f_1 $ is always very high, implying that the data are not compatible with a companion.  However, these results are limited by the fact that the $ O-C $ diagrams of 
these stars are 'contaminated' by other irregular variations, presumably due 
to other reasons like non-linear interactions between different pulsation 
modes, for example, and therefore these constraints to the orbital period and mass of a companion must be taken with some caution. For the $ f_2 $ and $ f_3 $ measurements, the mean difference to the synthetic data is smaller in sigma units (because of the larger uncertainties) and very uniform. The uncertainties of the O-C measurements are not small enough to favour a set of models in the period-mass parameter space.

For $ f_1 $ of DW~Lyn, there is a significant minimum at about $ \unit[1450]{d} $ ($\sim\unit[4]{years}$) and $\sim\unit[5]{M_{\jupiter}} $\footnote{$ \unit[1]{M_{\jupiter}} $ (Jupiter mass) $ = 1.899 \cdot \unit[10^{27}]{kg} $}, which is also well visible in the $ O-C $ diagram of Fig.~\ref{fig:dwlyn-oc}. This periodicity is not visible in the second frequency $f_2$ which, however, has much larger error bars due to the much lower amplitude of $f_2$ with respect to $f_1$.

\citet{lutz_exotime_2011} described a periodicity at 80 days, detected for $ f_2 $. We can recover this signal, however, with a low significance. This would correspond to a light-travel time amplitude of $ \unit[4]{s} $ (for $ m\,\sin i \approx \unit[15]{M_{\jupiter}} $), which is smaller than the $ \unit[15]{s} $ measured by \citet{lutz_exotime_2011}. Nevertheless, this signal is not confirmed by $ f_1 $. Thus, we rule out a companion induced signal in the arrival times due to the lack of simultaneous signals in $ f_1 $ and $ f_2 $ with similar amplitude. The tentative signal in $ f_2 $ is better explained by mode beating, as already described in Sect.~\ref{sec:results-dwlyn}. The variations seen in the first 200 days of the $ O-C $ diagram in Fig.~\ref{fig:dwlyn-oc} correspond to a periodicity of about 80 days and are accompanied by variations in the amplitude of the pulsation.

For V1636 Ori, \citet{lutz_search_2011} predicted a period at $ \unit[160]{d} $ and amplitude of $ \unit[12]{s} $. This can not be confirmed as a companion-induced signal. A periodic signal with an amplitude of $ \unit[6.5]{s} $ (for $ m\,\sin i \approx \unit[15]{M_{\jupiter}} $) is indicated in the analysis of $ f_1, $ but at a low significance and accompanied by many other signals of similar significance. This periodicity is not confirmed by a significant signal in the measurements of $ f_2 $.

\begin{figure*}[tp]
        \centering
        \begin{subfigure}[b]{\textwidth}
                \centering
\begingroup
  \makeatletter
  \providecommand\color[2][]{%
    \GenericError{(gnuplot) \space\space\space\@spaces}{%
      Package color not loaded in conjunction with
      terminal option `colourtext'%
    }{See the gnuplot documentation for explanation.%
    }{Either use 'blacktext' in gnuplot or load the package
      color.sty in LaTeX.}%
    \renewcommand\color[2][]{}%
  }%
  \providecommand\includegraphics[2][]{%
    \GenericError{(gnuplot) \space\space\space\@spaces}{%
      Package graphicx or graphics not loaded%
    }{See the gnuplot documentation for explanation.%
    }{The gnuplot epslatex terminal needs graphicx.sty or graphics.sty.}
  }%
  \providecommand\rotatebox[2]{#2}%
  \@ifundefined{ifGPcolor}{%
    \newif\ifGPcolor
    \GPcolortrue
  }{}%
  \@ifundefined{ifGPblacktext}{%
    \newif\ifGPblacktext
    \GPblacktextfalse
  }{}%
  \let\gplgaddtomacro\g@addto@macro
  \gdef\gplbacktext{}%
  \gdef\gplfronttext{}%
  \makeatother
  \ifGPblacktext
    \def\colorrgb#1{}%
    \def\colorgray#1{}%
  \else
    \ifGPcolor
      \def\colorrgb#1{\color[rgb]{#1}}%
      \def\colorgray#1{\color[gray]{#1}}%
      \expandafter\def\csname LTw\endcsname{\color{white}}%
      \expandafter\def\csname LTb\endcsname{\color{black}}%
      \expandafter\def\csname LTa\endcsname{\color{black}}%
      \expandafter\def\csname LT0\endcsname{\color[rgb]{1,0,0}}%
      \expandafter\def\csname LT1\endcsname{\color[rgb]{0,1,0}}%
      \expandafter\def\csname LT2\endcsname{\color[rgb]{0,0,1}}%
      \expandafter\def\csname LT3\endcsname{\color[rgb]{1,0,1}}%
      \expandafter\def\csname LT4\endcsname{\color[rgb]{0,1,1}}%
      \expandafter\def\csname LT5\endcsname{\color[rgb]{1,1,0}}%
      \expandafter\def\csname LT6\endcsname{\color[rgb]{0,0,0}}%
      \expandafter\def\csname LT7\endcsname{\color[rgb]{1,0.3,0}}%
      \expandafter\def\csname LT8\endcsname{\color[rgb]{0.5,0.5,0.5}}%
    \else
      \def\colorrgb#1{\color{black}}%
      \def\colorgray#1{\color[gray]{#1}}%
      \expandafter\def\csname LTw\endcsname{\color{white}}%
      \expandafter\def\csname LTb\endcsname{\color{black}}%
      \expandafter\def\csname LTa\endcsname{\color{black}}%
      \expandafter\def\csname LT0\endcsname{\color{black}}%
      \expandafter\def\csname LT1\endcsname{\color{black}}%
      \expandafter\def\csname LT2\endcsname{\color{black}}%
      \expandafter\def\csname LT3\endcsname{\color{black}}%
      \expandafter\def\csname LT4\endcsname{\color{black}}%
      \expandafter\def\csname LT5\endcsname{\color{black}}%
      \expandafter\def\csname LT6\endcsname{\color{black}}%
      \expandafter\def\csname LT7\endcsname{\color{black}}%
      \expandafter\def\csname LT8\endcsname{\color{black}}%
    \fi
  \fi
    \setlength{\unitlength}{0.0500bp}%
    \ifx\gptboxheight\undefined%
      \newlength{\gptboxheight}%
      \newlength{\gptboxwidth}%
      \newsavebox{\gptboxtext}%
    \fi%
    \setlength{\fboxrule}{0.5pt}%
    \setlength{\fboxsep}{1pt}%
\begin{picture}(10430.00,3600.00)%
    \gplgaddtomacro\gplbacktext{%
    }%
    \gplgaddtomacro\gplfronttext{%
      \csname LTb\endcsname%
      \put(1056,477){\makebox(0,0){\strut{}$1$}}%
      \put(2365,477){\makebox(0,0){\strut{}$10$}}%
      \put(3672,477){\makebox(0,0){\strut{}$100$}}%
      \put(4981,477){\makebox(0,0){\strut{}$1000$}}%
      \put(3201,147){\makebox(0,0){\strut{}$\unit[P/]{d}$}}%
      \put(778,881){\makebox(0,0)[r]{\strut{}$0$}}%
      \put(778,1185){\makebox(0,0)[r]{\strut{}$2$}}%
      \put(778,1489){\makebox(0,0)[r]{\strut{}$4$}}%
      \put(778,1793){\makebox(0,0)[r]{\strut{}$6$}}%
      \put(778,2095){\makebox(0,0)[r]{\strut{}$8$}}%
      \put(778,2399){\makebox(0,0)[r]{\strut{}$10$}}%
      \put(778,2703){\makebox(0,0)[r]{\strut{}$12$}}%
      \put(778,3007){\makebox(0,0)[r]{\strut{}$14$}}%
      \put(350,2020){\rotatebox{-270}{\makebox(0,0){\strut{}$\unit[m \sin i/]{M_{\jupiter}}$}}}%
      \put(3612,1185){\makebox(0,0)[l]{\strut{}$\unit[80]{d}$}}%
    }%
    \gplgaddtomacro\gplbacktext{%
    }%
    \gplgaddtomacro\gplfronttext{%
      \csname LTb\endcsname%
      \put(5347,477){\makebox(0,0){\strut{}$1$}}%
      \put(6656,477){\makebox(0,0){\strut{}$10$}}%
      \put(7963,477){\makebox(0,0){\strut{}$100$}}%
      \put(9272,477){\makebox(0,0){\strut{}$1000$}}%
      \put(7492,147){\makebox(0,0){\strut{}$\unit[P/]{d}$}}%
      \put(10091,881){\makebox(0,0)[l]{\strut{}$1$}}%
      \put(10091,1640){\makebox(0,0)[l]{\strut{}$2$}}%
      \put(10091,2399){\makebox(0,0)[l]{\strut{}$4$}}%
      \put(10091,3159){\makebox(0,0)[l]{\strut{}$8$}}%
      \put(10289,2020){\rotatebox{-270}{\makebox(0,0){\strut{}$\sigma$}}}%
      \put(7903,1178){\makebox(0,0)[l]{\strut{}$\unit[80]{d}$}}%
    }%
    \gplbacktext
    \put(0,0){\includegraphics[scale=0.5]{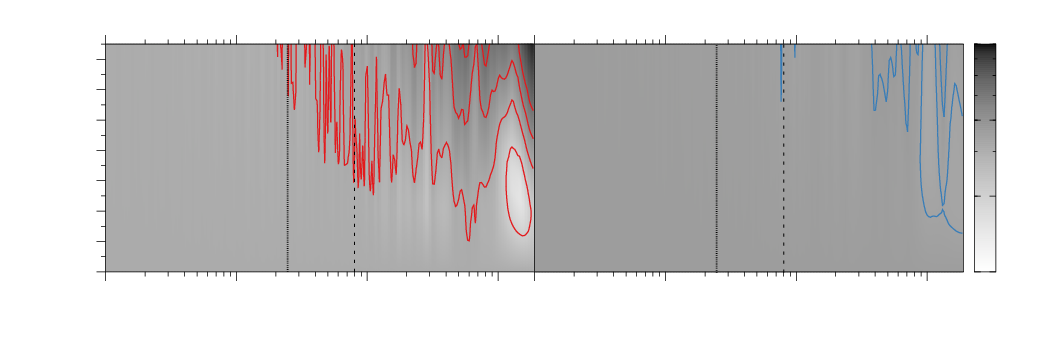}}%
    \gplfronttext{
        \tiny
        \color{cb-red}%
        \put(4970,1500){\makebox(0,0){\strut{}2}}%
        \put(4800,1900){\makebox(0,0){\strut{}3}}%
        \put(4950,2900){\makebox(0,0){\strut{}4}}%
        \put(5230,3000){\makebox(0,0){\strut{}5}}%
        \color{cb-blue}%
        \put(7950,2850){\makebox(0,0){\strut{}3.25}}%
        \put(9220,1300){\makebox(0,0){\strut{}3.25}}%
        \put(8850,2500){\makebox(0,0){\strut{}3.5}}%
        \put(9510,2880){\makebox(0,0){\strut{}3.5}}%
        \normalsize
        }%
  \end{picture}%
\endgroup


                \caption{DW Lyn. Contour lines for $ f_1 $ are placed at 2, 3, 4, and $ \unit[5]{\sigma} $ (left panel), and for $ f_2 $ at 3.25 and $ \unit[3.5]{\sigma} $ (right panel), as indicated by their labels. The planetary signal proposed by \citet{lutz_exotime_2011} at a period of $ \unit[80]{d} $ is indicated as dashed line.}
                \label{fig:dwlyn-rv-planet}
        \end{subfigure}
        \begin{subfigure}[b]{\textwidth}
                \centering
\begingroup
  \makeatletter
  \providecommand\color[2][]{%
    \GenericError{(gnuplot) \space\space\space\@spaces}{%
      Package color not loaded in conjunction with
      terminal option `colourtext'%
    }{See the gnuplot documentation for explanation.%
    }{Either use 'blacktext' in gnuplot or load the package
      color.sty in LaTeX.}%
    \renewcommand\color[2][]{}%
  }%
  \providecommand\includegraphics[2][]{%
    \GenericError{(gnuplot) \space\space\space\@spaces}{%
      Package graphicx or graphics not loaded%
    }{See the gnuplot documentation for explanation.%
    }{The gnuplot epslatex terminal needs graphicx.sty or graphics.sty.}
  }%
  \providecommand\rotatebox[2]{#2}%
  \@ifundefined{ifGPcolor}{%
    \newif\ifGPcolor
    \GPcolortrue
  }{}%
  \@ifundefined{ifGPblacktext}{%
    \newif\ifGPblacktext
    \GPblacktextfalse
  }{}%
  \let\gplgaddtomacro\g@addto@macro
  \gdef\gplbacktext{}%
  \gdef\gplfronttext{}%
  \makeatother
  \ifGPblacktext
    \def\colorrgb#1{}%
    \def\colorgray#1{}%
  \else
    \ifGPcolor
      \def\colorrgb#1{\color[rgb]{#1}}%
      \def\colorgray#1{\color[gray]{#1}}%
      \expandafter\def\csname LTw\endcsname{\color{white}}%
      \expandafter\def\csname LTb\endcsname{\color{black}}%
      \expandafter\def\csname LTa\endcsname{\color{black}}%
      \expandafter\def\csname LT0\endcsname{\color[rgb]{1,0,0}}%
      \expandafter\def\csname LT1\endcsname{\color[rgb]{0,1,0}}%
      \expandafter\def\csname LT2\endcsname{\color[rgb]{0,0,1}}%
      \expandafter\def\csname LT3\endcsname{\color[rgb]{1,0,1}}%
      \expandafter\def\csname LT4\endcsname{\color[rgb]{0,1,1}}%
      \expandafter\def\csname LT5\endcsname{\color[rgb]{1,1,0}}%
      \expandafter\def\csname LT6\endcsname{\color[rgb]{0,0,0}}%
      \expandafter\def\csname LT7\endcsname{\color[rgb]{1,0.3,0}}%
      \expandafter\def\csname LT8\endcsname{\color[rgb]{0.5,0.5,0.5}}%
    \else
      \def\colorrgb#1{\color{black}}%
      \def\colorgray#1{\color[gray]{#1}}%
      \expandafter\def\csname LTw\endcsname{\color{white}}%
      \expandafter\def\csname LTb\endcsname{\color{black}}%
      \expandafter\def\csname LTa\endcsname{\color{black}}%
      \expandafter\def\csname LT0\endcsname{\color{black}}%
      \expandafter\def\csname LT1\endcsname{\color{black}}%
      \expandafter\def\csname LT2\endcsname{\color{black}}%
      \expandafter\def\csname LT3\endcsname{\color{black}}%
      \expandafter\def\csname LT4\endcsname{\color{black}}%
      \expandafter\def\csname LT5\endcsname{\color{black}}%
      \expandafter\def\csname LT6\endcsname{\color{black}}%
      \expandafter\def\csname LT7\endcsname{\color{black}}%
      \expandafter\def\csname LT8\endcsname{\color{black}}%
    \fi
  \fi
    \setlength{\unitlength}{0.0500bp}%
    \ifx\gptboxheight\undefined%
      \newlength{\gptboxheight}%
      \newlength{\gptboxwidth}%
      \newsavebox{\gptboxtext}%
    \fi%
    \setlength{\fboxrule}{0.5pt}%
    \setlength{\fboxsep}{1pt}%
\begin{picture}(10430.00,3600.00)%
    \gplgaddtomacro\gplbacktext{%
    }%
    \gplgaddtomacro\gplfronttext{%
      \csname LTb\endcsname%
      \put(1056,477){\makebox(0,0){\strut{}$1$}}%
      \put(2319,477){\makebox(0,0){\strut{}$10$}}%
      \put(3581,477){\makebox(0,0){\strut{}$100$}}%
      \put(4843,477){\makebox(0,0){\strut{}$1000$}}%
      \put(3201,147){\makebox(0,0){\strut{}$\unit[P/]{d}$}}%
      \put(778,881){\makebox(0,0)[r]{\strut{}$0$}}%
      \put(778,1185){\makebox(0,0)[r]{\strut{}$2$}}%
      \put(778,1489){\makebox(0,0)[r]{\strut{}$4$}}%
      \put(778,1793){\makebox(0,0)[r]{\strut{}$6$}}%
      \put(778,2095){\makebox(0,0)[r]{\strut{}$8$}}%
      \put(778,2399){\makebox(0,0)[r]{\strut{}$10$}}%
      \put(778,2703){\makebox(0,0)[r]{\strut{}$12$}}%
      \put(778,3007){\makebox(0,0)[r]{\strut{}$14$}}%
      \put(350,2020){\rotatebox{-270}{\makebox(0,0){\strut{}$\unit[m \sin i/]{M_{\jupiter}}$}}}%
      \colorrgb{1.00,1.00,1.00}%
      \put(3904,1185){\makebox(0,0)[l]{\strut{}$\unit[160]{d}$}}%
    }%
    \gplgaddtomacro\gplbacktext{%
    }%
    \gplgaddtomacro\gplfronttext{%
      \csname LTb\endcsname%
      \put(5347,477){\makebox(0,0){\strut{}$1$}}%
      \put(6610,477){\makebox(0,0){\strut{}$10$}}%
      \put(7872,477){\makebox(0,0){\strut{}$100$}}%
      \put(9134,477){\makebox(0,0){\strut{}$1000$}}%
      \put(7492,147){\makebox(0,0){\strut{}$\unit[P/]{d}$}}%
      \put(10091,881){\makebox(0,0)[l]{\strut{}$1$}}%
      \put(10091,1464){\makebox(0,0)[l]{\strut{}$2$}}%
      \put(10091,2047){\makebox(0,0)[l]{\strut{}$4$}}%
      \put(10091,2630){\makebox(0,0)[l]{\strut{}$8$}}%
      \put(10289,2020){\rotatebox{-270}{\makebox(0,0){\strut{}$\sigma$}}}%
      \colorrgb{0.00,0.00,0.00}%
      \put(8195,1178){\makebox(0,0)[l]{\strut{}$\unit[160]{d}$}}%
    }%
    \gplbacktext
    \put(0,0){\includegraphics[scale=0.5]{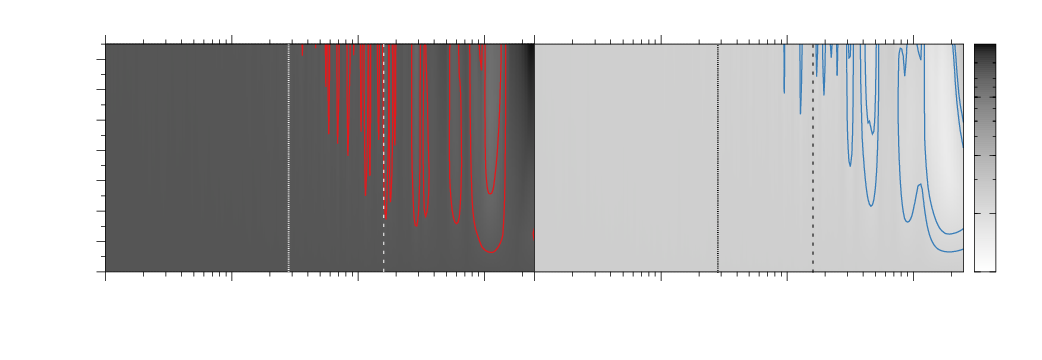}}%
    \gplfronttext{
        \tiny
        \color{cb-red}%
        \put(4850,1520){\makebox(0,0){\strut{}9}}%
        \put(4850,950){\makebox(0,0){\strut{}10}}%
        \put(4480,1200){\makebox(0,0){\strut{}10}}%
        \put(3970,2200){\makebox(0,0){\strut{}10}}%
        \color{cb-blue}%
        \put(9130,2100){\makebox(0,0){\strut{}2}}%
        \put(9050,1250){\makebox(0,0){\strut{}2.2}}%
        \put(8650,2125){\makebox(0,0){\strut{}2}}%
        \put(8650,1365){\makebox(0,0){\strut{}2.2}}%
        \put(8250,2415){\makebox(0,0){\strut{}2.2}}%
        \normalsize
        }%
  \end{picture}%
\endgroup


                \caption{V1636 Ori. Contour lines for $ f_1 $ are placed at 9 and $ \unit[10]{\sigma} $ (left panel), and for $ f_2 $ at 2 and $ \unit[2.2]{\sigma} $ (right panel), as indicated by their labels. The planetary signal proposed by \citet{lutz_exotime_2011} at a period of $ \unit[160]{d} $ is indicated as dashed line.}
                \label{fig:v1636ori-rv-planet}
        \end{subfigure}
        \begin{subfigure}[b]{\textwidth}
                \centering
\begingroup
  \makeatletter
  \providecommand\color[2][]{%
    \GenericError{(gnuplot) \space\space\space\@spaces}{%
      Package color not loaded in conjunction with
      terminal option `colourtext'%
    }{See the gnuplot documentation for explanation.%
    }{Either use 'blacktext' in gnuplot or load the package
      color.sty in LaTeX.}%
    \renewcommand\color[2][]{}%
  }%
  \providecommand\includegraphics[2][]{%
    \GenericError{(gnuplot) \space\space\space\@spaces}{%
      Package graphicx or graphics not loaded%
    }{See the gnuplot documentation for explanation.%
    }{The gnuplot epslatex terminal needs graphicx.sty or graphics.sty.}
  }%
  \providecommand\rotatebox[2]{#2}%
  \@ifundefined{ifGPcolor}{%
    \newif\ifGPcolor
    \GPcolortrue
  }{}%
  \@ifundefined{ifGPblacktext}{%
    \newif\ifGPblacktext
    \GPblacktextfalse
  }{}%
  \let\gplgaddtomacro\g@addto@macro
  \gdef\gplbacktext{}%
  \gdef\gplfronttext{}%
  \makeatother
  \ifGPblacktext
    \def\colorrgb#1{}%
    \def\colorgray#1{}%
  \else
    \ifGPcolor
      \def\colorrgb#1{\color[rgb]{#1}}%
      \def\colorgray#1{\color[gray]{#1}}%
      \expandafter\def\csname LTw\endcsname{\color{white}}%
      \expandafter\def\csname LTb\endcsname{\color{black}}%
      \expandafter\def\csname LTa\endcsname{\color{black}}%
      \expandafter\def\csname LT0\endcsname{\color[rgb]{1,0,0}}%
      \expandafter\def\csname LT1\endcsname{\color[rgb]{0,1,0}}%
      \expandafter\def\csname LT2\endcsname{\color[rgb]{0,0,1}}%
      \expandafter\def\csname LT3\endcsname{\color[rgb]{1,0,1}}%
      \expandafter\def\csname LT4\endcsname{\color[rgb]{0,1,1}}%
      \expandafter\def\csname LT5\endcsname{\color[rgb]{1,1,0}}%
      \expandafter\def\csname LT6\endcsname{\color[rgb]{0,0,0}}%
      \expandafter\def\csname LT7\endcsname{\color[rgb]{1,0.3,0}}%
      \expandafter\def\csname LT8\endcsname{\color[rgb]{0.5,0.5,0.5}}%
    \else
      \def\colorrgb#1{\color{black}}%
      \def\colorgray#1{\color[gray]{#1}}%
      \expandafter\def\csname LTw\endcsname{\color{white}}%
      \expandafter\def\csname LTb\endcsname{\color{black}}%
      \expandafter\def\csname LTa\endcsname{\color{black}}%
      \expandafter\def\csname LT0\endcsname{\color{black}}%
      \expandafter\def\csname LT1\endcsname{\color{black}}%
      \expandafter\def\csname LT2\endcsname{\color{black}}%
      \expandafter\def\csname LT3\endcsname{\color{black}}%
      \expandafter\def\csname LT4\endcsname{\color{black}}%
      \expandafter\def\csname LT5\endcsname{\color{black}}%
      \expandafter\def\csname LT6\endcsname{\color{black}}%
      \expandafter\def\csname LT7\endcsname{\color{black}}%
      \expandafter\def\csname LT8\endcsname{\color{black}}%
    \fi
  \fi
    \setlength{\unitlength}{0.0500bp}%
    \ifx\gptboxheight\undefined%
      \newlength{\gptboxheight}%
      \newlength{\gptboxwidth}%
      \newsavebox{\gptboxtext}%
    \fi%
    \setlength{\fboxrule}{0.5pt}%
    \setlength{\fboxsep}{1pt}%
\begin{picture}(10430.00,3600.00)%
    \gplgaddtomacro\gplbacktext{%
    }%
    \gplgaddtomacro\gplfronttext{%
      \csname LTb\endcsname%
      \put(1057,477){\makebox(0,0){\strut{}$1$}}%
      \put(1837,477){\makebox(0,0){\strut{}$10$}}%
      \put(2617,477){\makebox(0,0){\strut{}$100$}}%
      \put(3398,477){\makebox(0,0){\strut{}$1000$}}%
      \put(2486,147){\makebox(0,0){\strut{}$\unit[P/]{d}$}}%
      \put(779,881){\makebox(0,0)[r]{\strut{}$0$}}%
      \put(779,1185){\makebox(0,0)[r]{\strut{}$2$}}%
      \put(779,1489){\makebox(0,0)[r]{\strut{}$4$}}%
      \put(779,1793){\makebox(0,0)[r]{\strut{}$6$}}%
      \put(779,2095){\makebox(0,0)[r]{\strut{}$8$}}%
      \put(779,2399){\makebox(0,0)[r]{\strut{}$10$}}%
      \put(779,2703){\makebox(0,0)[r]{\strut{}$12$}}%
      \put(779,3007){\makebox(0,0)[r]{\strut{}$14$}}%
      \put(351,2020){\rotatebox{-270}{\makebox(0,0){\strut{}$\unit[m \sin i/]{M_{\jupiter}}$}}}%
    }%
    \gplgaddtomacro\gplbacktext{%
    }%
    \gplgaddtomacro\gplfronttext{%
      \csname LTb\endcsname%
      \put(3918,477){\makebox(0,0){\strut{}$1$}}%
      \put(4698,477){\makebox(0,0){\strut{}$10$}}%
      \put(5478,477){\makebox(0,0){\strut{}$100$}}%
      \put(6259,477){\makebox(0,0){\strut{}$1000$}}%
      \put(5347,147){\makebox(0,0){\strut{}$\unit[P/]{d}$}}%
    }%
    \gplgaddtomacro\gplbacktext{%
    }%
    \gplgaddtomacro\gplfronttext{%
      \csname LTb\endcsname%
      \put(6778,477){\makebox(0,0){\strut{}$1$}}%
      \put(7558,477){\makebox(0,0){\strut{}$10$}}%
      \put(8338,477){\makebox(0,0){\strut{}$100$}}%
      \put(9119,477){\makebox(0,0){\strut{}$1000$}}%
      \put(8207,147){\makebox(0,0){\strut{}$\unit[P/]{d}$}}%
      \put(9983,881){\makebox(0,0)[l]{\strut{}$0.5$}}%
      \put(9983,1255){\makebox(0,0)[l]{\strut{}$1$}}%
      \put(9983,1629){\makebox(0,0)[l]{\strut{}$2$}}%
      \put(9983,2003){\makebox(0,0)[l]{\strut{}$4$}}%
      \put(9983,2377){\makebox(0,0)[l]{\strut{}$8$}}%
      \put(9983,2752){\makebox(0,0)[l]{\strut{}$16$}}%
      \put(9983,3126){\makebox(0,0)[l]{\strut{}$32$}}%
      \put(10445,2020){\rotatebox{-270}{\makebox(0,0){\strut{}$\sigma$}}}%
    }%
    \gplbacktext
    \put(0,0){\includegraphics[scale=0.5]{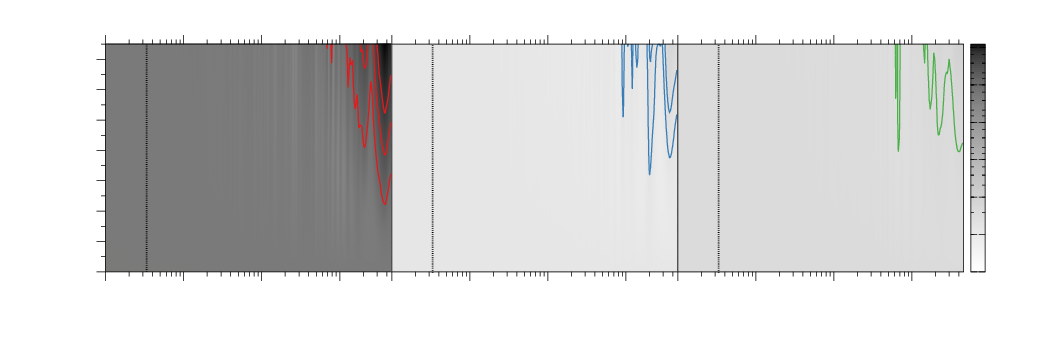}}%
    \gplfronttext{
        \tiny
        \color{cb-red}%
        \put(3790,1400){\makebox(0,0){\strut{}15}}%
        \put(3790,1920){\makebox(0,0){\strut{}20}}%
        \put(3790,2420){\makebox(0,0){\strut{}25}}%
        \color{cb-blue}%
        \put(6600,1920){\makebox(0,0){\strut{}1.1}}%
        \put(6650,2430){\makebox(0,0){\strut{}1.3}}%
        \color{cb-green}%
        \put(9080,2400){\makebox(0,0){\strut{}1.3}}%
        \put(9350,2065){\makebox(0,0){\strut{}1.5}}%
        \normalsize
        }%
  \end{picture}%
\endgroup


                \caption{QQ Vir. Contour lines for $ f_1 $ are placed at 15, 20, and $ \unit[25]{\sigma} $ (left panel), for $ f_2 $ at 1.1 and $ \unit[1.3]{\sigma} $ (middle panel), and for $ f_3 $ at 1.3 and $ \unit[1.5]{\sigma} $ (right panel), as indicated by their labels.}
                \label{fig:qqvir-rv-planet}
        \end{subfigure}
        \caption{Minimum companion mass as a function of orbital period. Greyscale shows the difference between the O-C measurements and artificial O-C data generated for a given combination of companion mass and orbit. We note that at this stage, the phase optimisation of the artificial data is done independently for each pulsation frequency. The median of gaps in between the epochs is indicated by a vertical dotted line. See text for more details.}
        \label{fig:rv-planet}
\end{figure*}
\begin{figure*}[tp]
        \centering
        \begin{subfigure}[b]{\textwidth}
                \centering
\begingroup
  \makeatletter
  \providecommand\color[2][]{%
    \GenericError{(gnuplot) \space\space\space\@spaces}{%
      Package color not loaded in conjunction with
      terminal option `colourtext'%
    }{See the gnuplot documentation for explanation.%
    }{Either use 'blacktext' in gnuplot or load the package
      color.sty in LaTeX.}%
    \renewcommand\color[2][]{}%
  }%
  \providecommand\includegraphics[2][]{%
    \GenericError{(gnuplot) \space\space\space\@spaces}{%
      Package graphicx or graphics not loaded%
    }{See the gnuplot documentation for explanation.%
    }{The gnuplot epslatex terminal needs graphicx.sty or graphics.sty.}
  }%
  \providecommand\rotatebox[2]{#2}%
  \@ifundefined{ifGPcolor}{%
    \newif\ifGPcolor
    \GPcolortrue
  }{}%
  \@ifundefined{ifGPblacktext}{%
    \newif\ifGPblacktext
    \GPblacktextfalse
  }{}%
  \let\gplgaddtomacro\g@addto@macro
  \gdef\gplbacktext{}%
  \gdef\gplfronttext{}%
  \makeatother
  \ifGPblacktext
    \def\colorrgb#1{}%
    \def\colorgray#1{}%
  \else
    \ifGPcolor
      \def\colorrgb#1{\color[rgb]{#1}}%
      \def\colorgray#1{\color[gray]{#1}}%
      \expandafter\def\csname LTw\endcsname{\color{white}}%
      \expandafter\def\csname LTb\endcsname{\color{black}}%
      \expandafter\def\csname LTa\endcsname{\color{black}}%
      \expandafter\def\csname LT0\endcsname{\color[rgb]{1,0,0}}%
      \expandafter\def\csname LT1\endcsname{\color[rgb]{0,1,0}}%
      \expandafter\def\csname LT2\endcsname{\color[rgb]{0,0,1}}%
      \expandafter\def\csname LT3\endcsname{\color[rgb]{1,0,1}}%
      \expandafter\def\csname LT4\endcsname{\color[rgb]{0,1,1}}%
      \expandafter\def\csname LT5\endcsname{\color[rgb]{1,1,0}}%
      \expandafter\def\csname LT6\endcsname{\color[rgb]{0,0,0}}%
      \expandafter\def\csname LT7\endcsname{\color[rgb]{1,0.3,0}}%
      \expandafter\def\csname LT8\endcsname{\color[rgb]{0.5,0.5,0.5}}%
    \else
      \def\colorrgb#1{\color{black}}%
      \def\colorgray#1{\color[gray]{#1}}%
      \expandafter\def\csname LTw\endcsname{\color{white}}%
      \expandafter\def\csname LTb\endcsname{\color{black}}%
      \expandafter\def\csname LTa\endcsname{\color{black}}%
      \expandafter\def\csname LT0\endcsname{\color{black}}%
      \expandafter\def\csname LT1\endcsname{\color{black}}%
      \expandafter\def\csname LT2\endcsname{\color{black}}%
      \expandafter\def\csname LT3\endcsname{\color{black}}%
      \expandafter\def\csname LT4\endcsname{\color{black}}%
      \expandafter\def\csname LT5\endcsname{\color{black}}%
      \expandafter\def\csname LT6\endcsname{\color{black}}%
      \expandafter\def\csname LT7\endcsname{\color{black}}%
      \expandafter\def\csname LT8\endcsname{\color{black}}%
    \fi
  \fi
    \setlength{\unitlength}{0.0500bp}%
    \ifx\gptboxheight\undefined%
      \newlength{\gptboxheight}%
      \newlength{\gptboxwidth}%
      \newsavebox{\gptboxtext}%
    \fi%
    \setlength{\fboxrule}{0.5pt}%
    \setlength{\fboxsep}{1pt}%
\begin{picture}(10430.00,3600.00)%
    \gplgaddtomacro\gplbacktext{%
    }%
    \gplgaddtomacro\gplfronttext{%
      \csname LTb\endcsname%
      \put(1056,477){\makebox(0,0){\strut{}$1$}}%
      \put(2247,477){\makebox(0,0){\strut{}$10$}}%
      \put(3438,477){\makebox(0,0){\strut{}$100$}}%
      \put(4629,477){\makebox(0,0){\strut{}$1000$}}%
      \put(3201,147){\makebox(0,0){\strut{}$\unit[P/]{d}$}}%
      \put(778,881){\makebox(0,0)[r]{\strut{}$0$}}%
      \put(778,1185){\makebox(0,0)[r]{\strut{}$2$}}%
      \put(778,1489){\makebox(0,0)[r]{\strut{}$4$}}%
      \put(778,1793){\makebox(0,0)[r]{\strut{}$6$}}%
      \put(778,2095){\makebox(0,0)[r]{\strut{}$8$}}%
      \put(778,2399){\makebox(0,0)[r]{\strut{}$10$}}%
      \put(778,2703){\makebox(0,0)[r]{\strut{}$12$}}%
      \put(778,3007){\makebox(0,0)[r]{\strut{}$14$}}%
      \put(350,2020){\rotatebox{-270}{\makebox(0,0){\strut{}$\unit[m \sin i/]{M_{\jupiter}}$}}}%
    }%
    \gplgaddtomacro\gplbacktext{%
    }%
    \gplgaddtomacro\gplfronttext{%
      \csname LTb\endcsname%
      \put(5347,477){\makebox(0,0){\strut{}$1$}}%
      \put(6538,477){\makebox(0,0){\strut{}$10$}}%
      \put(7729,477){\makebox(0,0){\strut{}$100$}}%
      \put(8920,477){\makebox(0,0){\strut{}$1000$}}%
      \put(7492,147){\makebox(0,0){\strut{}$\unit[P/]{d}$}}%
      \put(10091,881){\makebox(0,0)[l]{\strut{}$1$}}%
      \put(10091,1496){\makebox(0,0)[l]{\strut{}$2$}}%
      \put(10091,2112){\makebox(0,0)[l]{\strut{}$4$}}%
      \put(10091,2727){\makebox(0,0)[l]{\strut{}$8$}}%
      \put(10289,2020){\rotatebox{-270}{\makebox(0,0){\strut{}$\sigma$}}}%
    }%
    \gplbacktext
    \put(0,0){\includegraphics[scale=0.5]{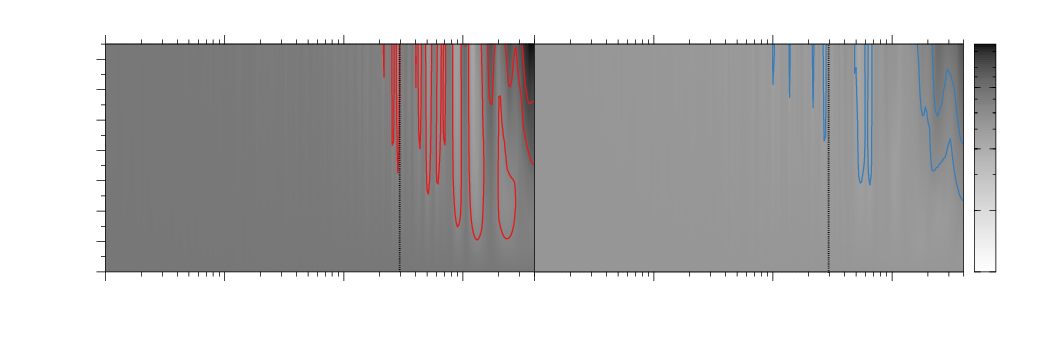}}%
    \gplfronttext{
        \tiny
        \color{cb-red}%
        \put(4020,2200){\makebox(0,0){\strut{}6}}%
        \put(4700,1060){\makebox(0,0){\strut{}6}}%
        \put(5000,1050){\makebox(0,0){\strut{}6}}%
        \put(4960,2820){\makebox(0,0){\strut{}8}}%
        \put(5225,2440){\makebox(0,0){\strut{}10}}%
        \color{black}
        \put(9380,2380){\makebox(0,0){\strut{}6.5}}%
        \put(9245,1770){\makebox(0,0){\strut{}5.5}}%
        \put(8600,1600){\makebox(0,0){\strut{}4.5}}%
        \put(8050,2400){\makebox(0,0){\strut{}4.5}}%
        \color{cb-blue}%
        \normalsize
    }%
  \end{picture}%
\endgroup


                \caption{V541 Hya. Contour lines for $ f_1 $ are placed at 6, 8, and $ \unit[10]{\sigma} $ (left panel), and for $ f_2 $ at 4.5, 5.5, and $ \unit[6.5]{\sigma} $ (right panel), as indicated by their labels.}
                \label{fig:v541hya-rv-planet}
        \end{subfigure}
        \caption{Continuation of Fig.~\ref{fig:rv-planet}.}
        \label{fig:rv-planet2}
\end{figure*}

\section{Summary and conclusion}

In this work, we present ground-based multi-site observations for the four sdBs, DW Lyn, V1636 Ori, QQ Vir, and V541 Hya. We investigated variations in the arrival times of their dominant stellar pulsation modes to draw conclusions about secular period drifts and possible sub-stellar companions. All light curves are analysed homogeneously. 

From the $ O-C $ measurements, we derive an evolutionary timescale from the change in period $ \dot{P} $. Comparing to model calculations from \citet{charpinet_adiabatic_2002}, we infer the evolutionary phase of the target. Although some $ \dot{P} $ measurements are influenced by mode splitting, we can tell from the sign of $ \dot{P_1} $ of DW~Lyn that the star is likely still in the stage of central \element{He} burning. We can draw a similar conclusion from the sign of $ \dot{P} $ of QQ~Vir. The $ \dot{P} $ measurements of V1636~Ori are likely affected by mode splitting, making it difficult to interpret the results in the context of stellar evolution. V541~Hya shows $ \dot{P} $ measurements close to zero, which indicates the star being at the transition phase between \element{He} burning and contraction due to the depletion of \element{He} in the core.
        
Comparing the atmospheric properties from Table~\ref{tab:stellar} with the evolutionary tracks for different models from Fig.~1 in \citet{charpinet_adiabatic_2002}, we can confirm the hypothesis that DW~Lyn and QQ~Vir are in their \element{He} burning phase. V541~Hya agrees within $ \unit[2]{\sigma} $ of the $ \log g $ measurement with one model at the turning point between the two evolutionary stages.
        
However, we can not exclude frequency and amplitude variations on smaller timescales than resolvable by our data set. Using temporally higher resolved \textit{Kepler}-data of KIC 3527751, \citet{zong_oscillation_2018} cautioned about long-term frequency or phase evolutions ascribing to non-linear amplitude and frequency modulations in pulsating sdBs. We see such effects already in our data set, even with a low temporal resolution compared to the \textit{Kepler} sampling with a duty cycle of more than 90 per cent.

Observations on DW Lyn and V1636 Ori were published by \citet{lutz_light_2008,schuh_exotime_2010,lutz_search_2011,lutz_exotime_2011}. Our analysis of these observations, including extended data sets, do not confirm the tentative companion periods of 80 and 160 days, respectively. These signals more likely arise due to mode beating indicated by partly unresolved frequency multiplets and amplitude modulations. 

Almost all analysed pulsation modes show formal significant changes in arrival times, but the amplitudes of these periodic signals do not correlate with frequencies, excluding the light-travel time effect due to orbital reflex motions for such variations and thus giving upper limits on companion masses. Only DW~Lyn might have a planetary companion on a long orbital period, as indicated by one arrival time measurement. But this can not be confirmed with a second measurement, due to larger uncertainties. Additionally, more studies question the presence of already proposed companions, for example, \citet{krzesinski_planetary_2015,hutchens_new_2017}. Our unique sample of long-term observations shows a complex behaviour of mode- and amplitude interactions in sdBs which should be addressed in further studies. Until this has been addressed, caution is advised when interpreting $ O-C $ pulse arrival times in terms of companions.



\begin{acknowledgements}
        We thank Wen-Shan Hsiao for observing at the Lulin Observatory and Mike D. Reed for observing at the Baker Observatory and Elia Leibowitz for observing at the WISE observatory.
        F.M. conducted the work in this paper in the framework of the International Max-Planck Research School (IMPRS) for Solar System Science at the University of G\"ottingen (Volkswagen Foundation project grant number VWZN3020).
        DK thanks the SAAO for generous allocations of telescope time and the National Research Foundation of South Africa and the University of the Western Cape for financial support.
        TDO gratefully acknowledges support from the U.S. National Science Foundation grant AST-0807919.
        L.M.\ was supported by the Premium Postdoctoral Research Program of the Hungarian Academy of Sciences. This project has been supported by the Lend\"ulet Program  of the Hungarian Academy of Sciences, project No. LP2018-7/2019.
        Based on observations made with the Italian Telescopio Nazionale Galileo (TNG) operated on the island of La Palma by the Fundación Galileo Galilei of the INAF (Istituto Nazionale di Astrofisica) at the Spanish Observatorio del Roque de los Muchachos of the Instituto de Astrofisica de Canarias.
        Based on observations collected at the Centro Astronómico Hisp\'{a}nico en Andaluc\'{i}a (CAHA) at Calar Alto, operated jointly by the Andalusian Universities and the Instituto de Astrof\'{i}sica de Andaluc\'{i}a (CSIC).
        Based on observations collected at Copernico telescope (Asiago, Italy) of the INAF - Osservatorio Astronomico di Padova.
        Based on observations made with the Nordic Optical Telescope, operated by the Nordic Optical Telescope Scientific Association at the Observatorio del Roque de los Muchachos, La Palma, Spain, of the Instituto de Astrofisica de Canarias.
        This paper includes data collected by the TESS mission. Funding for the TESS mission is provided by the NASA Explorer Program.
\end{acknowledgements}



\begin{appendix}

\section{TESS data}
\begin{figure*}[ht!]
        \centering
        \begin{subfigure}[b]{0.49\textwidth}
                \centering
\begingroup
  \makeatletter
  \providecommand\color[2][]{%
    \GenericError{(gnuplot) \space\space\space\@spaces}{%
      Package color not loaded in conjunction with
      terminal option `colourtext'%
    }{See the gnuplot documentation for explanation.%
    }{Either use 'blacktext' in gnuplot or load the package
      color.sty in LaTeX.}%
    \renewcommand\color[2][]{}%
  }%
  \providecommand\includegraphics[2][]{%
    \GenericError{(gnuplot) \space\space\space\@spaces}{%
      Package graphicx or graphics not loaded%
    }{See the gnuplot documentation for explanation.%
    }{The gnuplot epslatex terminal needs graphicx.sty or graphics.sty.}
  }%
  \providecommand\rotatebox[2]{#2}%
  \@ifundefined{ifGPcolor}{%
    \newif\ifGPcolor
    \GPcolortrue
  }{}%
  \@ifundefined{ifGPblacktext}{%
    \newif\ifGPblacktext
    \GPblacktexttrue
  }{}%
  \let\gplgaddtomacro\g@addto@macro
  \gdef\gplbacktext{}%
  \gdef\gplfronttext{}%
  \makeatother
  \ifGPblacktext
    \def\colorrgb#1{}%
    \def\colorgray#1{}%
  \else
    \ifGPcolor
      \def\colorrgb#1{\color[rgb]{#1}}%
      \def\colorgray#1{\color[gray]{#1}}%
      \expandafter\def\csname LTw\endcsname{\color{white}}%
      \expandafter\def\csname LTb\endcsname{\color{black}}%
      \expandafter\def\csname LTa\endcsname{\color{black}}%
      \expandafter\def\csname LT0\endcsname{\color[rgb]{1,0,0}}%
      \expandafter\def\csname LT1\endcsname{\color[rgb]{0,1,0}}%
      \expandafter\def\csname LT2\endcsname{\color[rgb]{0,0,1}}%
      \expandafter\def\csname LT3\endcsname{\color[rgb]{1,0,1}}%
      \expandafter\def\csname LT4\endcsname{\color[rgb]{0,1,1}}%
      \expandafter\def\csname LT5\endcsname{\color[rgb]{1,1,0}}%
      \expandafter\def\csname LT6\endcsname{\color[rgb]{0,0,0}}%
      \expandafter\def\csname LT7\endcsname{\color[rgb]{1,0.3,0}}%
      \expandafter\def\csname LT8\endcsname{\color[rgb]{0.5,0.5,0.5}}%
    \else
      \def\colorrgb#1{\color{black}}%
      \def\colorgray#1{\color[gray]{#1}}%
      \expandafter\def\csname LTw\endcsname{\color{white}}%
      \expandafter\def\csname LTb\endcsname{\color{black}}%
      \expandafter\def\csname LTa\endcsname{\color{black}}%
      \expandafter\def\csname LT0\endcsname{\color{black}}%
      \expandafter\def\csname LT1\endcsname{\color{black}}%
      \expandafter\def\csname LT2\endcsname{\color{black}}%
      \expandafter\def\csname LT3\endcsname{\color{black}}%
      \expandafter\def\csname LT4\endcsname{\color{black}}%
      \expandafter\def\csname LT5\endcsname{\color{black}}%
      \expandafter\def\csname LT6\endcsname{\color{black}}%
      \expandafter\def\csname LT7\endcsname{\color{black}}%
      \expandafter\def\csname LT8\endcsname{\color{black}}%
    \fi
  \fi
    \setlength{\unitlength}{0.0500bp}%
    \ifx\gptboxheight\undefined%
      \newlength{\gptboxheight}%
      \newlength{\gptboxwidth}%
      \newsavebox{\gptboxtext}%
    \fi%
    \setlength{\fboxrule}{0.5pt}%
    \setlength{\fboxsep}{1pt}%
\begin{picture}(5102.00,3152.00)%
    \gplgaddtomacro\gplbacktext{%
      \csname LTb\endcsname%
      \put(682,901){\makebox(0,0)[r]{\strut{}0.7}}%
      \put(682,1170){\makebox(0,0)[r]{\strut{}0.8}}%
      \put(682,1439){\makebox(0,0)[r]{\strut{}0.9}}%
      \put(682,1708){\makebox(0,0)[r]{\strut{}1.0}}%
      \put(682,1976){\makebox(0,0)[r]{\strut{}1.1}}%
      \put(682,2245){\makebox(0,0)[r]{\strut{}1.2}}%
      \put(682,2514){\makebox(0,0)[r]{\strut{}1.3}}%
      \put(1515,484){\makebox(0,0){\strut{}$58440$}}%
      \put(2791,484){\makebox(0,0){\strut{}$58450$}}%
      \put(4067,484){\makebox(0,0){\strut{}$58460$}}%
    }%
    \gplgaddtomacro\gplfronttext{%
      \csname LTb\endcsname%
      \put(176,1707){\rotatebox{-270}{\makebox(0,0){\strut{}relative flux $F$}}}%
      \put(2791,154){\makebox(0,0){\strut{}$\unit[t/]{MBJD_{TDB}}$}}%
    }%
    \gplbacktext
    \put(0,0){\includegraphics[scale=0.5]{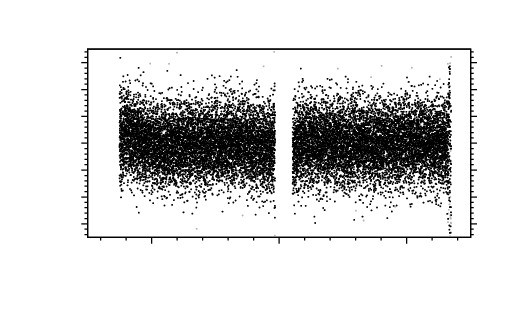}}%
    \gplfronttext
  \end{picture}%
\endgroup


                \caption{V1636 Ori.}
                \label{fig:v1636ori-lc-tess}
        \end{subfigure}
        \hfil
        \begin{subfigure}[b]{0.49\textwidth}
                \centering
\begingroup
  \makeatletter
  \providecommand\color[2][]{%
    \GenericError{(gnuplot) \space\space\space\@spaces}{%
      Package color not loaded in conjunction with
      terminal option `colourtext'%
    }{See the gnuplot documentation for explanation.%
    }{Either use 'blacktext' in gnuplot or load the package
      color.sty in LaTeX.}%
    \renewcommand\color[2][]{}%
  }%
  \providecommand\includegraphics[2][]{%
    \GenericError{(gnuplot) \space\space\space\@spaces}{%
      Package graphicx or graphics not loaded%
    }{See the gnuplot documentation for explanation.%
    }{The gnuplot epslatex terminal needs graphicx.sty or graphics.sty.}
  }%
  \providecommand\rotatebox[2]{#2}%
  \@ifundefined{ifGPcolor}{%
    \newif\ifGPcolor
    \GPcolortrue
  }{}%
  \@ifundefined{ifGPblacktext}{%
    \newif\ifGPblacktext
    \GPblacktexttrue
  }{}%
  \let\gplgaddtomacro\g@addto@macro
  \gdef\gplbacktext{}%
  \gdef\gplfronttext{}%
  \makeatother
  \ifGPblacktext
    \def\colorrgb#1{}%
    \def\colorgray#1{}%
  \else
    \ifGPcolor
      \def\colorrgb#1{\color[rgb]{#1}}%
      \def\colorgray#1{\color[gray]{#1}}%
      \expandafter\def\csname LTw\endcsname{\color{white}}%
      \expandafter\def\csname LTb\endcsname{\color{black}}%
      \expandafter\def\csname LTa\endcsname{\color{black}}%
      \expandafter\def\csname LT0\endcsname{\color[rgb]{1,0,0}}%
      \expandafter\def\csname LT1\endcsname{\color[rgb]{0,1,0}}%
      \expandafter\def\csname LT2\endcsname{\color[rgb]{0,0,1}}%
      \expandafter\def\csname LT3\endcsname{\color[rgb]{1,0,1}}%
      \expandafter\def\csname LT4\endcsname{\color[rgb]{0,1,1}}%
      \expandafter\def\csname LT5\endcsname{\color[rgb]{1,1,0}}%
      \expandafter\def\csname LT6\endcsname{\color[rgb]{0,0,0}}%
      \expandafter\def\csname LT7\endcsname{\color[rgb]{1,0.3,0}}%
      \expandafter\def\csname LT8\endcsname{\color[rgb]{0.5,0.5,0.5}}%
    \else
      \def\colorrgb#1{\color{black}}%
      \def\colorgray#1{\color[gray]{#1}}%
      \expandafter\def\csname LTw\endcsname{\color{white}}%
      \expandafter\def\csname LTb\endcsname{\color{black}}%
      \expandafter\def\csname LTa\endcsname{\color{black}}%
      \expandafter\def\csname LT0\endcsname{\color{black}}%
      \expandafter\def\csname LT1\endcsname{\color{black}}%
      \expandafter\def\csname LT2\endcsname{\color{black}}%
      \expandafter\def\csname LT3\endcsname{\color{black}}%
      \expandafter\def\csname LT4\endcsname{\color{black}}%
      \expandafter\def\csname LT5\endcsname{\color{black}}%
      \expandafter\def\csname LT6\endcsname{\color{black}}%
      \expandafter\def\csname LT7\endcsname{\color{black}}%
      \expandafter\def\csname LT8\endcsname{\color{black}}%
    \fi
  \fi
    \setlength{\unitlength}{0.0500bp}%
    \ifx\gptboxheight\undefined%
      \newlength{\gptboxheight}%
      \newlength{\gptboxwidth}%
      \newsavebox{\gptboxtext}%
    \fi%
    \setlength{\fboxrule}{0.5pt}%
    \setlength{\fboxsep}{1pt}%
\begin{picture}(5102.00,3152.00)%
    \gplgaddtomacro\gplbacktext{%
      \csname LTb\endcsname%
      \put(682,767){\makebox(0,0)[r]{\strut{}0.7}}%
      \put(682,1081){\makebox(0,0)[r]{\strut{}0.8}}%
      \put(682,1394){\makebox(0,0)[r]{\strut{}0.9}}%
      \put(682,1708){\makebox(0,0)[r]{\strut{}1.0}}%
      \put(682,2021){\makebox(0,0)[r]{\strut{}1.1}}%
      \put(682,2335){\makebox(0,0)[r]{\strut{}1.2}}%
      \put(682,2648){\makebox(0,0)[r]{\strut{}1.3}}%
      \put(1515,484){\makebox(0,0){\strut{}$58520$}}%
      \put(2791,484){\makebox(0,0){\strut{}$58530$}}%
      \put(4067,484){\makebox(0,0){\strut{}$58540$}}%
    }%
    \gplgaddtomacro\gplfronttext{%
      \csname LTb\endcsname%
      \put(176,1707){\rotatebox{-270}{\makebox(0,0){\strut{}relative flux $F$}}}%
      \put(2791,154){\makebox(0,0){\strut{}$\unit[t/]{MBJD_{TDB}}$}}%
    }%
    \gplbacktext
    \put(0,0){\includegraphics[scale=0.5]{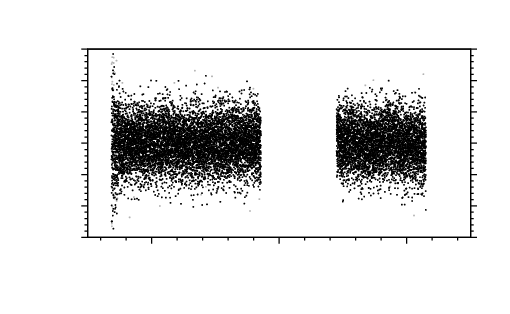}}%
    \gplfronttext
  \end{picture}%
\endgroup


                \caption{V541 Hya.}
                \label{fig:v541hya-lc-tess}
        \end{subfigure}
        \caption{Light curves of the TESS observations. Grey points are considered outliers and partially exceeding the plotting range.}
        \label{fig:lc-tess}
\end{figure*}
\begin{figure*}[ht!]
        \centering
\begingroup
\definecolor{r}{HTML}{e41a1c}
  \makeatletter
  \providecommand\color[2][]{%
    \GenericError{(gnuplot) \space\space\space\@spaces}{%
      Package color not loaded in conjunction with
      terminal option `colourtext'%
    }{See the gnuplot documentation for explanation.%
    }{Either use 'blacktext' in gnuplot or load the package
      color.sty in LaTeX.}%
    \renewcommand\color[2][]{}%
  }%
  \providecommand\includegraphics[2][]{%
    \GenericError{(gnuplot) \space\space\space\@spaces}{%
      Package graphicx or graphics not loaded%
    }{See the gnuplot documentation for explanation.%
    }{The gnuplot epslatex terminal needs graphicx.sty or graphics.sty.}
  }%
  \providecommand\rotatebox[2]{#2}%
  \@ifundefined{ifGPcolor}{%
    \newif\ifGPcolor
    \GPcolortrue
  }{}%
  \@ifundefined{ifGPblacktext}{%
    \newif\ifGPblacktext
    \GPblacktexttrue
  }{}%
  \let\gplgaddtomacro\g@addto@macro
  \gdef\gplbacktext{}%
  \gdef\gplfronttext{}%
  \makeatother
  \ifGPblacktext
    \def\colorrgb#1{}%
    \def\colorgray#1{}%
  \else
    \ifGPcolor
      \def\colorrgb#1{\color[rgb]{#1}}%
      \def\colorgray#1{\color[gray]{#1}}%
      \expandafter\def\csname LTw\endcsname{\color{white}}%
      \expandafter\def\csname LTb\endcsname{\color{black}}%
      \expandafter\def\csname LTa\endcsname{\color{black}}%
      \expandafter\def\csname LT0\endcsname{\color[rgb]{1,0,0}}%
      \expandafter\def\csname LT1\endcsname{\color[rgb]{0,1,0}}%
      \expandafter\def\csname LT2\endcsname{\color[rgb]{0,0,1}}%
      \expandafter\def\csname LT3\endcsname{\color[rgb]{1,0,1}}%
      \expandafter\def\csname LT4\endcsname{\color[rgb]{0,1,1}}%
      \expandafter\def\csname LT5\endcsname{\color[rgb]{1,1,0}}%
      \expandafter\def\csname LT6\endcsname{\color[rgb]{0,0,0}}%
      \expandafter\def\csname LT7\endcsname{\color[rgb]{1,0.3,0}}%
      \expandafter\def\csname LT8\endcsname{\color[rgb]{0.5,0.5,0.5}}%
    \else
      \def\colorrgb#1{\color{black}}%
      \def\colorgray#1{\color[gray]{#1}}%
      \expandafter\def\csname LTw\endcsname{\color{white}}%
      \expandafter\def\csname LTb\endcsname{\color{black}}%
      \expandafter\def\csname LTa\endcsname{\color{black}}%
      \expandafter\def\csname LT0\endcsname{\color{black}}%
      \expandafter\def\csname LT1\endcsname{\color{black}}%
      \expandafter\def\csname LT2\endcsname{\color{black}}%
      \expandafter\def\csname LT3\endcsname{\color{black}}%
      \expandafter\def\csname LT4\endcsname{\color{black}}%
      \expandafter\def\csname LT5\endcsname{\color{black}}%
      \expandafter\def\csname LT6\endcsname{\color{black}}%
      \expandafter\def\csname LT7\endcsname{\color{black}}%
      \expandafter\def\csname LT8\endcsname{\color{black}}%
    \fi
  \fi
    \setlength{\unitlength}{0.0500bp}%
    \ifx\gptboxheight\undefined%
      \newlength{\gptboxheight}%
      \newlength{\gptboxwidth}%
      \newsavebox{\gptboxtext}%
    \fi%
    \setlength{\fboxrule}{0.5pt}%
    \setlength{\fboxsep}{1pt}%
\begin{picture}(10420.00,6440.00)%
    \gplgaddtomacro\gplbacktext{%
      \csname LTb\endcsname%
      \put(742,3220){\makebox(0,0)[r]{\strut{}$0$}}%
      \csname LTb\endcsname%
      \put(742,3664){\makebox(0,0)[r]{\strut{}$0.1$}}%
      \csname LTb\endcsname%
      \put(742,4107){\makebox(0,0)[r]{\strut{}$0.2$}}%
      \csname LTb\endcsname%
      \put(742,4551){\makebox(0,0)[r]{\strut{}$0.3$}}%
      \csname LTb\endcsname%
      \put(742,4994){\makebox(0,0)[r]{\strut{}$0.4$}}%
      \csname LTb\endcsname%
      \put(742,5438){\makebox(0,0)[r]{\strut{}$0.5$}}%
      \csname LTb\endcsname%
      \put(742,5881){\makebox(0,0)[r]{\strut{}$0.6$}}%
      \csname LTb\endcsname%
      \put(918,6141){\makebox(0,0){\strut{}$0$}}%
      \csname LTb\endcsname%
      \put(1387,6141){\makebox(0,0){\strut{}$500$}}%
      \csname LTb\endcsname%
      \put(1856,6141){\makebox(0,0){\strut{}$1000$}}%
      \csname LTb\endcsname%
      \put(2325,6141){\makebox(0,0){\strut{}$1500$}}%
      \csname LTb\endcsname%
      \put(2794,6141){\makebox(0,0){\strut{}$2000$}}%
      \csname LTb\endcsname%
      \put(3263,6141){\makebox(0,0){\strut{}$2500$}}%
      \csname LTb\endcsname%
      \put(3732,6141){\makebox(0,0){\strut{}$3000$}}%
      \csname LTb\endcsname%
      \put(4201,6141){\makebox(0,0){\strut{}$3500$}}%
      \csname LTb\endcsname%
      \put(4670,6141){\makebox(0,0){\strut{}$4000$}}%
      \csname LTb\endcsname%
      \put(5139,6141){\makebox(0,0){\strut{}$4500$}}%
      \csname LTb\endcsname%
      \put(5608,6141){\makebox(0,0){\strut{}$5000$}}%
      \csname LTb\endcsname%
      \put(6077,6141){\makebox(0,0){\strut{}$5500$}}%
      \csname LTb\endcsname%
      \put(6546,6141){\makebox(0,0){\strut{}$6000$}}%
      \csname LTb\endcsname%
      \put(7015,6141){\makebox(0,0){\strut{}$6500$}}%
      \csname LTb\endcsname%
      \put(7484,6141){\makebox(0,0){\strut{}$7000$}}%
      \csname LTb\endcsname%
      \put(7953,6141){\makebox(0,0){\strut{}$7500$}}%
      \csname LTb\endcsname%
      \put(8422,6141){\makebox(0,0){\strut{}$8000$}}%
      \csname LTb\endcsname%
      \put(8891,6141){\makebox(0,0){\strut{}$8500$}}%
      \csname LTb\endcsname%
      \put(9360,6141){\makebox(0,0){\strut{}$9000$}}%
    }%
    \gplgaddtomacro\gplfronttext{%
      \csname LTb\endcsname%
      \put(241,4550){\rotatebox{-270}{\makebox(0,0){\strut{}$\unit[A_{\text{obs}}/]{\%}$}}}%
      \csname LTb\endcsname%
      \put(5260,6419){\makebox(0,0){\strut{}$\unit[f /]{\mu Hz}$}}%
      \csname LTb\endcsname%
      \put(1938,5714){\makebox(0,0)[r]{\strut{}V1636 Ori}}%
    }%
    \gplgaddtomacro\gplbacktext{%
      \csname LTb\endcsname%
      \put(742,3219){\makebox(0,0)[r]{\strut{}}}%
      \csname LTb\endcsname%
      \put(742,558){\makebox(0,0)[r]{\strut{}$0$}}%
      \csname LTb\endcsname%
      \put(742,1002){\makebox(0,0)[r]{\strut{}$0.1$}}%
      \csname LTb\endcsname%
      \put(742,1445){\makebox(0,0)[r]{\strut{}$0.2$}}%
      \csname LTb\endcsname%
      \put(742,1889){\makebox(0,0)[r]{\strut{}$0.3$}}%
      \csname LTb\endcsname%
      \put(742,2332){\makebox(0,0)[r]{\strut{}$0.4$}}%
      \csname LTb\endcsname%
      \put(742,2776){\makebox(0,0)[r]{\strut{}$0.5$}}%
      \csname LTb\endcsname%
      \put(918,298){\makebox(0,0){\strut{}$0$}}%
      \csname LTb\endcsname%
      \put(1461,298){\makebox(0,0){\strut{}$50$}}%
      \csname LTb\endcsname%
      \put(2004,298){\makebox(0,0){\strut{}$100$}}%
      \csname LTb\endcsname%
      \put(2546,298){\makebox(0,0){\strut{}$150$}}%
      \csname LTb\endcsname%
      \put(3089,298){\makebox(0,0){\strut{}$200$}}%
      \csname LTb\endcsname%
      \put(3632,298){\makebox(0,0){\strut{}$250$}}%
      \csname LTb\endcsname%
      \put(4175,298){\makebox(0,0){\strut{}$300$}}%
      \csname LTb\endcsname%
      \put(4718,298){\makebox(0,0){\strut{}$350$}}%
      \csname LTb\endcsname%
      \put(5261,298){\makebox(0,0){\strut{}$400$}}%
      \csname LTb\endcsname%
      \put(5803,298){\makebox(0,0){\strut{}$450$}}%
      \csname LTb\endcsname%
      \put(6346,298){\makebox(0,0){\strut{}$500$}}%
      \csname LTb\endcsname%
      \put(6889,298){\makebox(0,0){\strut{}$550$}}%
      \csname LTb\endcsname%
      \put(7432,298){\makebox(0,0){\strut{}$600$}}%
      \csname LTb\endcsname%
      \put(7975,298){\makebox(0,0){\strut{}$650$}}%
      \csname LTb\endcsname%
      \put(8517,298){\makebox(0,0){\strut{}$700$}}%
      \csname LTb\endcsname%
      \put(9060,298){\makebox(0,0){\strut{}$750$}}%
      \csname LTb\endcsname%
      \put(9603,298){\makebox(0,0){\strut{}$800$}}%
    }%
    \gplgaddtomacro\gplfronttext{%
      \csname LTb\endcsname%
      \put(241,1888){\rotatebox{-270}{\makebox(0,0){\strut{}$\unit[A_{\text{obs}}/]{\%}$}}}%
      \csname LTb\endcsname%
      \put(5260,19){\makebox(0,0){\strut{}$\unit[f /]{d^{-1}}$}}%
      \csname LTb\endcsname%
      \put(1836,3052){\makebox(0,0)[r]{\strut{}V541 Hya}}%
    }%
    \gplbacktext
    \put(0,0){\includegraphics[scale=0.5]{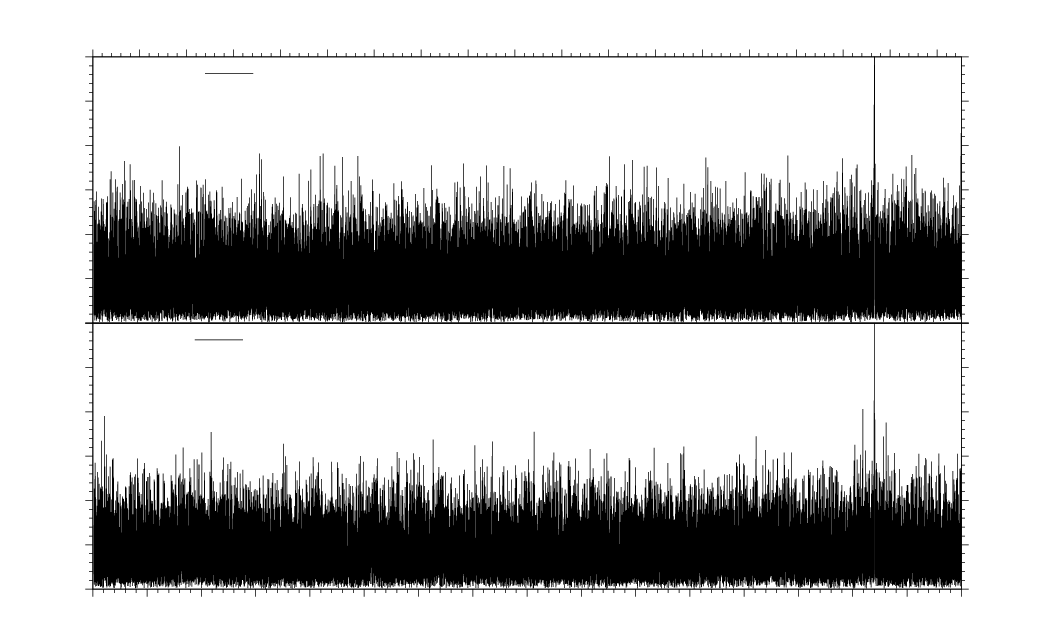}}%
    \gplfronttext
  \end{picture}%
\endgroup


        \caption{Amplitude $ A $ spectrum of the TESS observations. The upper panel shows the spectrum of V1636 Ori, the lower panel the spectrum of V541 Hya. The only peak above the noise level is the $ \unit[120]{s} $ alias due to the cadence of the observations.}
        \label{fig:tess-power}
\end{figure*}

\section{Amplitude spectra}

\begin{table}[h]
        \centering
        \caption{Additional pulsation modes identified for our targets not used in the $ O-C $ analysis due to their low S/N.}
        \label{tab:add-freq}
        \begin{tabular}{lr@{.}lr@{.}l}
                \toprule
                Target & \multicolumn{2}{c}{$ \unit[f/]{d^{-1}} $} & \multicolumn{2}{c}{$ \unit[A/]{\%} $}  \\
                \midrule
                DW Lyn & 475 & 8231(2) & 0 & 09(18) \\
                 & 319 & 4042(3) & 0 & 06(12) \\
                 & 463 & 0100(6) & 0 & 03(18) \\
                V1636 Ori & 566 & 24031(3) & 0 & 6(3) \\
                QQ Vir & 733 & 0704(1) & 0 & 3(1) \\
                 & 664 & 4886(1) & 0 & 2(1) \\
                 & 572 & 73611(5) & 0 & 19(9) \\
                 & 664 & 7122(1) & 0 & 1(1) \\
                 & 434 & 1522(6) & 0 & 01(7) \\
                 & 502 & 410(2) & 0 & 01(9) \\
                V541 Hya  & 531 & 16759(16) & 0 & 03(7) \\
                 & 603 & 88741(6) & 0 & 03(8) \\
                \bottomrule
        \end{tabular}
\end{table}
\begin{figure*}[h]
        \centering
\begingroup
\definecolor{r}{HTML}{e41a1c}
  \makeatletter
  \providecommand\color[2][]{%
    \GenericError{(gnuplot) \space\space\space\@spaces}{%
      Package color not loaded in conjunction with
      terminal option `colourtext'%
    }{See the gnuplot documentation for explanation.%
    }{Either use 'blacktext' in gnuplot or load the package
      color.sty in LaTeX.}%
    \renewcommand\color[2][]{}%
  }%
  \providecommand\includegraphics[2][]{%
    \GenericError{(gnuplot) \space\space\space\@spaces}{%
      Package graphicx or graphics not loaded%
    }{See the gnuplot documentation for explanation.%
    }{The gnuplot epslatex terminal needs graphicx.sty or graphics.sty.}
  }%
  \providecommand\rotatebox[2]{#2}%
  \@ifundefined{ifGPcolor}{%
    \newif\ifGPcolor
    \GPcolortrue
  }{}%
  \@ifundefined{ifGPblacktext}{%
    \newif\ifGPblacktext
    \GPblacktexttrue
  }{}%
  \let\gplgaddtomacro\g@addto@macro
  \gdef\gplbacktext{}%
  \gdef\gplfronttext{}%
  \makeatother
  \ifGPblacktext
    \def\colorrgb#1{}%
    \def\colorgray#1{}%
  \else
    \ifGPcolor
      \def\colorrgb#1{\color[rgb]{#1}}%
      \def\colorgray#1{\color[gray]{#1}}%
      \expandafter\def\csname LTw\endcsname{\color{white}}%
      \expandafter\def\csname LTb\endcsname{\color{black}}%
      \expandafter\def\csname LTa\endcsname{\color{black}}%
      \expandafter\def\csname LT0\endcsname{\color[rgb]{1,0,0}}%
      \expandafter\def\csname LT1\endcsname{\color[rgb]{0,1,0}}%
      \expandafter\def\csname LT2\endcsname{\color[rgb]{0,0,1}}%
      \expandafter\def\csname LT3\endcsname{\color[rgb]{1,0,1}}%
      \expandafter\def\csname LT4\endcsname{\color[rgb]{0,1,1}}%
      \expandafter\def\csname LT5\endcsname{\color[rgb]{1,1,0}}%
      \expandafter\def\csname LT6\endcsname{\color[rgb]{0,0,0}}%
      \expandafter\def\csname LT7\endcsname{\color[rgb]{1,0.3,0}}%
      \expandafter\def\csname LT8\endcsname{\color[rgb]{0.5,0.5,0.5}}%
    \else
      \def\colorrgb#1{\color{black}}%
      \def\colorgray#1{\color[gray]{#1}}%
      \expandafter\def\csname LTw\endcsname{\color{white}}%
      \expandafter\def\csname LTb\endcsname{\color{black}}%
      \expandafter\def\csname LTa\endcsname{\color{black}}%
      \expandafter\def\csname LT0\endcsname{\color{black}}%
      \expandafter\def\csname LT1\endcsname{\color{black}}%
      \expandafter\def\csname LT2\endcsname{\color{black}}%
      \expandafter\def\csname LT3\endcsname{\color{black}}%
      \expandafter\def\csname LT4\endcsname{\color{black}}%
      \expandafter\def\csname LT5\endcsname{\color{black}}%
      \expandafter\def\csname LT6\endcsname{\color{black}}%
      \expandafter\def\csname LT7\endcsname{\color{black}}%
      \expandafter\def\csname LT8\endcsname{\color{black}}%
    \fi
  \fi
    \setlength{\unitlength}{0.0500bp}%
    \ifx\gptboxheight\undefined%
      \newlength{\gptboxheight}%
      \newlength{\gptboxwidth}%
      \newsavebox{\gptboxtext}%
    \fi%
    \setlength{\fboxrule}{0.5pt}%
    \setlength{\fboxsep}{1pt}%
\begin{picture}(10420.00,6440.00)%
    \gplgaddtomacro\gplbacktext{%
      \csname LTb\endcsname%
      \put(640,3220){\makebox(0,0)[r]{\strut{}$0$}}%
      \csname LTb\endcsname%
      \put(640,3752){\makebox(0,0)[r]{\strut{}$0.5$}}%
      \csname LTb\endcsname%
      \put(640,4284){\makebox(0,0)[r]{\strut{}$1$}}%
      \csname LTb\endcsname%
      \put(640,4817){\makebox(0,0)[r]{\strut{}$1.5$}}%
      \csname LTb\endcsname%
      \put(640,5349){\makebox(0,0)[r]{\strut{}$2$}}%
      \csname LTb\endcsname%
      \put(640,5881){\makebox(0,0)[r]{\strut{}$2.5$}}%
      \csname LTb\endcsname%
      \put(816,6141){\makebox(0,0){\strut{}$0$}}%
      \csname LTb\endcsname%
      \put(1575,6141){\makebox(0,0){\strut{}$500$}}%
      \csname LTb\endcsname%
      \put(2334,6141){\makebox(0,0){\strut{}$1000$}}%
      \csname LTb\endcsname%
      \put(3094,6141){\makebox(0,0){\strut{}$1500$}}%
      \csname LTb\endcsname%
      \put(3853,6141){\makebox(0,0){\strut{}$2000$}}%
      \csname LTb\endcsname%
      \put(4612,6141){\makebox(0,0){\strut{}$2500$}}%
      \csname LTb\endcsname%
      \put(5371,6141){\makebox(0,0){\strut{}$3000$}}%
      \csname LTb\endcsname%
      \put(6130,6141){\makebox(0,0){\strut{}$3500$}}%
      \csname LTb\endcsname%
      \put(6890,6141){\makebox(0,0){\strut{}$4000$}}%
      \csname LTb\endcsname%
      \put(7649,6141){\makebox(0,0){\strut{}$4500$}}%
      \csname LTb\endcsname%
      \put(8408,6141){\makebox(0,0){\strut{}$5000$}}%
      \csname LTb\endcsname%
      \put(9167,6141){\makebox(0,0){\strut{}$5500$}}%
    }%
    \gplgaddtomacro\gplfronttext{%
      \csname LTb\endcsname%
      \put(139,4550){\rotatebox{-270}{\makebox(0,0){\strut{}$\unit[A_{\text{obs}}/]{\%}$}}}%
      \csname LTb\endcsname%
      \put(5209,6419){\makebox(0,0){\strut{}$\unit[f /]{\mu Hz}$}}%
      \csname LTb\endcsname%
      \put(8815,5714){\makebox(0,0)[r]{\strut{}observations}}%
    }%
    \gplgaddtomacro\gplbacktext{%
      \csname LTb\endcsname%
      \put(640,3219){\makebox(0,0)[r]{\strut{}}}%
      \csname LTb\endcsname%
      \put(640,558){\makebox(0,0)[r]{\strut{}$0$}}%
      \csname LTb\endcsname%
      \put(640,1223){\makebox(0,0)[r]{\strut{}$0.05$}}%
      \csname LTb\endcsname%
      \put(640,1889){\makebox(0,0)[r]{\strut{}$0.1$}}%
      \csname LTb\endcsname%
      \put(640,2554){\makebox(0,0)[r]{\strut{}$0.15$}}%
      \csname LTb\endcsname%
      \put(816,298){\makebox(0,0){\strut{}$0$}}%
      \csname LTb\endcsname%
      \put(1695,298){\makebox(0,0){\strut{}$50$}}%
      \csname LTb\endcsname%
      \put(2573,298){\makebox(0,0){\strut{}$100$}}%
      \csname LTb\endcsname%
      \put(3452,298){\makebox(0,0){\strut{}$150$}}%
      \csname LTb\endcsname%
      \put(4331,298){\makebox(0,0){\strut{}$200$}}%
      \csname LTb\endcsname%
      \put(5210,298){\makebox(0,0){\strut{}$250$}}%
      \csname LTb\endcsname%
      \put(6088,298){\makebox(0,0){\strut{}$300$}}%
      \csname LTb\endcsname%
      \put(6967,298){\makebox(0,0){\strut{}$350$}}%
      \csname LTb\endcsname%
      \put(7846,298){\makebox(0,0){\strut{}$400$}}%
      \csname LTb\endcsname%
      \put(8724,298){\makebox(0,0){\strut{}$450$}}%
      \csname LTb\endcsname%
      \put(9603,298){\makebox(0,0){\strut{}$500$}}%
    }%
    \gplgaddtomacro\gplfronttext{%
      \csname LTb\endcsname%
      \put(37,1888){\rotatebox{-270}{\makebox(0,0){\strut{}$\unit[A_{\text{res}}/]{\%}$}}}%
      \csname LTb\endcsname%
      \put(5209,19){\makebox(0,0){\strut{}$\unit[f /]{d^{-1}}$}}%
      \csname LTb\endcsname%
      \put(8815,3052){\makebox(0,0)[r]{\strut{}residuals}}%
    }%
    \gplbacktext
    \put(0,0){\includegraphics[scale=0.5]{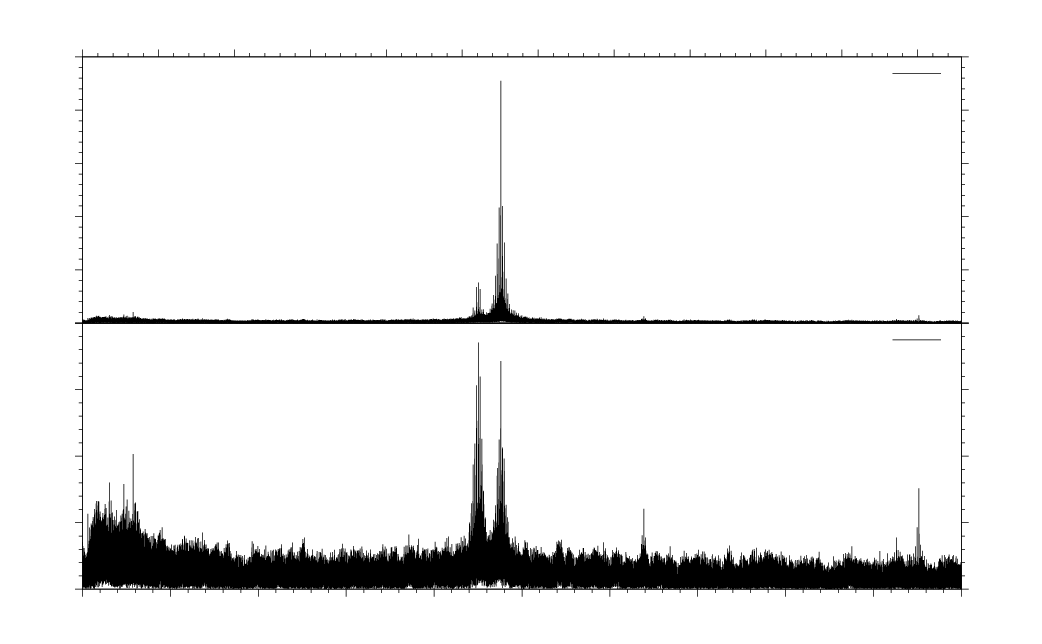}}%
    \gplfronttext
  \end{picture}%
\endgroup


        \caption{Amplitude spectrum of DW Lyn. The upper panel shows the observations $ A_{obs} $. The lower panel shows the residuals $ A_{res} $ after subtracting the light curve models from the observations.}
        \label{fig:dwlyn-power}
\end{figure*}
\begin{figure*}[h]
        \centering
\begingroup
\definecolor{r}{HTML}{e41a1c}
  \makeatletter
  \providecommand\color[2][]{%
    \GenericError{(gnuplot) \space\space\space\@spaces}{%
      Package color not loaded in conjunction with
      terminal option `colourtext'%
    }{See the gnuplot documentation for explanation.%
    }{Either use 'blacktext' in gnuplot or load the package
      color.sty in LaTeX.}%
    \renewcommand\color[2][]{}%
  }%
  \providecommand\includegraphics[2][]{%
    \GenericError{(gnuplot) \space\space\space\@spaces}{%
      Package graphicx or graphics not loaded%
    }{See the gnuplot documentation for explanation.%
    }{The gnuplot epslatex terminal needs graphicx.sty or graphics.sty.}
  }%
  \providecommand\rotatebox[2]{#2}%
  \@ifundefined{ifGPcolor}{%
    \newif\ifGPcolor
    \GPcolortrue
  }{}%
  \@ifundefined{ifGPblacktext}{%
    \newif\ifGPblacktext
    \GPblacktexttrue
  }{}%
  \let\gplgaddtomacro\g@addto@macro
  \gdef\gplbacktext{}%
  \gdef\gplfronttext{}%
  \makeatother
  \ifGPblacktext
    \def\colorrgb#1{}%
    \def\colorgray#1{}%
  \else
    \ifGPcolor
      \def\colorrgb#1{\color[rgb]{#1}}%
      \def\colorgray#1{\color[gray]{#1}}%
      \expandafter\def\csname LTw\endcsname{\color{white}}%
      \expandafter\def\csname LTb\endcsname{\color{black}}%
      \expandafter\def\csname LTa\endcsname{\color{black}}%
      \expandafter\def\csname LT0\endcsname{\color[rgb]{1,0,0}}%
      \expandafter\def\csname LT1\endcsname{\color[rgb]{0,1,0}}%
      \expandafter\def\csname LT2\endcsname{\color[rgb]{0,0,1}}%
      \expandafter\def\csname LT3\endcsname{\color[rgb]{1,0,1}}%
      \expandafter\def\csname LT4\endcsname{\color[rgb]{0,1,1}}%
      \expandafter\def\csname LT5\endcsname{\color[rgb]{1,1,0}}%
      \expandafter\def\csname LT6\endcsname{\color[rgb]{0,0,0}}%
      \expandafter\def\csname LT7\endcsname{\color[rgb]{1,0.3,0}}%
      \expandafter\def\csname LT8\endcsname{\color[rgb]{0.5,0.5,0.5}}%
    \else
      \def\colorrgb#1{\color{black}}%
      \def\colorgray#1{\color[gray]{#1}}%
      \expandafter\def\csname LTw\endcsname{\color{white}}%
      \expandafter\def\csname LTb\endcsname{\color{black}}%
      \expandafter\def\csname LTa\endcsname{\color{black}}%
      \expandafter\def\csname LT0\endcsname{\color{black}}%
      \expandafter\def\csname LT1\endcsname{\color{black}}%
      \expandafter\def\csname LT2\endcsname{\color{black}}%
      \expandafter\def\csname LT3\endcsname{\color{black}}%
      \expandafter\def\csname LT4\endcsname{\color{black}}%
      \expandafter\def\csname LT5\endcsname{\color{black}}%
      \expandafter\def\csname LT6\endcsname{\color{black}}%
      \expandafter\def\csname LT7\endcsname{\color{black}}%
      \expandafter\def\csname LT8\endcsname{\color{black}}%
    \fi
  \fi
    \setlength{\unitlength}{0.0500bp}%
    \ifx\gptboxheight\undefined%
      \newlength{\gptboxheight}%
      \newlength{\gptboxwidth}%
      \newsavebox{\gptboxtext}%
    \fi%
    \setlength{\fboxrule}{0.5pt}%
    \setlength{\fboxsep}{1pt}%
\begin{picture}(10420.00,6440.00)%
    \gplgaddtomacro\gplbacktext{%
      \csname LTb\endcsname%
      \put(640,3220){\makebox(0,0)[r]{\strut{}$0$}}%
      \csname LTb\endcsname%
      \put(640,3664){\makebox(0,0)[r]{\strut{}$0.1$}}%
      \csname LTb\endcsname%
      \put(640,4107){\makebox(0,0)[r]{\strut{}$0.2$}}%
      \csname LTb\endcsname%
      \put(640,4551){\makebox(0,0)[r]{\strut{}$0.3$}}%
      \csname LTb\endcsname%
      \put(640,4994){\makebox(0,0)[r]{\strut{}$0.4$}}%
      \csname LTb\endcsname%
      \put(640,5438){\makebox(0,0)[r]{\strut{}$0.5$}}%
      \csname LTb\endcsname%
      \put(640,5881){\makebox(0,0)[r]{\strut{}$0.6$}}%
      \csname LTb\endcsname%
      \put(816,6141){\makebox(0,0){\strut{}$0$}}%
      \csname LTb\endcsname%
      \put(1290,6141){\makebox(0,0){\strut{}$500$}}%
      \csname LTb\endcsname%
      \put(1765,6141){\makebox(0,0){\strut{}$1000$}}%
      \csname LTb\endcsname%
      \put(2239,6141){\makebox(0,0){\strut{}$1500$}}%
      \csname LTb\endcsname%
      \put(2714,6141){\makebox(0,0){\strut{}$2000$}}%
      \csname LTb\endcsname%
      \put(3188,6141){\makebox(0,0){\strut{}$2500$}}%
      \csname LTb\endcsname%
      \put(3663,6141){\makebox(0,0){\strut{}$3000$}}%
      \csname LTb\endcsname%
      \put(4137,6141){\makebox(0,0){\strut{}$3500$}}%
      \csname LTb\endcsname%
      \put(4612,6141){\makebox(0,0){\strut{}$4000$}}%
      \csname LTb\endcsname%
      \put(5086,6141){\makebox(0,0){\strut{}$4500$}}%
      \csname LTb\endcsname%
      \put(5561,6141){\makebox(0,0){\strut{}$5000$}}%
      \csname LTb\endcsname%
      \put(6035,6141){\makebox(0,0){\strut{}$5500$}}%
      \csname LTb\endcsname%
      \put(6510,6141){\makebox(0,0){\strut{}$6000$}}%
      \csname LTb\endcsname%
      \put(6984,6141){\makebox(0,0){\strut{}$6500$}}%
      \csname LTb\endcsname%
      \put(7459,6141){\makebox(0,0){\strut{}$7000$}}%
      \csname LTb\endcsname%
      \put(7933,6141){\makebox(0,0){\strut{}$7500$}}%
      \csname LTb\endcsname%
      \put(8408,6141){\makebox(0,0){\strut{}$8000$}}%
      \csname LTb\endcsname%
      \put(8882,6141){\makebox(0,0){\strut{}$8500$}}%
      \csname LTb\endcsname%
      \put(9357,6141){\makebox(0,0){\strut{}$9000$}}%
    }%
    \gplgaddtomacro\gplfronttext{%
      \csname LTb\endcsname%
      \put(139,4550){\rotatebox{-270}{\makebox(0,0){\strut{}$\unit[A_{\text{obs}}/]{\%}$}}}%
      \csname LTb\endcsname%
      \put(5209,6419){\makebox(0,0){\strut{}$\unit[f /]{\mu Hz}$}}%
      \csname LTb\endcsname%
      \put(8815,5714){\makebox(0,0)[r]{\strut{}observations}}%
    }%
    \gplgaddtomacro\gplbacktext{%
      \csname LTb\endcsname%
      \put(640,3219){\makebox(0,0)[r]{\strut{}}}%
      \csname LTb\endcsname%
      \put(640,558){\makebox(0,0)[r]{\strut{}$0$}}%
      \csname LTb\endcsname%
      \put(640,854){\makebox(0,0)[r]{\strut{}$0.05$}}%
      \csname LTb\endcsname%
      \put(640,1149){\makebox(0,0)[r]{\strut{}$0.1$}}%
      \csname LTb\endcsname%
      \put(640,1445){\makebox(0,0)[r]{\strut{}$0.15$}}%
      \csname LTb\endcsname%
      \put(640,1741){\makebox(0,0)[r]{\strut{}$0.2$}}%
      \csname LTb\endcsname%
      \put(640,2036){\makebox(0,0)[r]{\strut{}$0.25$}}%
      \csname LTb\endcsname%
      \put(640,2332){\makebox(0,0)[r]{\strut{}$0.3$}}%
      \csname LTb\endcsname%
      \put(640,2628){\makebox(0,0)[r]{\strut{}$0.35$}}%
      \csname LTb\endcsname%
      \put(640,2923){\makebox(0,0)[r]{\strut{}$0.4$}}%
      \csname LTb\endcsname%
      \put(816,298){\makebox(0,0){\strut{}$0$}}%
      \csname LTb\endcsname%
      \put(1365,298){\makebox(0,0){\strut{}$50$}}%
      \csname LTb\endcsname%
      \put(1914,298){\makebox(0,0){\strut{}$100$}}%
      \csname LTb\endcsname%
      \put(2464,298){\makebox(0,0){\strut{}$150$}}%
      \csname LTb\endcsname%
      \put(3013,298){\makebox(0,0){\strut{}$200$}}%
      \csname LTb\endcsname%
      \put(3562,298){\makebox(0,0){\strut{}$250$}}%
      \csname LTb\endcsname%
      \put(4111,298){\makebox(0,0){\strut{}$300$}}%
      \csname LTb\endcsname%
      \put(4660,298){\makebox(0,0){\strut{}$350$}}%
      \csname LTb\endcsname%
      \put(5210,298){\makebox(0,0){\strut{}$400$}}%
      \csname LTb\endcsname%
      \put(5759,298){\makebox(0,0){\strut{}$450$}}%
      \csname LTb\endcsname%
      \put(6308,298){\makebox(0,0){\strut{}$500$}}%
      \csname LTb\endcsname%
      \put(6857,298){\makebox(0,0){\strut{}$550$}}%
      \csname LTb\endcsname%
      \put(7406,298){\makebox(0,0){\strut{}$600$}}%
      \csname LTb\endcsname%
      \put(7955,298){\makebox(0,0){\strut{}$650$}}%
      \csname LTb\endcsname%
      \put(8505,298){\makebox(0,0){\strut{}$700$}}%
      \csname LTb\endcsname%
      \put(9054,298){\makebox(0,0){\strut{}$750$}}%
      \csname LTb\endcsname%
      \put(9603,298){\makebox(0,0){\strut{}$800$}}%
    }%
    \gplgaddtomacro\gplfronttext{%
      \csname LTb\endcsname%
      \put(37,1888){\rotatebox{-270}{\makebox(0,0){\strut{}$\unit[A_{\text{res}}/]{\%}$}}}%
      \csname LTb\endcsname%
      \put(5209,19){\makebox(0,0){\strut{}$\unit[f /]{d^{-1}}$}}%
      \csname LTb\endcsname%
      \put(8815,3052){\makebox(0,0)[r]{\strut{}residuals}}%
    }%
    \gplbacktext
    \put(0,0){\includegraphics[scale=0.5]{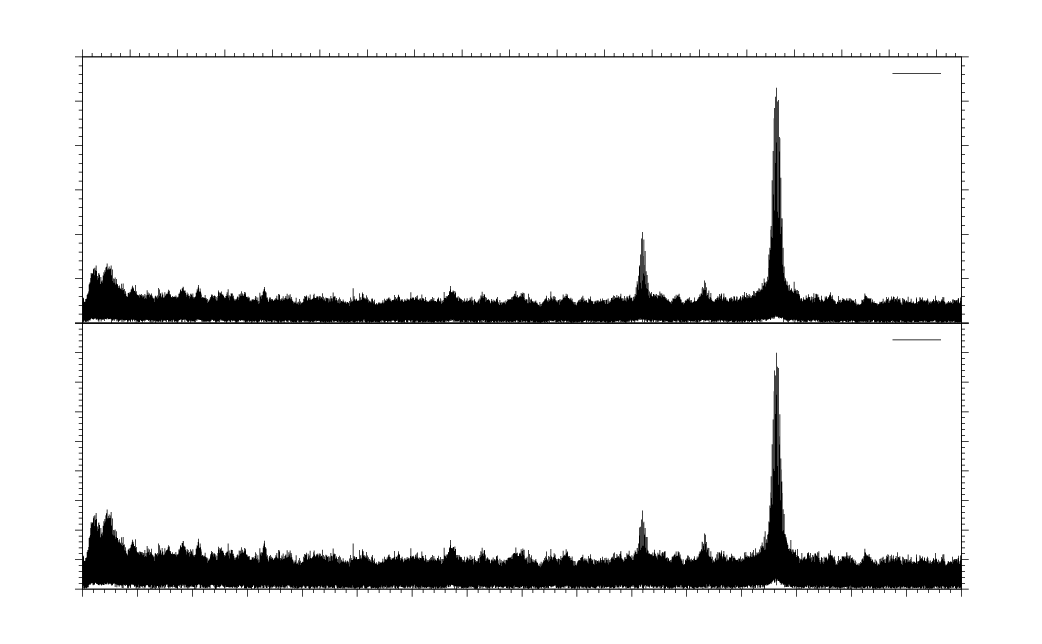}}%
    \gplfronttext
  \end{picture}%
\endgroup


        \caption{Same as Fig.~\ref{fig:dwlyn-power} but for V1636 Ori.}
        \label{fig:v1636ori-power}
\end{figure*}
\begin{figure*}[h]
        \centering
\begingroup
\definecolor{r}{HTML}{e41a1c}
  \makeatletter
  \providecommand\color[2][]{%
    \GenericError{(gnuplot) \space\space\space\@spaces}{%
      Package color not loaded in conjunction with
      terminal option `colourtext'%
    }{See the gnuplot documentation for explanation.%
    }{Either use 'blacktext' in gnuplot or load the package
      color.sty in LaTeX.}%
    \renewcommand\color[2][]{}%
  }%
  \providecommand\includegraphics[2][]{%
    \GenericError{(gnuplot) \space\space\space\@spaces}{%
      Package graphicx or graphics not loaded%
    }{See the gnuplot documentation for explanation.%
    }{The gnuplot epslatex terminal needs graphicx.sty or graphics.sty.}
  }%
  \providecommand\rotatebox[2]{#2}%
  \@ifundefined{ifGPcolor}{%
    \newif\ifGPcolor
    \GPcolortrue
  }{}%
  \@ifundefined{ifGPblacktext}{%
    \newif\ifGPblacktext
    \GPblacktexttrue
  }{}%
  \let\gplgaddtomacro\g@addto@macro
  \gdef\gplbacktext{}%
  \gdef\gplfronttext{}%
  \makeatother
  \ifGPblacktext
    \def\colorrgb#1{}%
    \def\colorgray#1{}%
  \else
    \ifGPcolor
      \def\colorrgb#1{\color[rgb]{#1}}%
      \def\colorgray#1{\color[gray]{#1}}%
      \expandafter\def\csname LTw\endcsname{\color{white}}%
      \expandafter\def\csname LTb\endcsname{\color{black}}%
      \expandafter\def\csname LTa\endcsname{\color{black}}%
      \expandafter\def\csname LT0\endcsname{\color[rgb]{1,0,0}}%
      \expandafter\def\csname LT1\endcsname{\color[rgb]{0,1,0}}%
      \expandafter\def\csname LT2\endcsname{\color[rgb]{0,0,1}}%
      \expandafter\def\csname LT3\endcsname{\color[rgb]{1,0,1}}%
      \expandafter\def\csname LT4\endcsname{\color[rgb]{0,1,1}}%
      \expandafter\def\csname LT5\endcsname{\color[rgb]{1,1,0}}%
      \expandafter\def\csname LT6\endcsname{\color[rgb]{0,0,0}}%
      \expandafter\def\csname LT7\endcsname{\color[rgb]{1,0.3,0}}%
      \expandafter\def\csname LT8\endcsname{\color[rgb]{0.5,0.5,0.5}}%
    \else
      \def\colorrgb#1{\color{black}}%
      \def\colorgray#1{\color[gray]{#1}}%
      \expandafter\def\csname LTw\endcsname{\color{white}}%
      \expandafter\def\csname LTb\endcsname{\color{black}}%
      \expandafter\def\csname LTa\endcsname{\color{black}}%
      \expandafter\def\csname LT0\endcsname{\color{black}}%
      \expandafter\def\csname LT1\endcsname{\color{black}}%
      \expandafter\def\csname LT2\endcsname{\color{black}}%
      \expandafter\def\csname LT3\endcsname{\color{black}}%
      \expandafter\def\csname LT4\endcsname{\color{black}}%
      \expandafter\def\csname LT5\endcsname{\color{black}}%
      \expandafter\def\csname LT6\endcsname{\color{black}}%
      \expandafter\def\csname LT7\endcsname{\color{black}}%
      \expandafter\def\csname LT8\endcsname{\color{black}}%
    \fi
  \fi
    \setlength{\unitlength}{0.0500bp}%
    \ifx\gptboxheight\undefined%
      \newlength{\gptboxheight}%
      \newlength{\gptboxwidth}%
      \newsavebox{\gptboxtext}%
    \fi%
    \setlength{\fboxrule}{0.5pt}%
    \setlength{\fboxsep}{1pt}%
\begin{picture}(10420.00,6440.00)%
    \gplgaddtomacro\gplbacktext{%
      \csname LTb\endcsname%
      \put(640,3220){\makebox(0,0)[r]{\strut{}$0$}}%
      \csname LTb\endcsname%
      \put(640,3752){\makebox(0,0)[r]{\strut{}$0.5$}}%
      \csname LTb\endcsname%
      \put(640,4284){\makebox(0,0)[r]{\strut{}$1$}}%
      \csname LTb\endcsname%
      \put(640,4817){\makebox(0,0)[r]{\strut{}$1.5$}}%
      \csname LTb\endcsname%
      \put(640,5349){\makebox(0,0)[r]{\strut{}$2$}}%
      \csname LTb\endcsname%
      \put(640,5881){\makebox(0,0)[r]{\strut{}$2.5$}}%
      \csname LTb\endcsname%
      \put(816,6141){\makebox(0,0){\strut{}$0$}}%
      \csname LTb\endcsname%
      \put(1290,6141){\makebox(0,0){\strut{}$500$}}%
      \csname LTb\endcsname%
      \put(1765,6141){\makebox(0,0){\strut{}$1000$}}%
      \csname LTb\endcsname%
      \put(2239,6141){\makebox(0,0){\strut{}$1500$}}%
      \csname LTb\endcsname%
      \put(2714,6141){\makebox(0,0){\strut{}$2000$}}%
      \csname LTb\endcsname%
      \put(3188,6141){\makebox(0,0){\strut{}$2500$}}%
      \csname LTb\endcsname%
      \put(3663,6141){\makebox(0,0){\strut{}$3000$}}%
      \csname LTb\endcsname%
      \put(4137,6141){\makebox(0,0){\strut{}$3500$}}%
      \csname LTb\endcsname%
      \put(4612,6141){\makebox(0,0){\strut{}$4000$}}%
      \csname LTb\endcsname%
      \put(5086,6141){\makebox(0,0){\strut{}$4500$}}%
      \csname LTb\endcsname%
      \put(5561,6141){\makebox(0,0){\strut{}$5000$}}%
      \csname LTb\endcsname%
      \put(6035,6141){\makebox(0,0){\strut{}$5500$}}%
      \csname LTb\endcsname%
      \put(6510,6141){\makebox(0,0){\strut{}$6000$}}%
      \csname LTb\endcsname%
      \put(6984,6141){\makebox(0,0){\strut{}$6500$}}%
      \csname LTb\endcsname%
      \put(7459,6141){\makebox(0,0){\strut{}$7000$}}%
      \csname LTb\endcsname%
      \put(7933,6141){\makebox(0,0){\strut{}$7500$}}%
      \csname LTb\endcsname%
      \put(8408,6141){\makebox(0,0){\strut{}$8000$}}%
      \csname LTb\endcsname%
      \put(8882,6141){\makebox(0,0){\strut{}$8500$}}%
      \csname LTb\endcsname%
      \put(9357,6141){\makebox(0,0){\strut{}$9000$}}%
    }%
    \gplgaddtomacro\gplfronttext{%
      \csname LTb\endcsname%
      \put(139,4550){\rotatebox{-270}{\makebox(0,0){\strut{}$\unit[A_{\text{obs}}/]{\%}$}}}%
      \csname LTb\endcsname%
      \put(5209,6419){\makebox(0,0){\strut{}$\unit[f /]{\mu Hz}$}}%
      \csname LTb\endcsname%
      \put(8815,5714){\makebox(0,0)[r]{\strut{}observations}}%
    }%
    \gplgaddtomacro\gplbacktext{%
      \csname LTb\endcsname%
      \put(640,3219){\makebox(0,0)[r]{\strut{}}}%
      \csname LTb\endcsname%
      \put(640,558){\makebox(0,0)[r]{\strut{}$0$}}%
      \csname LTb\endcsname%
      \put(640,854){\makebox(0,0)[r]{\strut{}$0.1$}}%
      \csname LTb\endcsname%
      \put(640,1149){\makebox(0,0)[r]{\strut{}$0.2$}}%
      \csname LTb\endcsname%
      \put(640,1445){\makebox(0,0)[r]{\strut{}$0.3$}}%
      \csname LTb\endcsname%
      \put(640,1741){\makebox(0,0)[r]{\strut{}$0.4$}}%
      \csname LTb\endcsname%
      \put(640,2036){\makebox(0,0)[r]{\strut{}$0.5$}}%
      \csname LTb\endcsname%
      \put(640,2332){\makebox(0,0)[r]{\strut{}$0.6$}}%
      \csname LTb\endcsname%
      \put(640,2628){\makebox(0,0)[r]{\strut{}$0.7$}}%
      \csname LTb\endcsname%
      \put(640,2923){\makebox(0,0)[r]{\strut{}$0.8$}}%
      \csname LTb\endcsname%
      \put(816,298){\makebox(0,0){\strut{}$0$}}%
      \csname LTb\endcsname%
      \put(1365,298){\makebox(0,0){\strut{}$50$}}%
      \csname LTb\endcsname%
      \put(1914,298){\makebox(0,0){\strut{}$100$}}%
      \csname LTb\endcsname%
      \put(2464,298){\makebox(0,0){\strut{}$150$}}%
      \csname LTb\endcsname%
      \put(3013,298){\makebox(0,0){\strut{}$200$}}%
      \csname LTb\endcsname%
      \put(3562,298){\makebox(0,0){\strut{}$250$}}%
      \csname LTb\endcsname%
      \put(4111,298){\makebox(0,0){\strut{}$300$}}%
      \csname LTb\endcsname%
      \put(4660,298){\makebox(0,0){\strut{}$350$}}%
      \csname LTb\endcsname%
      \put(5210,298){\makebox(0,0){\strut{}$400$}}%
      \csname LTb\endcsname%
      \put(5759,298){\makebox(0,0){\strut{}$450$}}%
      \csname LTb\endcsname%
      \put(6308,298){\makebox(0,0){\strut{}$500$}}%
      \csname LTb\endcsname%
      \put(6857,298){\makebox(0,0){\strut{}$550$}}%
      \csname LTb\endcsname%
      \put(7406,298){\makebox(0,0){\strut{}$600$}}%
      \csname LTb\endcsname%
      \put(7955,298){\makebox(0,0){\strut{}$650$}}%
      \csname LTb\endcsname%
      \put(8505,298){\makebox(0,0){\strut{}$700$}}%
      \csname LTb\endcsname%
      \put(9054,298){\makebox(0,0){\strut{}$750$}}%
      \csname LTb\endcsname%
      \put(9603,298){\makebox(0,0){\strut{}$800$}}%
    }%
    \gplgaddtomacro\gplfronttext{%
      \csname LTb\endcsname%
      \put(139,1888){\rotatebox{-270}{\makebox(0,0){\strut{}$\unit[A_{\text{res}}/]{\%}$}}}%
      \csname LTb\endcsname%
      \put(5209,19){\makebox(0,0){\strut{}$\unit[f /]{d^{-1}}$}}%
      \csname LTb\endcsname%
      \put(8815,3052){\makebox(0,0)[r]{\strut{}residuals}}%
    }%
    \gplbacktext
    \put(0,0){\includegraphics[scale=0.5]{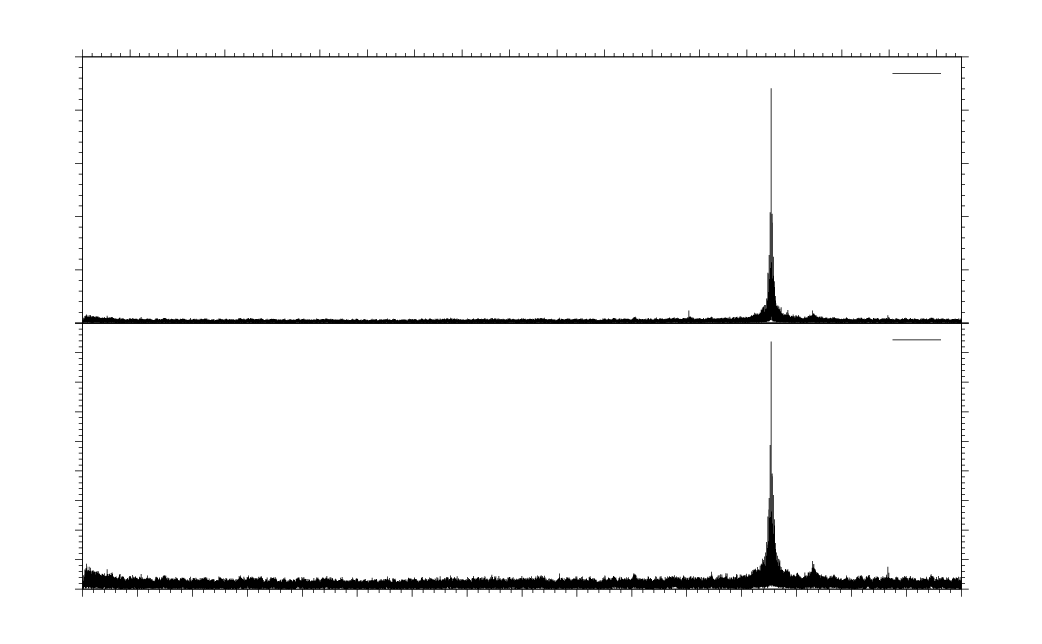}}%
    \gplfronttext
  \end{picture}%
\endgroup


        \caption{Same as Fig.~\ref{fig:dwlyn-power} but for QQ Vir.}
        \label{fig:qqvir-power}
\end{figure*}
\begin{figure*}[h]
        \centering
\begingroup
\definecolor{r}{HTML}{e41a1c}
  \makeatletter
  \providecommand\color[2][]{%
    \GenericError{(gnuplot) \space\space\space\@spaces}{%
      Package color not loaded in conjunction with
      terminal option `colourtext'%
    }{See the gnuplot documentation for explanation.%
    }{Either use 'blacktext' in gnuplot or load the package
      color.sty in LaTeX.}%
    \renewcommand\color[2][]{}%
  }%
  \providecommand\includegraphics[2][]{%
    \GenericError{(gnuplot) \space\space\space\@spaces}{%
      Package graphicx or graphics not loaded%
    }{See the gnuplot documentation for explanation.%
    }{The gnuplot epslatex terminal needs graphicx.sty or graphics.sty.}
  }%
  \providecommand\rotatebox[2]{#2}%
  \@ifundefined{ifGPcolor}{%
    \newif\ifGPcolor
    \GPcolortrue
  }{}%
  \@ifundefined{ifGPblacktext}{%
    \newif\ifGPblacktext
    \GPblacktexttrue
  }{}%
  \let\gplgaddtomacro\g@addto@macro
  \gdef\gplbacktext{}%
  \gdef\gplfronttext{}%
  \makeatother
  \ifGPblacktext
    \def\colorrgb#1{}%
    \def\colorgray#1{}%
  \else
    \ifGPcolor
      \def\colorrgb#1{\color[rgb]{#1}}%
      \def\colorgray#1{\color[gray]{#1}}%
      \expandafter\def\csname LTw\endcsname{\color{white}}%
      \expandafter\def\csname LTb\endcsname{\color{black}}%
      \expandafter\def\csname LTa\endcsname{\color{black}}%
      \expandafter\def\csname LT0\endcsname{\color[rgb]{1,0,0}}%
      \expandafter\def\csname LT1\endcsname{\color[rgb]{0,1,0}}%
      \expandafter\def\csname LT2\endcsname{\color[rgb]{0,0,1}}%
      \expandafter\def\csname LT3\endcsname{\color[rgb]{1,0,1}}%
      \expandafter\def\csname LT4\endcsname{\color[rgb]{0,1,1}}%
      \expandafter\def\csname LT5\endcsname{\color[rgb]{1,1,0}}%
      \expandafter\def\csname LT6\endcsname{\color[rgb]{0,0,0}}%
      \expandafter\def\csname LT7\endcsname{\color[rgb]{1,0.3,0}}%
      \expandafter\def\csname LT8\endcsname{\color[rgb]{0.5,0.5,0.5}}%
    \else
      \def\colorrgb#1{\color{black}}%
      \def\colorgray#1{\color[gray]{#1}}%
      \expandafter\def\csname LTw\endcsname{\color{white}}%
      \expandafter\def\csname LTb\endcsname{\color{black}}%
      \expandafter\def\csname LTa\endcsname{\color{black}}%
      \expandafter\def\csname LT0\endcsname{\color{black}}%
      \expandafter\def\csname LT1\endcsname{\color{black}}%
      \expandafter\def\csname LT2\endcsname{\color{black}}%
      \expandafter\def\csname LT3\endcsname{\color{black}}%
      \expandafter\def\csname LT4\endcsname{\color{black}}%
      \expandafter\def\csname LT5\endcsname{\color{black}}%
      \expandafter\def\csname LT6\endcsname{\color{black}}%
      \expandafter\def\csname LT7\endcsname{\color{black}}%
      \expandafter\def\csname LT8\endcsname{\color{black}}%
    \fi
  \fi
    \setlength{\unitlength}{0.0500bp}%
    \ifx\gptboxheight\undefined%
      \newlength{\gptboxheight}%
      \newlength{\gptboxwidth}%
      \newsavebox{\gptboxtext}%
    \fi%
    \setlength{\fboxrule}{0.5pt}%
    \setlength{\fboxsep}{1pt}%
\begin{picture}(10420.00,6440.00)%
    \gplgaddtomacro\gplbacktext{%
      \csname LTb\endcsname%
      \put(640,3220){\makebox(0,0)[r]{\strut{}$0$}}%
      \csname LTb\endcsname%
      \put(640,3664){\makebox(0,0)[r]{\strut{}$0.05$}}%
      \csname LTb\endcsname%
      \put(640,4107){\makebox(0,0)[r]{\strut{}$0.1$}}%
      \csname LTb\endcsname%
      \put(640,4551){\makebox(0,0)[r]{\strut{}$0.15$}}%
      \csname LTb\endcsname%
      \put(640,4994){\makebox(0,0)[r]{\strut{}$0.2$}}%
      \csname LTb\endcsname%
      \put(640,5438){\makebox(0,0)[r]{\strut{}$0.25$}}%
      \csname LTb\endcsname%
      \put(640,5881){\makebox(0,0)[r]{\strut{}$0.3$}}%
      \csname LTb\endcsname%
      \put(816,6141){\makebox(0,0){\strut{}$0$}}%
      \csname LTb\endcsname%
      \put(1290,6141){\makebox(0,0){\strut{}$500$}}%
      \csname LTb\endcsname%
      \put(1765,6141){\makebox(0,0){\strut{}$1000$}}%
      \csname LTb\endcsname%
      \put(2239,6141){\makebox(0,0){\strut{}$1500$}}%
      \csname LTb\endcsname%
      \put(2714,6141){\makebox(0,0){\strut{}$2000$}}%
      \csname LTb\endcsname%
      \put(3188,6141){\makebox(0,0){\strut{}$2500$}}%
      \csname LTb\endcsname%
      \put(3663,6141){\makebox(0,0){\strut{}$3000$}}%
      \csname LTb\endcsname%
      \put(4137,6141){\makebox(0,0){\strut{}$3500$}}%
      \csname LTb\endcsname%
      \put(4612,6141){\makebox(0,0){\strut{}$4000$}}%
      \csname LTb\endcsname%
      \put(5086,6141){\makebox(0,0){\strut{}$4500$}}%
      \csname LTb\endcsname%
      \put(5561,6141){\makebox(0,0){\strut{}$5000$}}%
      \csname LTb\endcsname%
      \put(6035,6141){\makebox(0,0){\strut{}$5500$}}%
      \csname LTb\endcsname%
      \put(6510,6141){\makebox(0,0){\strut{}$6000$}}%
      \csname LTb\endcsname%
      \put(6984,6141){\makebox(0,0){\strut{}$6500$}}%
      \csname LTb\endcsname%
      \put(7459,6141){\makebox(0,0){\strut{}$7000$}}%
      \csname LTb\endcsname%
      \put(7933,6141){\makebox(0,0){\strut{}$7500$}}%
      \csname LTb\endcsname%
      \put(8408,6141){\makebox(0,0){\strut{}$8000$}}%
      \csname LTb\endcsname%
      \put(8882,6141){\makebox(0,0){\strut{}$8500$}}%
      \csname LTb\endcsname%
      \put(9357,6141){\makebox(0,0){\strut{}$9000$}}%
    }%
    \gplgaddtomacro\gplfronttext{%
      \csname LTb\endcsname%
      \put(37,4550){\rotatebox{-270}{\makebox(0,0){\strut{}$\unit[A_{\text{obs}}/]{\%}$}}}%
      \csname LTb\endcsname%
      \put(5209,6419){\makebox(0,0){\strut{}$\unit[f /]{\mu Hz}$}}%
      \csname LTb\endcsname%
      \put(8815,5714){\makebox(0,0)[r]{\strut{}observations}}%
    }%
    \gplgaddtomacro\gplbacktext{%
      \csname LTb\endcsname%
      \put(640,3219){\makebox(0,0)[r]{\strut{}}}%
      \csname LTb\endcsname%
      \put(640,558){\makebox(0,0)[r]{\strut{}$0$}}%
      \csname LTb\endcsname%
      \put(640,1223){\makebox(0,0)[r]{\strut{}$0.05$}}%
      \csname LTb\endcsname%
      \put(640,1889){\makebox(0,0)[r]{\strut{}$0.1$}}%
      \csname LTb\endcsname%
      \put(640,2554){\makebox(0,0)[r]{\strut{}$0.15$}}%
      \csname LTb\endcsname%
      \put(816,298){\makebox(0,0){\strut{}$0$}}%
      \csname LTb\endcsname%
      \put(1365,298){\makebox(0,0){\strut{}$50$}}%
      \csname LTb\endcsname%
      \put(1914,298){\makebox(0,0){\strut{}$100$}}%
      \csname LTb\endcsname%
      \put(2464,298){\makebox(0,0){\strut{}$150$}}%
      \csname LTb\endcsname%
      \put(3013,298){\makebox(0,0){\strut{}$200$}}%
      \csname LTb\endcsname%
      \put(3562,298){\makebox(0,0){\strut{}$250$}}%
      \csname LTb\endcsname%
      \put(4111,298){\makebox(0,0){\strut{}$300$}}%
      \csname LTb\endcsname%
      \put(4660,298){\makebox(0,0){\strut{}$350$}}%
      \csname LTb\endcsname%
      \put(5210,298){\makebox(0,0){\strut{}$400$}}%
      \csname LTb\endcsname%
      \put(5759,298){\makebox(0,0){\strut{}$450$}}%
      \csname LTb\endcsname%
      \put(6308,298){\makebox(0,0){\strut{}$500$}}%
      \csname LTb\endcsname%
      \put(6857,298){\makebox(0,0){\strut{}$550$}}%
      \csname LTb\endcsname%
      \put(7406,298){\makebox(0,0){\strut{}$600$}}%
      \csname LTb\endcsname%
      \put(7955,298){\makebox(0,0){\strut{}$650$}}%
      \csname LTb\endcsname%
      \put(8505,298){\makebox(0,0){\strut{}$700$}}%
      \csname LTb\endcsname%
      \put(9054,298){\makebox(0,0){\strut{}$750$}}%
      \csname LTb\endcsname%
      \put(9603,298){\makebox(0,0){\strut{}$800$}}%
    }%
    \gplgaddtomacro\gplfronttext{%
      \csname LTb\endcsname%
      \put(37,1888){\rotatebox{-270}{\makebox(0,0){\strut{}$\unit[A_{\text{res}}/]{\%}$}}}%
      \csname LTb\endcsname%
      \put(5209,19){\makebox(0,0){\strut{}$\unit[f /]{d^{-1}}$}}%
      \csname LTb\endcsname%
      \put(8815,3052){\makebox(0,0)[r]{\strut{}residuals}}%
    }%
    \gplbacktext
    \put(0,0){\includegraphics[scale=0.5]{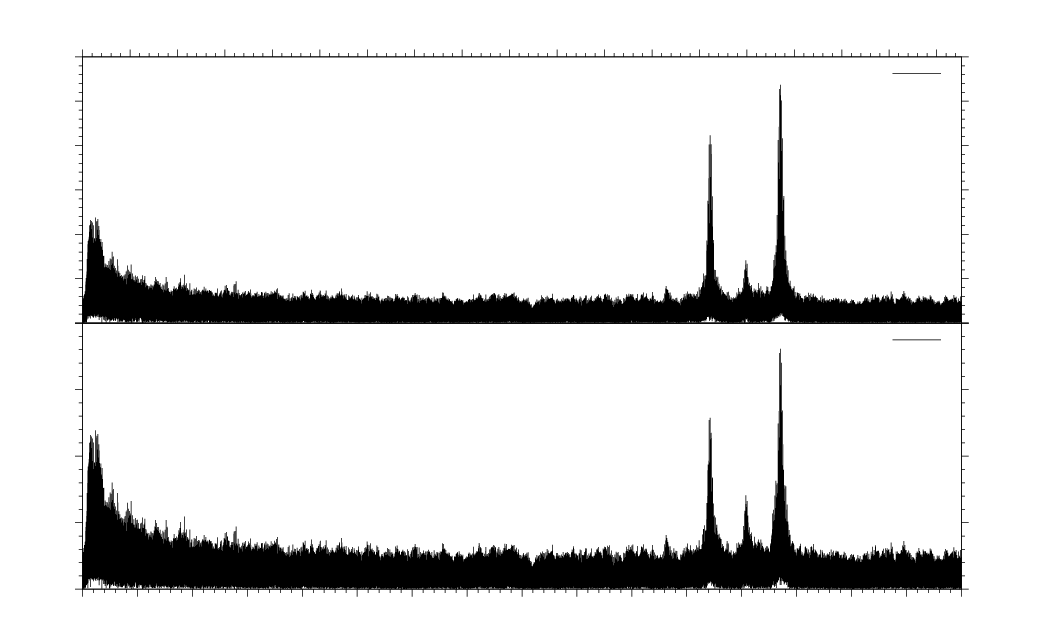}}%
    \gplfronttext
  \end{picture}%
\endgroup


        \caption{Same as Fig.~\ref{fig:dwlyn-power} but for V541 Hya.}
        \label{fig:v541hya-power}
\end{figure*}

\end{appendix}


\end{document}